\documentclass[12pt]{article}

\usepackage{amsmath,amssymb,graphicx,multirow,xspace}
\usepackage[colorlinks=true,urlcolor=blue,anchorcolor=blue,citecolor=blue,filecolor=blue,linkcolor=blue,menucolor=blue,pagecolor=blue]{hyperref}
\usepackage[compress,numbers]{natbib}
\usepackage{subfigure, placeins}
\usepackage{booktabs}
\usepackage{blindtext, rotating}
\usepackage{afterpage}
\usepackage{enumitem}
\usepackage{ marvosym }
\usepackage{verbatim}
\usepackage{authblk}

\makeatletter
\renewcommand{\@dotsep}{1000}
\makeatother

\newcommand{\captionfonts}{\small}
\makeatletter
\long\def\@makecaption#1#2{%
  \vskip\abovecaptionskip
  \sbox\@tempboxa{{\captionfonts #1: #2}}%
  \ifdim \wd\@tempboxa >\hsize
    {\captionfonts #1: #2\par}
  \else
    \hbox to\hsize{\hfil\box\@tempboxa\hfil}%
  \fi
  \vskip\belowcaptionskip}
\makeatother

\newcommand{\Sla}[1]%
{\kern0.12em{\raise.15ex\hbox{$/$}\kern-.74em #1}}

\long\def\symbolfootnote[#1]#2{\begingroup%
\def\thefootnote{\fnsymbol{footnote}}\footnote[#1]{#2}\endgroup}

\newcommand{\be}{\begin{equation}}
\newcommand{\ee}{\end{equation}}

\newcommand{\bea}{\begin{eqnarray}}
\newcommand{\eea}{\end{eqnarray}}
\newcommand{\mat}{\begin{pmatrix}}
\newcommand{\rix}{\end{pmatrix}}

\renewcommand{\bar}{\overline}
\renewcommand{\slash}[1]{#1\!\!\!/}

\newcommand{\go}{{\tilde g}}

\newcommand{\Ho}{{\tilde H}}

\newcommand{\cho}{{\tilde \chi}}

\newcommand{\st}{{\tilde t}}

\newcommand{\snu}{{\tilde\nu}}
\newcommand{\stau}{{\tilde\tau}}

\newcommand{\qq}{\qquad}

\newcommand{\beqa}{\begin{eqnarray}}
\newcommand{\eeqa}{\end{eqnarray}}
\newcommand{\lp}{\left(}
\newcommand{\rp}{\right)}

\newcommand{\veva}[1]{\left< #1\right>}

\newcommand{\beq}{\begin{equation}}
\newcommand{\eeq}{\end{equation}}
\newcommand{\hc}{\mbox{h.c.}}

\newcommand{\abs}[1]{\left\vert#1\right\vert}

\newcommand{\order}[1]{{\cal O}\left(#1\right)}

\newcommand{\met}{{\slash E_T}}

\newcommand{\search}{CMS displaced $e\mu$}
\newcommand{\thesearch}{the CMS displaced $e\mu$ search}

\begin{document}

\title{Long-Lived Staus and Displaced Leptons at the LHC}

\date{\today}

\author{Jared A.~Evans}
\author{Jessie Shelton}

\affil{\small{Department  of Physics\\ 
University of Illinois at Urbana-Champaign\\ 
Urbana, IL 61801}}

\maketitle

\begin{abstract}

  As the majority of LHC searches are focused on prompt signatures,
  specific long-lived particles have the potential to be overlooked by
  the otherwise systematic new physics programs at ATLAS and CMS.
  While in many cases long-lived superparticles are now stringently 
  constrained by existing exotic searches, we point out that the highly 
  motivated model of gauge mediation with staus as the
  next-to-lightest superparticle (NLSP) is relatively far less tested.  
We recast LHC searches for heavy
  stable charged particles, disappearing tracks, and opposite-flavor
  leptons with large impact parameters to assess current constraints
  on a variety of spectra that contain an NLSP stau, and find that 
  portions of the parameter space motivated by naturalness are still 
  experimentally unexplored.    We additionally note a gap in the current 
  experimental search program: same-flavor leptons with large impact
  parameters evade the suite of existing searches for long-lived objects.  This gap
  is especially noteworthy as vetoes on displaced leptons in prompt
  new physics searches could be systematically discarding such
  events.  We discuss several motivated models that can exhibit same-flavor displaced leptons:
  gauge mediation with co-NLSP sleptons, extended gauge
  mediation, R-parity violation, and lepton-flavored dark matter that
  freezes in during a matter-dominated era of the early universe.  To
  address this gap, we propose a straightforward extension of the CMS
  search for leptons with large impact parameters, and project
  sensitivity to these scenarios at 13 TeV.  Throughout this analysis, we
  highlight several methods whereby  LHC searches for exotic long-lived
  objects could potentially improve their sensitivity to the displaced
  leptons originating from gauge mediation and beyond.

\end{abstract}

\newpage

\tableofcontents

\section{Introduction}
\label{sec:intro}

Run I at the LHC has been a phenomenal success.  However, despite
pressure from naturalness for new electroweak scale particles, no such
particles have been observed yet by the LHC experiments (see e.g.,
\cite{Halkiadakis:2014qda}).  Weak-scale supersymmetry (SUSY), the
long-standing front-runner to explain the stability of the electroweak
scale, has been subjected to more and more stringent constraints as
time progresses.  The parameter spaces for natural gluinos, stops, and
even electroweakinos are now highly constrained across a wide variety
of spectra.  The current absence of clear signals of new physics
confronts us with three logical possibilities concerning weak-scale
superpartners: (i) they are not there to be found; (ii) they are just
around the corner (and will hopefully be seen at Run II); or (iii)
they are hidden somehow from the existing array of searches.  A
variety of mechanisms have been devised over the years in order to
hide supersymmetry from collider searches, but natural spectra in the
majority of these scenarios are now under pressure from the
comprehensive search programs at ATLAS and CMS \cite{Evans:2015caa}.
Even mechanisms such as $R$-parity violation \cite{Barbier:2004ez} or
stealth \cite{Fan:2011yu} that can hide SUSY from traditional
$\met$-based searches are already significantly at odds with the
signals expected from natural Majorana gluinos \cite{Evans:2013jna}, 
although counter examples exist \cite{Evans:2013jna,Fan:2015mxp}.

New particles with macroscopic decay lengths, as can easily arise in
gauge-mediated SUSY breaking (GMSB) \cite{Giudice:1998bp}, $R$-parity
violating (RPV) SUSY, mini-split SUSY \cite{Arvanitaki:2012ps}, and other
models (e.g.,~\cite{Gupta:2007ui}), can potentially elude the selection criteria for the standard
suite of collider searches, but, thanks to dedicated searches for
long-lived objects, are now often more constrained than their prompt
counterparts \cite{Liu:2015bma,Csaki:2015uza,Zwane:2015bra}.  However,
models predicting \emph{solitary} displaced leptons
are less thoroughly constrained.  Solitary displaced leptons are 
leptons that originate from a displaced vertex that has no other 
visible decay products, i.e., from a long-lived particle that decays to an 
invisible particle and a lepton that does not point back to the
primary vertex.  While displaced leptons figure in many existing
searches, almost all of these searches look for a vertex containing a
displaced lepton together with at least one other object, such as
another lepton \cite{Aad:2014yea, CMS:2014hka,Aad:2015rba} or $\geq 4$
additional tracks \cite{Aad:2012zx,Aad:2015rba}; the only existing
search sensitive to solitary displaced
leptons is the CMS search for a displaced opposite-sign $e$-$\mu$ pair
\cite{Khachatryan:2014mea}.  Models predicting solitary 
 displaced leptons can be surprisingly invisible to current
searches, as lepton quality requirements in most prompt searches veto
leptons with impact parameters down to just a few hundred microns, and
often discard entire events with cosmic-ray-motivated vetoes on muons
with large impact parameters.
  
Solitary displaced leptons arise in many theories.  Perhaps the
best-motivated examples arise from theories of GMSB, which
frequently predict spectra where the right-handed stau is the
next-to-lightest superpartner (NLSP).
Although minimal models of GMSB have difficulty accommodating the
heavy Higgs mass of 125 GeV \cite{Aad:2012tfa,Chatrchyan:2012ufa}
without introducing large tuning \cite{Draper:2011aa}, extensions to
GMSB can readily account for this while approaching the minimal fine
tuning possible in the MSSM, see e.g.,
\cite{Craig:2012xp,Abdullah:2012tq,Byakti:2013ti,Evans:2013kxa,Calibbi:2013mka,Knapen:2013zla,Ding:2013pya,Basirnia:2015vga}.
Adding new fields to the MSSM can also raise $m_h$, either by GMSB-specific mechanisms, e.g.,
\cite{Fischler:2013tva,Liu:2013vaa,Allanach:2015cia,Delgado:2015bwa},
or with modular modifications to the SUSY Higgs sector, such as
non-decoupling $D$-terms
\cite{Batra:2003nj,Cheung:2012zq,Craig:2012bs,Lu:2013cta,McGarrie:2014xxa,Bertuzzo:2014sma}.
The lifetime of a slepton NLSP decaying via $\tilde \ell \to
\ell\tilde G$ can be expressed as
\beq
  c \tau\approx  100 \,\mu\mbox{m} \lp \frac{100\mbox{ GeV}}{m_\stau} \rp^5  \lp \frac{\sqrt F}{100\mbox{ TeV}} \rp^4.
  \label{eq:GMSB4life}
\eeq 
Even for a relatively low SUSY-breaking scale $\sqrt F\sim 100$ TeV,
the decay lengths of such sleptons border on being displaced at the
LHC.  For high SUSY-breaking scales, these sleptons become
detector-stable and fall under the purview of searches for heavy
stable charged particles (HSCPs).  In the intermediate regime, which
spans roughly four orders of magnitude in lifetime ($c\tau \sim 100
\,\mu$m\,--\,1m), where the slepton lives too briefly to survive the
detector, but long enough to be vetoed in many standard prompt SUSY
searches, the resulting signature is opposite-sign taus originating a
macroscopic distance away from the primary interaction point.  If the
lifetime of the slepton is sufficiently long (at the LHC,
$c\tau\sim\order{50\,\mbox{cm}}$), it is possible to search directly
for the track left by the slepton.  In the eyes of the tracking
algorithm, the slepton will disappear if it decays before reaching the
calorimeter, making the primary signal of a stau in this lifetime
range a kinked or disappearing track.  For shorter slepton
lifetimes ($100\,\mu\mathrm{m}\lesssim c\tau\lesssim
5\,\mbox{cm}$), it is the displaced daughter leptons that drive
search sensitivity.
At LEP2, dedicated long-lived slepton searches covered this entire range
of signatures, where OPAL \cite{Abbiendi:2005gc} sets the best
limits.  At the LHC, HSCP searches
\cite{Chatrchyan:2013oca,ATLAS:2014fka,Heisig:2012zq}, searches for
disappearing tracks \cite{Aad:2013yna,CMS:2014gxa}, and the CMS search
for displaced $e^\pm\mu^\mp$ \cite{Khachatryan:2014mea} (henceforth,
``\search'') together target the range of displaced slepton
signatures, but in general do not specifically target sleptons, and
are not optimized for them.

Moreover, several classes of theories can give rise exclusively to
displaced {\em same-flavor} leptons. In the context of SUSY, this
signature can arise with RPV couplings or in extended gauge mediation.
Outside of SUSY, lepton-flavored dark matter can provide an elegant
mechanism to produce such signatures.  All of these models can yield
displaced signatures that are currently not covered by the existing
array of LHC searches, highlighting an outstanding gap in the search
coverage for new physics. Additionally, as we will demonstrate, a
same-flavor displaced lepton search would significantly improve
sensitivity to long-lived staus as well as theories with long-lived
slepton co-NLSPs
\cite{Dimopoulos:1996vz,Ambrosanio:1997rv,Ruderman:2010kj}.

The aim of this paper is twofold.  In Section~\ref{sec:staus}, we will
establish existing collider constraints on long-lived staus by
recasting \thesearch, the disappearing track searches at ATLAS and
CMS, and a heavy stable charged particle search, thus obtaining a
clear picture of current sensitivity to displaced decays of a stau
NLSP.  In addition to direct stau production, we also consider
production in decay chains originating from gluinos, stops, and
Higgsinos.  In the course of this endeavor, we will discuss several
possible modifications to the existing searches that could enhance
sensitivity to displaced staus.  Next, we discuss several concrete
models in Section~\ref{sec:SFmodels} that give rise to same-flavor
displaced lepton signatures.  We propose a simple extension to the
existing \search\ search strategy to close the gap in LHC searches for
displaced same-flavor leptons, and estimate the resulting sensitivity
at 13 TeV for several models in Section~\ref{sec:SFsearch}.
Extensions and modifications of existing search strategies that could
potentially enhance sensitivity to displaced stau decays in particular
and solitary displaced lepton signatures in general are
summarized in the conclusions.

\section{LHC Sensitivity to Long-Lived Staus}
\label{sec:staus}

One of the most common predictions in models of GMSB is that the
$\stau_R$ is the NLSP (with the gravitino as the LSP).  Even at
relatively low SUSY-breaking scales, the tiny width for $\stau_R \to
\tau \tilde G$ can result in displaced leptons, as shown in
(\ref{eq:GMSB4life}).  The best LEP2 limits on direct NLSP stau
pair production come from OPAL, and exclude $\stau$ $(\tilde \mu)$
NLSPs below 87 GeV (94 GeV) and become more stringent for longer
lifetimes (up to 97 GeV for both particles) \cite{Abbiendi:2005gc}.
Depending on the stau lifetime, the resulting collider signatures may
yield:
\begin{itemize}
\item {\em Opposite-sign solitary displaced leptons.}
  A lepton's displacement is characterized by its impact parameter,
  which is typically defined as the minimum three-dimensional distance
  from the lepton track to the primary vertex, although in
  \thesearch~\cite{Khachatryan:2014mea} a two-dimensional impact
  parameter is used with respect to the center of the beampipe.  The
  \search\ search is the only existing LHC analysis with sensitivity
  to solitary displaced leptons.

\item {\em Disappearing tracks.} At longer lifetimes $c\tau
  \sim\order{50\,\mbox{cm}}$, the $\stau$ will have left a
  reconstructable, short, high-$p_T$ track in the tracker.  This
  places the stau in the territory of the disappearing track searches
  \cite{Aad:2013yna,CMS:2014gxa}.  The signature in this range is
  really a kinked track, as was directly searched for at LEP
  \cite{Abbiendi:2005gc}. In the busier environment offered by the
  LHC, however, the track associated with the daughter lepton is
  typically not reconstructed or may not be associated with the parent
  slepton track, and triggering on a kinked track is nearly impossible.
  The LHC disappearing track searches are only sensitive to the
  sleptons' visible decay products if they leave significant
  calorimeter deposits or make tracks in the muon chamber.  
  
\item {\em HSCPs.} Even longer lifetimes yield detector-stable charged
  particles, which have been directly searched for in
  Refs.~\cite{Chatrchyan:2013oca,ATLAS:2014fka}. While we expect the
  ATLAS and CMS HSCP searches to have similar sensitivity, we choose
  to recast the CMS search, as detailed efficiency maps to facilitate
  recasting are provided \cite{Khachatryan:2015lla}.\footnote{Searches
    targeting particles that decay in the calorimeters or the muon
    system either explicitly veto events where a charged track in the
    inner detector points to the displaced decay or require multiple
    charged tracks at a displaced vertex, and thus are not sensitive
    to this class of signals \cite{Aad:2015asa, Aad:2015uaa}.}

\end {itemize}

The main objective of this section is to establish the current LHC
coverage of long-lived staus 
from the various search strategies discussed above.  In addition to
direct stau production, we will consider staus produced in cascade
decays originating from gluinos, stops, or Higgsinos.  All of the
benchmarks we consider are generated in Madgraph 5
\cite{Alwall:2011uj} (using the TauDecay package
\cite{Hagiwara:2012vz}) and showered in Pythia 8
\cite{Sjostrand:2007gs}.  Hadrons are clustered according to the jet
algorithms of the individual searches, and a simple jet energy
smearing with resolution of $\sigma_E = 0.05 \sqrt{E(\mbox{GeV})}$
\cite{Khachatryan:2014gga} is applied.  The signal production
cross-sections are fixed to the nominal NLO+NLL value as provided by
Ref.~\cite{Kramer:2012bx} or as computed in Prospino2
\cite{Beenakker:1996ed} or Resummino \cite{Fuks:2013vua}.  In the
following subsections, we will describe the relevant searches in some
detail with emphasis on our recasting procedure (for details of
validation, see Appendix~\ref{sec:val}). The resulting constraints on
spectra with a long-lived $\stau_R$ NLSP are collected in
Section~\ref{sec:GMSBstaus}.

\subsection{CMS Heavy Stable Charged Particle Search}
\label{sec:CMSHSCP}

The CMS HSCP search \cite{Chatrchyan:2013oca} looks at a variety of
models containing heavy charged particles that survive the detector.
Of the various sub-analyses employed in this search, the most
pertinent for long-lived staus is the ``tracker + time-of-flight''
sub-analysis, which requires a track to be reconstructed in both the
inner tracker and the muon system.\footnote{Although the
  ``tracker only'' sub-analysis is motivated by charge-flipping
  scenarios, it could potentially provide greater sensitivity for
  shorter stau lifetimes.
  However, as the relevant lifetime window is expected to overlap with
  the range covered by disappearing track searches, and efficiency maps
  for this search region are not provided, we do not consider this
  possibility in detail here.}  In this signal region, events are
required to have at least one high-quality track with $\abs{\eta}<2.1$
and $p_T>70$ GeV.  This track must pass mild isolation criteria, with
the sum of all nearby tracks $\sum p^{trks}_{T,\Delta R<0.3}<50$ GeV
and $I^{calo,track}_{\Delta R<0.3}<0.3$, where
\begin{equation}
I^{C,X}_{\Delta R<R}<Y
\end{equation}
means that in a $\Delta R=R$ cone around object $X$, the sum of either
calorimeter deposits ($C=$ calo) or charged tracks ($C=$ trks) divided
by the $p_T$ of $X$ must be less than $Y$.  Additionally, this track
needs to exceed some average $dE/dx$ value (see
Ref.~\cite{Chatrchyan:2013oca} for details) and have a sufficiently low
velocity $\beta$ satisfying $1/\beta>1.225$.

While long-lived staus are one of the signal models considered in the
CMS HSCP search \cite{Chatrchyan:2013oca}, in order to understand the
sensitivity of this search for staus with general values of $(m_\stau ,
c\tau)$, and to establish results for staus appearing at the end of
cascade decays, we need to recast the search.  To do so, we follow the
detailed instructions provided in Ref.~\cite{Khachatryan:2015lla}
using only the 8 TeV data.  These instructions employ efficiency maps
\cite{CMSHSCPeffmap} that provide both an on- and off-line probability
$(P^{\mathrm{on}}, P^{\mathrm{off}})$ for an individual track to
satisfy the basic requirements of the search as a function of $p_T$,
$\eta$, $\beta$, and $m_\stau$.  We then scale $P^{\mathrm{on}}$ by
the probability that a track survived through the muon chamber,
$e^{-\frac{m_\stau x(\eta)}{p c\tau}}$, where $x(\eta)$ is a simple
geometric approximation for the size of the detector:
$x(\abs{\eta}\leq0.8)=900$ cm, $x(0.8<\abs{\eta}\leq1.1)=1000$ cm,
$x(1.1<\abs{\eta}\leq2.1)=1100$ cm.
As the efficiency maps do not include the effect of the isolation
cuts, we set $P^{\mathrm{off}}=0$ if either $\sum p^{trks}_{T,\Delta
  R<0.3}>50$ GeV or $I^{calo,track}_{\Delta R<0.3}>0.3$.  The on- and
off-line probabilities are combined to yield a total efficiency of
 \beq
 P^{net}=(P^{\mathrm{on}}_1+P^{\mathrm{on}}_2-P^{\mathrm{on}}_1\times P^{\mathrm{on}}_2)(P^{\mathrm{off}}_1+P^{\mathrm{off}}_2-P^{\mathrm{off}}_1\times P^{\mathrm{off}}_2),
\eeq
 where the two different probabilities correspond to the two different
$\stau$s in the event (with one track, it is simply
$P^{net}=P^{\mathrm{on}} P^{\mathrm{off}}$).  The signal falls into
one of four signal regions based on $m_\stau$, which leads to a
maximum number of allowed signal events used to set our 95\% exclusion contours:
\beq 
N_{95}=\left\{ \begin{tabular}{ll}
    21.6 & $m_\stau\leq166$ GeV \\
    8.3 & 166 GeV $<m_\stau\leq330$ GeV \\
    3.0 & 330 GeV $<m_\stau\leq500$ GeV \\
    3.0 & 500 GeV $<m_\stau$ 
 \end{tabular} \right. 
\eeq
Our modeling reliably 
reproduces the constraints from the search, so we assign the 
recommended 25\% uncertainty to our modeling of this search 
in Figures \ref{fig:direct} \& \ref{fig:stopgo}.  Further details of the 
validation of our modeling are given in Appendix \ref{sec:val}.

\subsection{Disappearing Track Searches}
\label{sec:DT}
   
Both ATLAS and CMS have searches for disappearing tracks, i.e., tracks
of high quality within the inner layers of the tracker that suddenly
vanish, leaving no hits in the outer layers of the tracker.
Disappearing track signals are characteristic of long-lived nearly
degenerate winos, as can arise in anomaly-mediated SUSY breaking
(AMSB) \cite{Randall:1998uk,Giudice:1998xp}, and the LHC searches are
optimized for this model.  These disappearing track searches can also
be sensitive to long-lived $\stau_R$s, as the tracking algorithms are
not directly sensitive to the visible decay products of charged
particles that decay in flight.

For readability in our figures, we will present only the strongest
limit from the two disappearing track searches.  A breakdown of the
individual sensitivities is presented in Appendix \ref{sec:val}.
Below we discuss the ATLAS \cite{Aad:2013yna} and CMS
\cite{CMS:2014gxa} searches in turn.

\subsubsection{ATLAS Disappearing Tracks}
\label{sec:ATLASDT}

The ATLAS disappearing track search \cite{Aad:2013yna} requires at
least one hard jet with $p_T>90$ GeV, large missing energy $\met>90$
GeV, and a minimal azimuthal separation between the $\met$ and the
hardest two jets of $\Delta\phi^{\mathrm{jet-}\met}>1.5$.  The search
additionally requires that there are no electron or muon candidates
(satisfying loose ID requirements) in the event.  Backgrounds
containing muons are further suppressed by requiring no tracks in the
muon calorimeter with $p_T>10$ GeV.

After this basic selection, the search requires a high $p_T$ ($>75$
GeV) track stub of good quality that leaves hits in the inner tracker
(pixel and silicon microstrip layers), but fewer than five hits in the
straw-tube transition radiation tracker (TRT) occupying the outer
tracker region with an inner (outer) radius of $56.3$ cm ($106.6$)
cm.\footnote{For comparison, a typical charged particle leaves an
  average of 32 hits in the TRT \cite{Aad:2013yna}.}  This disappearing track must be the
highest $p_T$ track in the event, sit within the range
$0.1<\abs\eta<1.9$, be isolated from other tracks,
$I^{trks,track}_{\Delta R<0.4}<0.04$, and separated from all jets with
$p_T> 45$ GeV by $\Delta R>0.4$.

In order to model the efficiency for a charged particle decaying
within the tracker to leave a track that passes the selection
requirements, we partition the tracker into 10-cm bins of radial
displacement, $L_{xy}$.  Each bin is weighted with the probability
$\mathcal P$ for the long-lived particle to decay within that bin.  As
the disappearing track is required to have at least two hits in the
silicon microstrip layers that begin near $L_{xy}=30$ cm, particles
that decay before this have zero identification efficiency,
$\epsilon_{ID}=0.0$.  Starting at 30 cm of radial displacement we have
$\epsilon_{ID}=1.0$ until the TRT starts at 56.3 cm.  After this, we
model the number of TRT hits using a Poisson distribution based on how
far the particle has propagated radially through the TRT.  We assign
an average of 25 hits to particles which survive the entire TRT;
however, we set an efficiency floor of $\epsilon_{ID,min}=0.1$,
following Figure 2 of \cite{Aad:2013yna}.  This floor allows for the
increased sensitivity to the larger values of $c\tau$ that ATLAS
observes.  Our resulting simple modeling of the track identification
efficiency is shown in Table~\ref{tab:AtlasEff}.  Variations on the
implementation of the track identification efficiency produced only
small modifications to the ultimate sensitivity.
The net probability that an event containing two
long-lived staus will possess a disappearing track that passes the
selection is
\beq
w = \sum_{x\in \mbox{\scriptsize bins}} \epsilon_{ID}(x) 
 \mathcal P_1(x)\mathcal P_2(<6\mbox{ m}) + \sum_{x \in \mbox{\scriptsize bins}}
     \epsilon_{ID}(x) \mathcal P_2(x) \mathcal P_1(<30\mbox{ cm}), 
\eeq
where $\mathcal{P}_1 \, (\mathcal{P}_2)$ refers to the higher (lower)
$p_T$ stau. The second term is the region where the lower-$p_T$ stau
yields the hardest track in the event.
 
This search has four non-exclusive signal regions, defined by the
$p_T$ of the disappearing track, $p_T>75,\,100,\,150,$ and $200$ GeV,
which have a maximum allowed number of signal events at 95\% CL of
$N_{95}=35.7,\,20.8,$ $12.6,$ and $8.9$ observed
($N_{95,exp}=28.8,\,21.3,$ $13.6,$ and $11.3$ expected), respectively.
At each point we use the observed sensitivity from the bin with the
best expected sensitivity to place constraints.  This search has been
validated on the AMSB wino model and has fairly good agreement, as
shown in Appendix~\ref{sec:val}.  Our recast yields weakened limits at
high values of $c\tau$ relative to the experimental result; however,
this limitation of our modeling will not be important for our
conclusions.
 
  \begin{table}[t]
\begin{center}
\begin{tabular}{|c||c|c|c|c|c|c|c|c|} \hline
{ $L_{xy}$ (cm)}  & $<30$ &$\!\!30\!-\!50\!\!$ & $\!\!50\!-\!60\!\!$ & $\!\!60\!-\!70\!\!$ & $\!\!70\!-\!80\!\!$ &  $\!\!80\!-\!90\!\!$ &$\!\!90\!-\!600\!\!$ &$>600$ \\ \hline
{ ID Efficiency $\epsilon_{ID}$} & $0.0$ & $1.0$ &  $0.99$& $0.91$ & $0.52$ &  $0.18$& $0.1$ & $0.0$ \\ \hline
\end{tabular}
\caption{Our simplified modeling of identification efficiencies in the
  ATLAS disappearing track search as a function of the radial
  displacement $L_{xy}$. We note that our limits are largely
  insensitive to the precise details of the modeling within the
  TRT.  \label{tab:AtlasEff}
}
\end{center}  
\end{table}
 
Of course, the benchmark models we are considering have an additional
layer of complication, namely that our $\stau_R$s do not simply
disappear, but yield an energetic decay product -- an electron, muon,
or hadronic tau -- that can deposit energy in the calorimeter, appear
as a jet, modify the $\met$ distribution,  and/or leave high $p_T$ tracks
in the muon calorimeter.  In order to simulate this, we model the
$\stau$ decays as occurring at the center of the 10-cm discrete bin of radial
displacement.  If the $\stau$ gives an electron or a hadronic tau {\it
  with} neutral pions\footnote{Hadronic taus with neutral pions also
  contain at least one charged pion, which typically deposits most of
  its energy into the HCAL.  For simplicity, in this class of hadronic
  tau, we deposit all energy at the center of the ECAL for the
  purposes of determining the resulting position of the jet.}
(hadronic tau {\it without} neutral pions), we deposit the energy of
the decay products at the point where the track emanating from the
center of this bin intersects a cylinder going roughly halfway through
the ECAL (HCAL) -- we use $\{R,Z\}=\{175\mbox{ cm},420\mbox{
  cm}\}\,(\{325\mbox{ cm},520\mbox{ cm}\})$ for ATLAS. This
calorimeter deposit is labeled as a jet (photons are not distinguished
from jets in this search), and the 90 GeV jet, recalculated $\met$,
and various isolation requirements are checked with these reprocessed
kinematics.  If the stau decays within the calorimeter, we simply
label it as a jet centered at the point where the track connected with
the calorimeter.  If a track survives far enough into the muon
chamber (we use a six-meter radius), the event is assumed to be vetoed
by the strict muon veto criteria of the ATLAS search.
  
Despite the successful validation of our recasting procedure for the
original wino signal model, we stress that the additional
complications due to the stau decay products greatly decrease the
reliability of our modeling.  Because of this, we present a 50\%
modeling uncertainty for this search.  Nonetheless, our modeling is
sufficiently accurate to demonstrate that disappearing track searches
also have good sensitivity to ``kinked track'' signals.  Implementing
a full GEANT-based detector simulation in order to accurately treat
this class of signals is important, but is best done by
experimentalists.

\subsubsection{CMS Disappearing Tracks}
\label{sec:CMSDT}

CMS also has a search looking for disappearing tracks, motivated by
nearly-degenerate winos in AMSB \cite{CMS:2014gxa}.  In this search,
CMS requires large missing energy of $\met>100$ GeV, and at least one
hard jet with $p_T>110$ GeV, $\abs\eta<2.4$, which has at least 20\%
of its energy in charged hadrons, less than 70\% in neutral hadrons or
photons, and less than 50\% in electrons. The hardest two jets and the
$\met$ must be azimuthally separated by $\Delta\phi^{\mathrm{jet-}\met}>0.5$,
and all jets with $p_T>30$ and $\abs{\eta}<4.5$ must be separated from
one another by $\Delta\phi^{jj}<2.5$ to reduce QCD background.
  
The candidate disappearing tracks are required to be of high quality,
have $p_T>50$ GeV, and fall within an $\eta$ range of either
$\abs\eta<0.15$, $0.35<\abs\eta<1.42$, or $1.85<\abs\eta<2.1$.  The
tracks are required to be isolated, with no jets of $p_T>30$ GeV
within $\Delta R$ of 0.5, $I^{trks,track}_{\Delta R<0.3}<0.05$, and
$E_{calo}^{\Delta R<0.5}<10$ GeV.  As in the ATLAS search, the tracks
are required to have left abnormally few hits in the outer layers of
the silicon tracker.  A simple efficiency map based on the radial
displacement is provided in the appendix of Ref.~\cite{CMS:2014gxa}
(table 8).  As this efficiency map does not factor out the $\eta$
requirements or isolation (both of which can affect our signal models
significantly), we rescale the efficiency values by an overall factor
of 1.50, and impose $\eta$ acceptance and isolation separately.  The
rescaling factor was was determined from the effect of these cuts on
our AMSB wino samples, and reliably fits with the data.  The observation of 2
events with $1.4\pm1.2$ expected gives a 95\% upper limit on
a new physics signal $\times$ acceptance of $N_{95}=5.3$ events.  Our
modeling reproduces the exclusion contour shown in \cite{CMS:2014gxa}
very accurately for all wino lifetimes (Appendix \ref{sec:val}).
  
As with the ATLAS disappearing track search, the fact that the stau
has hard visible decay products modifies the story significantly; we
once again model the impact of these decay products in 10-cm discrete
bins of radial displacement.  First, a stau which decays to a hard
hadronic tau very close to the interaction point can potentially
provide the event's hard jet.  We choose a transverse decay length of
$L<2$ cm to model this possibility simply; decay products originating
further away than this are assumed to not pass the charged hadron
fraction $>20\%$ requirement placed on a jet with $p_T>110$ GeV.
Second, the stau decay products can cause the track to fail the strict
10 GeV isolation requirement.  Third, $\stau$ decay products alter the
$\met$ and can affect the jet separation requirements.  As in the
ATLAS search, if the $\stau$ yields an electron or a hadronic tau {\it
  with} neutral pions (hadronic tau {\it without} neutral pions), we
deposit the energy of the decay products at the point where the track
emanating from the center of this bin intersects a cylinder passing
roughly halfway through the ECAL (HCAL) -- we use $\{R,Z\}=\{155\mbox{
  cm},240\mbox{ cm}\}\,(\{235\mbox{ cm},480\mbox{ cm}\})$ for CMS.  If
the stau decays within the calorimeter we deposit all of the decay
product energy there; if the stau decays in the muon chamber or
beyond, we assume no jet is reconstructed and all of this energy is
invisible.  These calorimeter deposits are labeled as jets (photons
are not distinguished from jets, except by the leading jet
requirement), and $\met$, isolation, and jet separation are checked
with these new objects.

Again, as in the ATLAS search, despite our very precise validation of
the AMSB wino model, we stress that the additional complications due
to the stau decay products greatly impact the reliability of our
modeling.  Because of this, we again assign a 50\% modeling
uncertainty in the results presented for this search.  A more detailed
treatment is best performed by experimentalists.

\subsection{CMS Displaced $e\mu$ Search}
\label{sec:CMSemu}

In \thesearch~\cite{Khachatryan:2014mea}, the benchmark model
considered is the direct pair production of stops that decay through
small lepton-flavor-universal RPV $\lambda'_{ijk}L_iQ_jD^c_k$
couplings ($\lambda'_{133}=\lambda'_{233}=\lambda'_{333}$) to yield
displaced $\st\to e b$, $\mu b$, and $\tau b$ decays.  In this search,
the leptons are required to be fairly hard, in the central region of
the detector, and isolated from jets, other calorimeter deposits, and
each other.  The most distinguishing preselection requirement in the
search is that the transverse impact parameter, $d_0$, with respect to
the primary vertex is required to be larger than 100 $\mu$m for both
leptons.  The impact parameter is actually not the point where the
parent object (e.g., $\stau$, $\tau$ or $b$) decays, but rather the
distance to the point of closest approach for the lepton's track
relative to the center of the beampipe (in most other searches, the
impact parameter is defined with respect to the primary vertex).  This
is especially important as backgrounds from $Z\to\tau\tau$ or heavy
flavor tend to result in leptons that are nearly collinear with the
parent due to a small mass-to-momentum ratio, and thus yield a small
impact parameter even with an abnormally long-lived parent.  After
imposing preselection requirements, events are divided into three
exclusive signal regions: SR3, where both leptons have transverse
impact parameters $d_e$ and $d_\mu$ between $0.1$ and $2.0$ cm; SR2,
with $d_e$ and $d_\mu$ between $0.05$ and $2.0$ cm, but not satisfying
the requirement of SR3; and SR1, with $d_e$ and $d_\mu$ between $0.02$
and $2.0$ cm, but not within SR2 or SR3.  These selection requirements
are summarized in Table \ref{tab:cuts}.

\begin{table}[t]
\begin{center}
{\small
\vspace{-1cm}

\begin{tabular}{|c|} \hline
{\bf Cut Summary of \search}    \\ \hline \hline
{\bf Preselection}    \\ \hline
{1 OS $e^\pm\mu^\mp$ pair} \\  
{$d_\ell>100\,\mu$m} \\
{$p_{T,\ell}>25$ GeV,  $\abs{\eta_\ell}<2.5$} \\  
{Reject $1.44 <\abs{\eta_e}<1.56$} \\  
{$I^{calo,e}_{\Delta R<0.3}<0.10$}, {$I^{calo,\mu}_{\Delta R<0.4}<0.12$} \\  
{ $\Delta R_{\ell j}>0.5\; \forall$ jets with $p_T>10$ GeV } \\
{$\Delta R_{e\mu}>0.5$} \\
{$v_{T,\tilde \ell} < 4 \,\mbox{cm}$, $ v_{Z,\tilde \ell} <  30 \,\mbox{cm}$}\\
{Veto additional leptons} \\
 \hline 
\end{tabular} \hspace{73mm} \; 
\put(-195,-110){\includegraphics[scale=0.57]{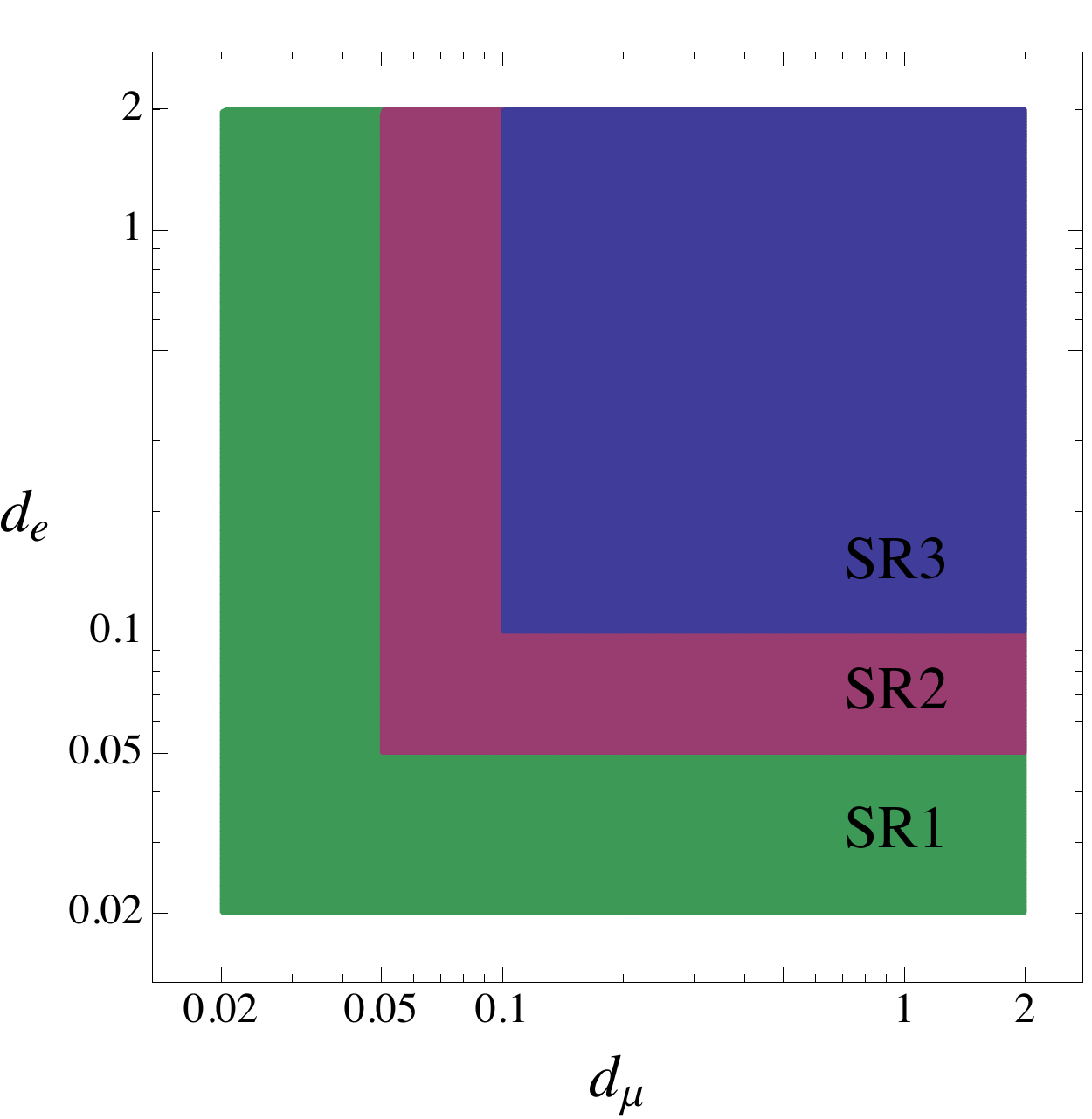}}
\caption{{\bf Left:} The preselection cuts used in
  \cite{Khachatryan:2014mea} (see also
  \cite{CMS:2014bra,CMSemuEfficiency}).  {\bf Right:} an illustration
  of the cuts on the transverse impact parameter that define the three
  exclusive signal regions.
  \label{tab:cuts}} }
\end{center}  
\end{table}
 
In addition to imposing the cuts of the search, we utilize the
recommended parameterization provided in \cite{CMSemuEfficiency} to
model the trigger, selection, and reconstruction efficiencies for each
species of lepton.  We also mandate $v_{T,\tilde \ell}\,(v_{Z,\tilde
  \ell}) < 4\,(30)\,\mbox{cm}$ \cite{CMSemuEfficiency}, where
$v_{T,\tilde \ell}\,(v_{Z,\tilde \ell})$ is the transverse
(longitudinal) position of the secondary vertex.  Beyond this range,
tracking fails.  To determine the 95\% CL exclusion contour, the
truth-level properties of the staus and their decay products are used
to derive a $c\tau$-dependent weight for each event to have the lepton
transverse impact parameters falling into one of the three signal
regions.\footnote{For numerical feasibility, the finite $\tau$
  lifetime of $c\tau_\tau=87 \mu m$ was neglected. This is a small
  effect on the lepton impact parameter because $m_\tau/m_\stau\ll1$,
  and thus the tau and lepton momentum are roughly collinear, i.e.,
  $\hat p_\tau\sim \hat p_\ell$.  At large $\stau$ lifetimes this is a
  very good approximation; at smaller lifetimes there can be a
  moderate increase in efficiency, especially for the population of
  SR1. This effect was verified to be $\lesssim 10\%$.}  The exclusion
confidence level from the combination of the three exclusive signal
region bins were derived using frequentist methods on the background
estimates provided in the search and assuming the nominal NLO+NLL
value for the cross-sections.

A validation against the displaced stop model considered in the CMS
study is presented in Appendix~\ref{sec:val}.  In the region of
highest sensitivity, the recast agrees excellently with the results
of the search.  Near $c\tau\sim1$ m or 100 $\mu$m, we expect our
modeling to underestimate the actual constraint slightly.  We assign
the recommended 25\% modeling uncertainty to the search in all
figures.

\subsection{Constraints on Long-lived Staus}
\label{sec:GMSBstaus}

\begin{figure}[!t]
\begin{center}
\includegraphics[scale=0.61]{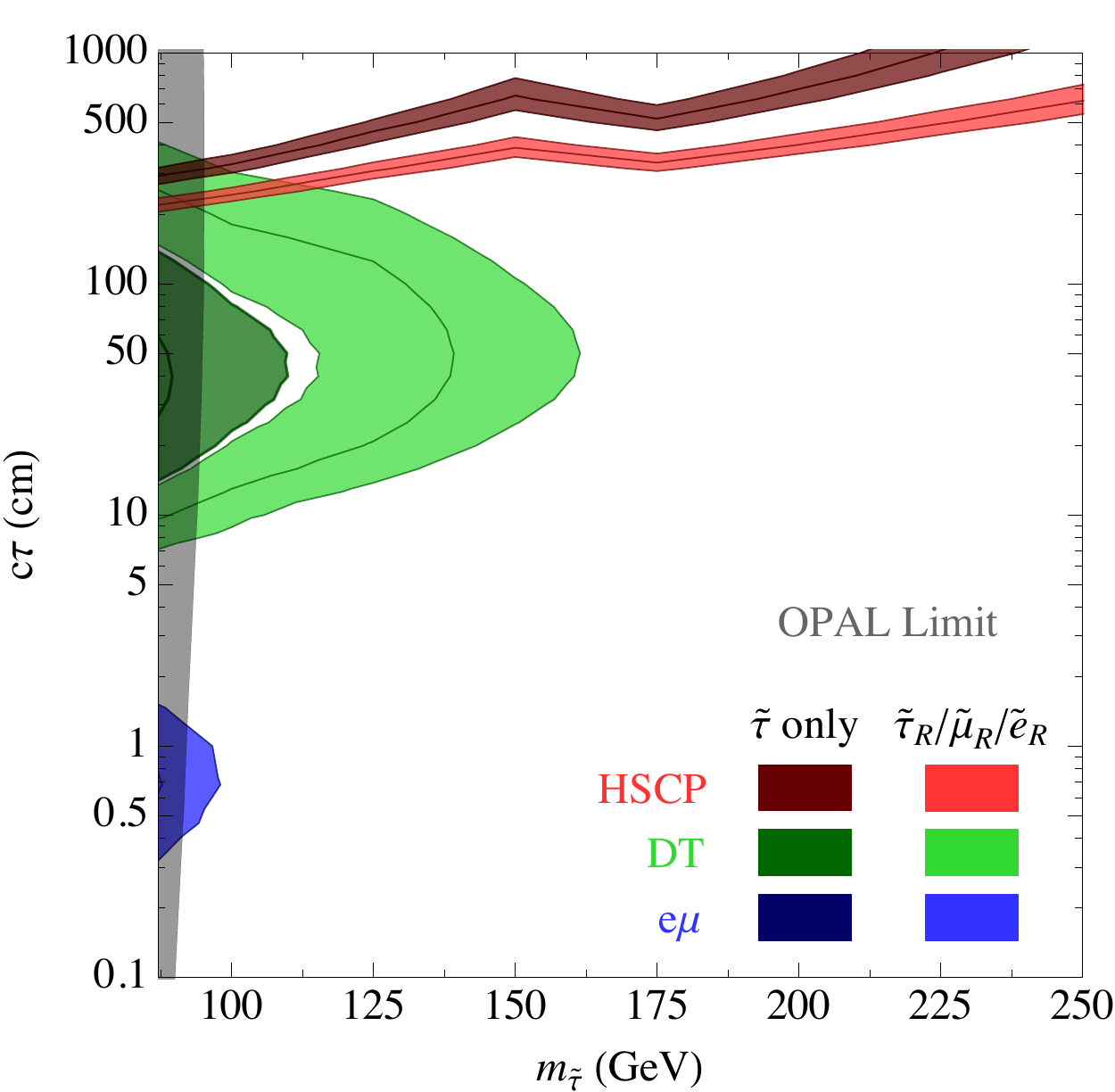}
\includegraphics[scale=0.61]{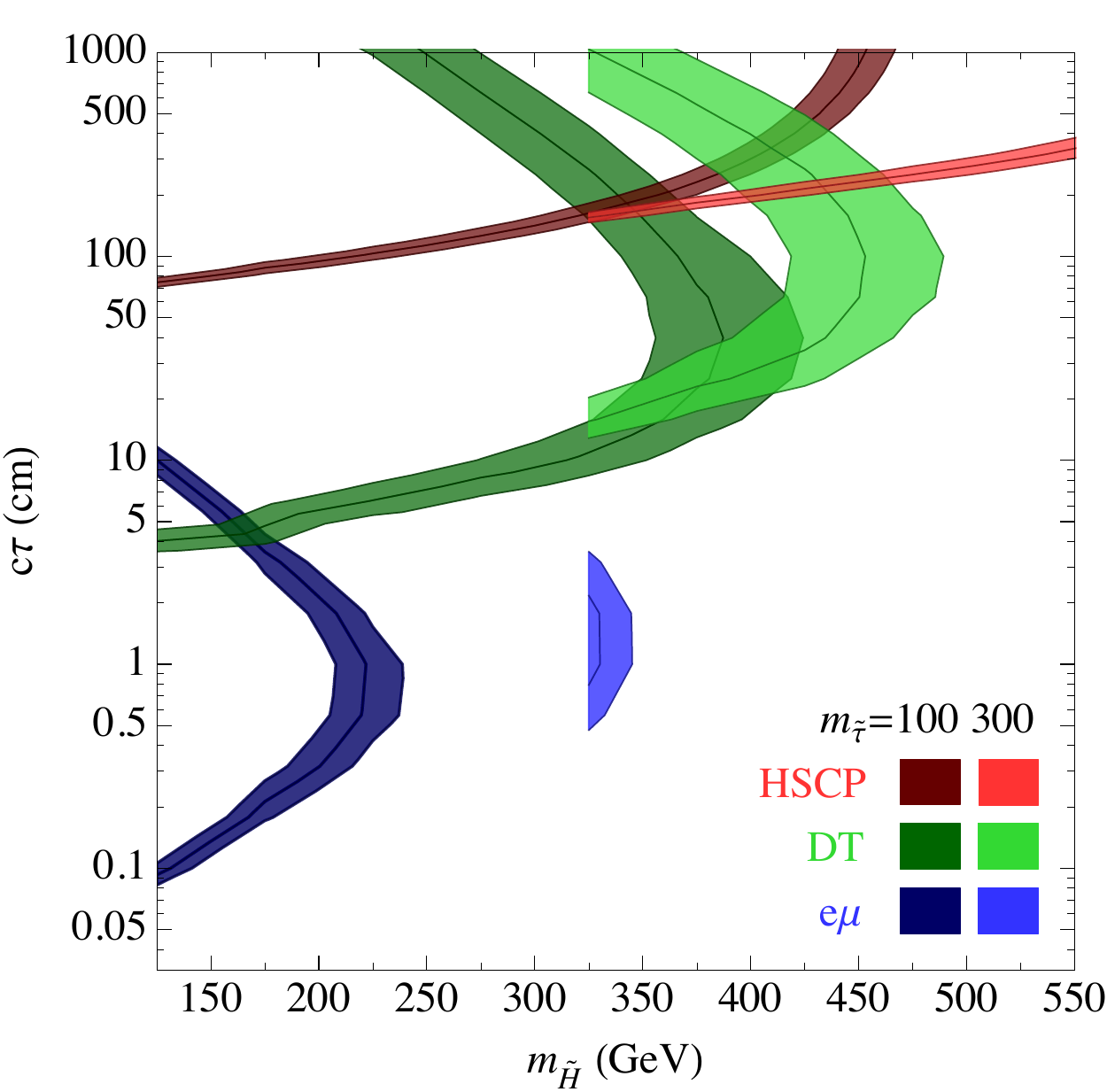}
\end{center}
\caption{{\bf Left:} Constraints on direct production for the case of
  a single isolated, light, right-handed stau NLSP (dark), as well as
  for the case of nearly degenerate three generations of right-handed
  sleptons (bright). Near $c\tau\sim1$ cm, \thesearch\ is most
  sensitive \cite{Khachatryan:2014mea} (blue).  (The $\stau$ only limit 
 from this search falls well below the LEP bound and is not shown).  
 Near $c\tau\sim50$ cm,
  the disappearing track searches at CMS \cite{CMS:2014gxa} and ATLAS
  \cite{Aad:2013yna} (green) are most sensitive; we show only the
  stronger of the two limits (for selected individual
  sensitivities, see Figure~\ref{fig:DTcomp}).  Above $c\tau\sim2$ m,
  the CMS heavy stable charged particle search
  \cite{Chatrchyan:2013oca} (red) sets powerful constraints.  The most
  stringent LEP2 bounds from OPAL \cite{Abbiendi:2005gc} are shown in
  light gray, ranging from 87 to 97 GeV.  {\bf Right:} Constraints on
  production of degenerate Higgsinos decaying as $\Ho^\pm \to
  \stau_R^\pm \nu /\Ho^0_{1,2} \to \stau_R^\pm \tau^\mp$.  Only direct
  production of the Higgsino is used for setting a limit.  A scenario
  with a 100 GeV (300 GeV) stau is shown in dark (bright) colors.  The
  minimum Higgsino mass shown is 125 GeV (325 GeV).  Search colors are
  as in Figure \ref{fig:direct} left.}
\label{fig:direct}
\end{figure}

\begin{figure}[!t]
\begin{center}
\includegraphics[scale=0.6]{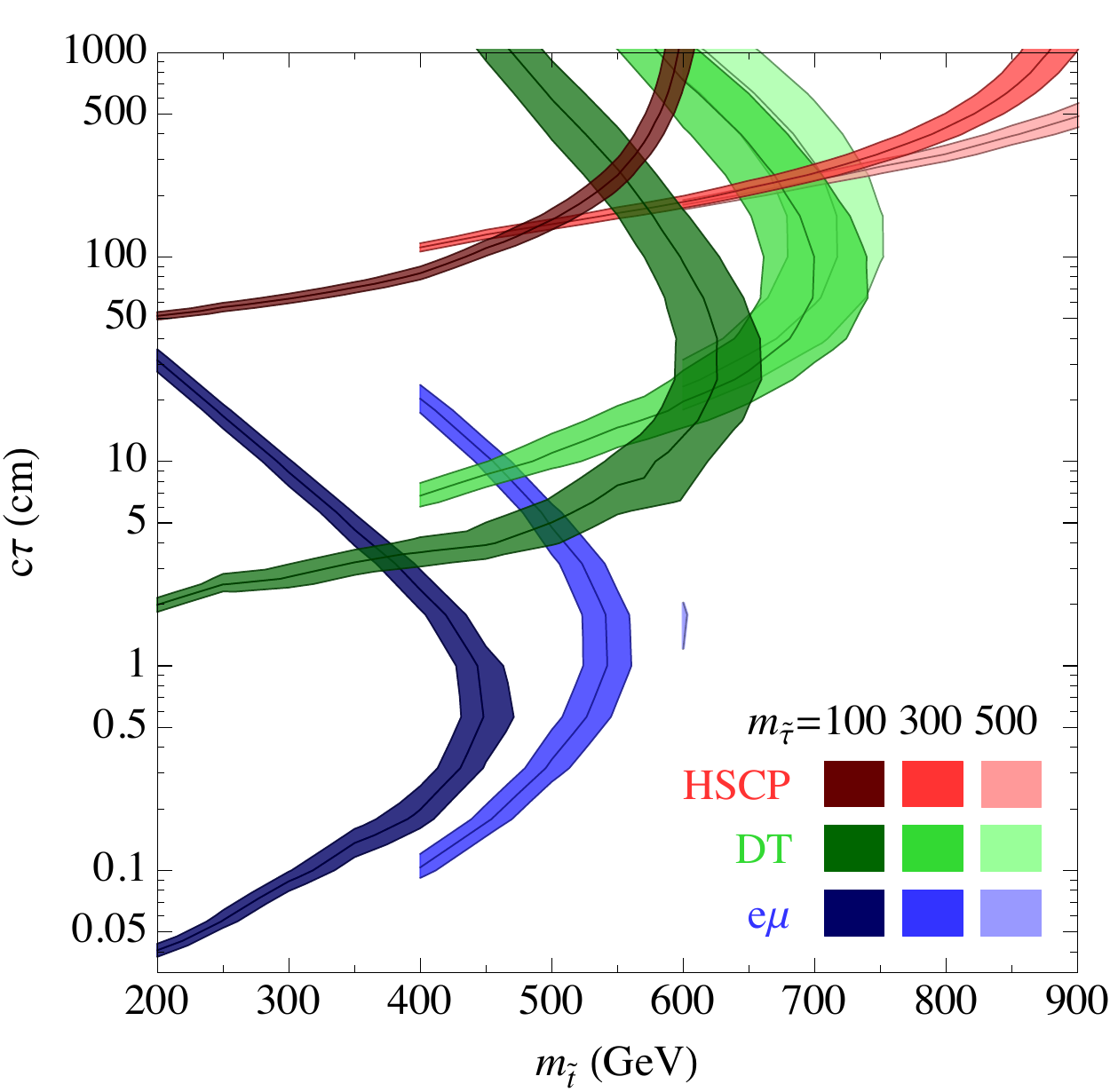}
\includegraphics[scale=0.6]{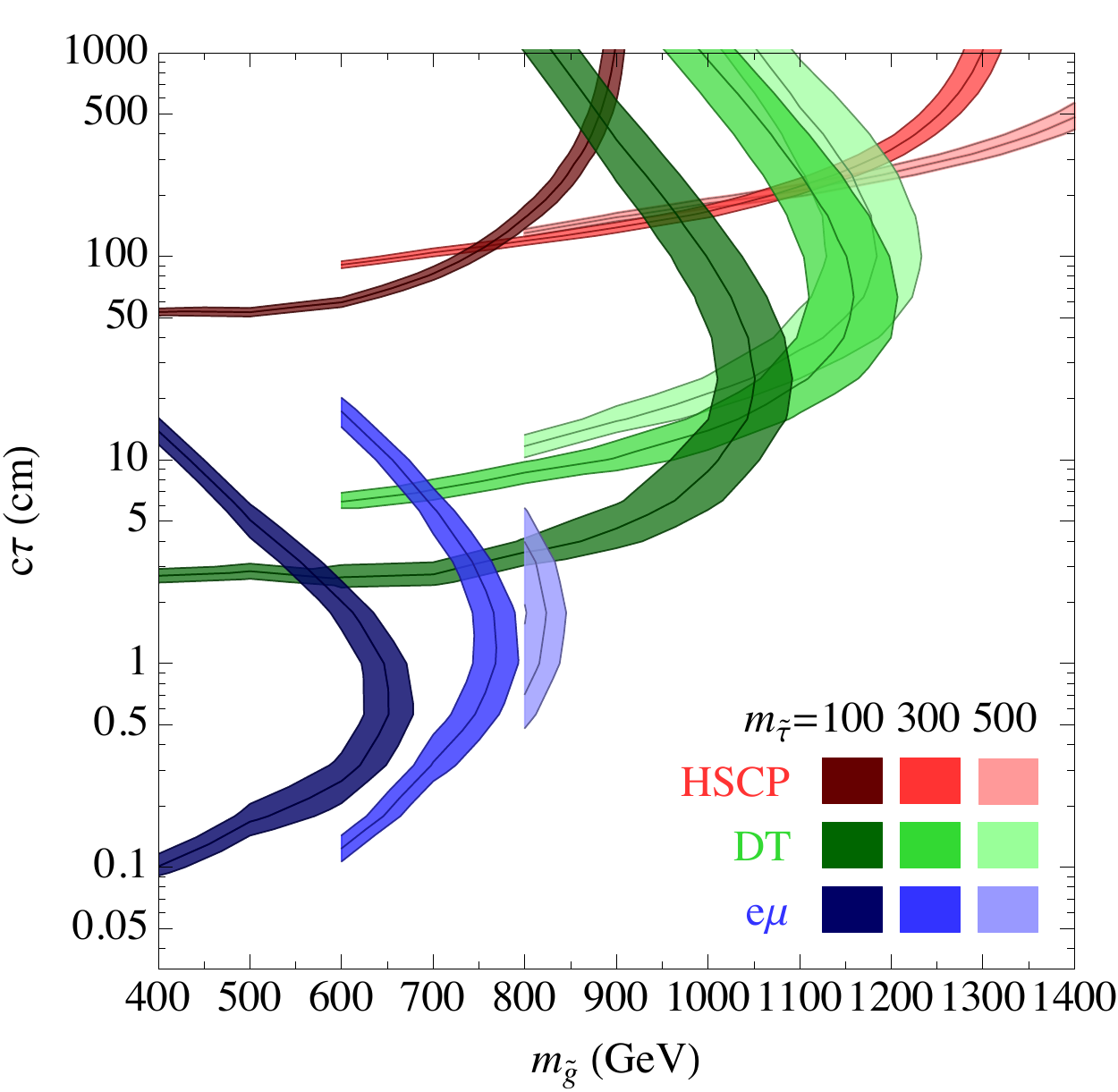}
\end{center}
\caption{{\bf Left:} Constraints on production of right-handed stops
  decaying as $\st \to b \Ho^+\to b \nu \stau_R^+$ with
  $m_{\Ho^+}=m_\st -50$ GeV.  
  Only the direct production of stops is used for setting a limit.  A
  scenario with a 100 (300 [500]) GeV stau is shown in dark (bright
  [light]) colors.  The minimum stop mass shown is 200 (400 [600])
  GeV.  Search colors are as in Figure \ref{fig:direct} left.  {\bf
    Right:} Constraints on production of Majorana gluinos decaying
  through stops and Higgsinos into a displaced stau final state.  Only
  the direct production of the gluinos is used to set a limit.  All
  coloration is as in the stop model Figure \ref{fig:stopgo} left.
  The minimum gluino mass shown is 400 (600 [800]) GeV.  Dirac gluinos
  \cite{Fox:2002bu}, which only give opposite-sign leptons, have a
  cross-section $\times$ efficiency that is four times larger than the
  results presented here.}
\label{fig:stopgo}
\end{figure}

In this subsection we show the constraints on long-lived staus found
from the searches described above and comment on potential avenues for
improvement.  To explore a wide variety of scenarios, we consider
several simplified benchmark models for the pair production of $\stau$
NLSPs.  In each model, the $\stau_R$ lifetime, $c\tau$, is treated as
a free parameter (for all lifetimes of interest, the gravitino is
effectively massless and has no influence on kinematics).  The models
considered are:
  \begin{itemize}
  \item Direct $\stau_R$ production where the $\stau_R$ is isolated at
    the bottom of the spectrum.  95\% CL limits are shown in Figure
    \ref{fig:direct} (left) in the $m_\stau$ -- $c\tau$ plane.  LEP2 bounds from 
    OPAL \cite{Abbiendi:2005gc} are shown in gray.
  \item Direct slepton production in the case where there are three nearly
    degenerate generations, $\tilde e_R$, $\tilde \mu_R$ and $\stau_R$
    ($m_{\tilde e_R}\!=\!m_{\tilde \mu_R}\!=\!m_{\tilde \tau_R}+10$
    GeV) with prompt decay $\tilde e_R,\tilde \mu_R \to \stau_R +
    \{\mbox{soft}\}$.  95\% CL limits are shown in Figure
    \ref{fig:direct} (left) in the $m_\stau$ -- $c\tau$ plane.
  \item Higgsino production with prompt decays $\Ho^\pm\to
    \stau_R^\pm \nu$, and $\Ho^0 \to \stau_R^\pm \tau^\mp$.  95\% CL
    limits are shown for $m_\stau=100$ and 300 GeV in Figure
    \ref{fig:direct} (right).
  \item Stop production with prompt decay $\st \to b \Ho^+ \to b \nu
    \stau_R^+$.  95\% CL limits on this scenario are shown for
    $m_\stau=100,$ 300, and 500 GeV in Figure \ref{fig:stopgo} (left)
    with $m_\Ho=m_\st-50$ GeV.
  \item Majorana gluino production with prompt decay $\go\to \st \bar t \to \bar
    t b \Ho^+ \to \bar t b \nu \stau_R^+$ and the conjugate decay.
    95\% CL limits are shown for $m_\stau=100,$ 300, and 500 GeV in
    Figure \ref{fig:stopgo} (right) for $m_\st = m_\go-200$ GeV and
    $m_\Ho=m_\st-50$ GeV.
  \end{itemize} 
 
As can be seen in Figures~\ref{fig:direct} and~\ref{fig:stopgo}, the
three search strategies -- HSCP, disappearing track, and displaced
$e\mu$ -- are complementary and constrain different lifetime regions.
For the disappearing track searches, we show only the search that sets
the best limit at each point, which is usually the CMS search (see
Figure~\ref{fig:DTcomp} for comparison of the two disappearing track 
searches in a variety of scenarios).  The HSCP search turns on sharply around
$c\tau\sim 1$ -- 3 m, and grows stronger at longer lifetimes.  The
disappearing track searches are most sensitive in the $c\tau\sim50$ cm
range.  The displaced $e\mu$ search peaks around $c\tau\sim1$ cm.
 
As the HSCP search is powerful and has a very high acceptance, we have
no suggestions for potential improvements to that search, short of
providing efficiency maps for the tracker-only search to enhance the
usability of the results.  We note that in cascade decay scenarios
with a large mass hierarchy between the initially produced parent particle
(e.g., a gluino) and the stau, the requirement that $\beta$ differ
from $c$ typically fails above a certain parent particle mass.  Of
course, in most of these scenarios direct production of the stau would
be constrained by the HSCP search.
 
The disappearing track searches use very different selections and
methods between the two experiments.  The CMS search has very high
background rejection, resulting in only two events in their signal
region, while the ATLAS search admits dozens of background events with a
more inclusive selection.  Between the two searches, CMS typically
sets the stronger limit; however, at shorter lifetimes the two are
nearly identical in strength.  Part of the reason that CMS is able to
set stronger limits at higher $c\tau$ is that ATLAS vetoes events with
activity in the muon chamber, thus losing a large fraction of events
that effectively contain a heavy stable charged particle, which are
very well constrained by the HSCP searches.  See
Appendix~\ref{sec:val} for more details and plots comparing the two
searches.  The disappearing track searches peak at a proper lifetime
of around 20 to 50 cm.
These searches are not able to constrain direct production of
$\stau_R$ beyond the limits set by OPAL, but production of three
near-degenerate species of right-handed sleptons are constrained up to $\sim140$ GeV.
Higgsinos are constrained by disappearing track searches up to 375-450
GeV, stops are constrained up to 600-700 GeV, and gluinos are
constrained up to 1050-1200 GeV.
 
Although modeling signal acceptance in these searches is challenging,
modeling the background for the searches is substantially more
difficult, so while we have some suggestions for modifications to
these searches that would enhance sensitivity to displaced slepton
decays, we stress that we cannot quantitatively assess how these
modifications will affect the backgrounds.  Thus our most important
suggestion for both experiments is to simply include the NLSP
$\stau_R$ benchmark model in the search.
   
For the ATLAS disappearing track search \cite{Aad:2013yna}, the
pointing and timing capabilities of the ATLAS ECAL \cite{Aad:2014gfa} 
could allow for ECAL deposits originating away from the IP and/or 
arriving later than the rest of the calorimeter activity (due to the 
slower speed of the staus and decay geometry)  to be distinguished 
from prompt jets in the search.  More importantly, these capabilities 
could potentially allow for an additional discriminant to improve
sensitivity to staus, i.e., substantial, late-time calorimeter deposits that
point toward the vicinity of where the disappearing track vanished.
If computationally feasible, the pointing information could even 
be utilized as a constraint to facilitate an off-line reconstruction of the 
kinked track in the TRT.  Using an additional discriminant
of this kind could allow for a relaxation of the harsh $p_T$ cuts while
maintaining, if not improving, background rejection.  We additionally
stress that providing efficiency maps would be invaluable for
recasting.
   
For the CMS disappearing track search \cite{CMS:2014gxa}, the very
strict isolation cut on the track, $E_{calo}^{\Delta R<0.5}<10$ GeV,
significantly reduces sensitivity to staus as the stau decay products
often fall within this cone (and do so more frequently at larger
displacements).  
As the CMS ECAL timing resolution is very good \cite{CMS:2015sjc}, 
energy deposits within the isolation cone that arrive later than expected 
could be dropped from $E_{calo}^{\Delta R<0.5}$.  The basic
preselection of this search with an added off-line kinked track requirement in
place of the stringent isolation requirement could provide background
rejection, but as the analysis techniques are very different, this may be 
regarded as a distinct search proposal. 
 
In the $c\tau\sim 1$ cm regime best covered by \thesearch, there are
no limits on the direct production of staus. Including the
production of nearly degenerate $\tilde \mu_R$ and $\tilde e_R$
increases the overall production cross-section enough to yield mild
constraints in a narrow lifetime window, but still below those set by
OPAL.  For Higgsino-, stop-, and gluino-initiated $\stau_R$
production, the reach extends to 225-325 GeV, 450-600 GeV, and 650-800
GeV for lifetimes of $\order{1\mbox{ cm}}$, with sensitivity dying off
for longer and shorter lifetimes.
  
As \thesearch\ uses the most recently designed strategy of the four
searches, it is not surprising that this is where we found the most
potential for improvement. 
Although much of the behavior of the sensitivities shown in Figures
\ref{fig:direct} \& \ref{fig:stopgo} is a straightforward result of
the falling production cross-section with mass and the experimentally
available window for lepton impact parameters, there are several other
factors in play that influence these results.  Here, we highlight
several important points:
   \begin {itemize}
   \item In GMSB, the NLSP is typically a \emph{right-handed} stau,
     which decays to a highly right-handed polarized $\tau$.  This is
     important because the tau polarization significantly affects the
     energy of the final state light lepton
     \cite{Tsai:1971vv,Hagiwara:1989fn}. The differential lepton
     energy distribution from a polarized stau decay can be written as
     \cite{Agashe:2014kda}
\beq 
\frac{1}{\Gamma} \frac{d\Gamma}{dx} =
     \frac23\left[ 5 - 36 x^2(1-x) + P_\stau\lp 1 - 36 x^2 +64 x^3\rp
     \right] ,
\eeq 
where $x\equiv E_\ell/m_\stau$, $P_\stau$ is $+1$ $(-1)$ for
right-handed (left-handed) staus, and we have neglected corrections of
order $(m_\ell/m_\tau)^2$ and $m_\tau/m_\stau$.  Right-handed staus
tend to suppress the energy given to the light lepton, while
left-handed staus enhance it (see Figure~\ref{fig:staupol}).  In the
rest frame of a pure right-handed stau, about 50\% of decays impart
less than 13\% of the stau energy to the light lepton.  As the CMS
search requires relatively hard leptons with $p_{T,\ell}>25$ GeV, this
preference for soft leptons greatly degrades acceptance, especially
for lighter staus.  Lowering the $p_T$ threshold for one or both
species of leptons would greatly increase the acceptance for $\stau_R$
NLSPs.  While lowering lepton $p_T$ thresholds may present
difficulties for triggering on direct stau production, it can make a
significant difference in benchmark models where the stau is produced
at the bottom of a cascade decay and other hard objects are available
for triggering.
 
    \begin{figure}[!t]
\begin{center}
\includegraphics[scale=0.9]{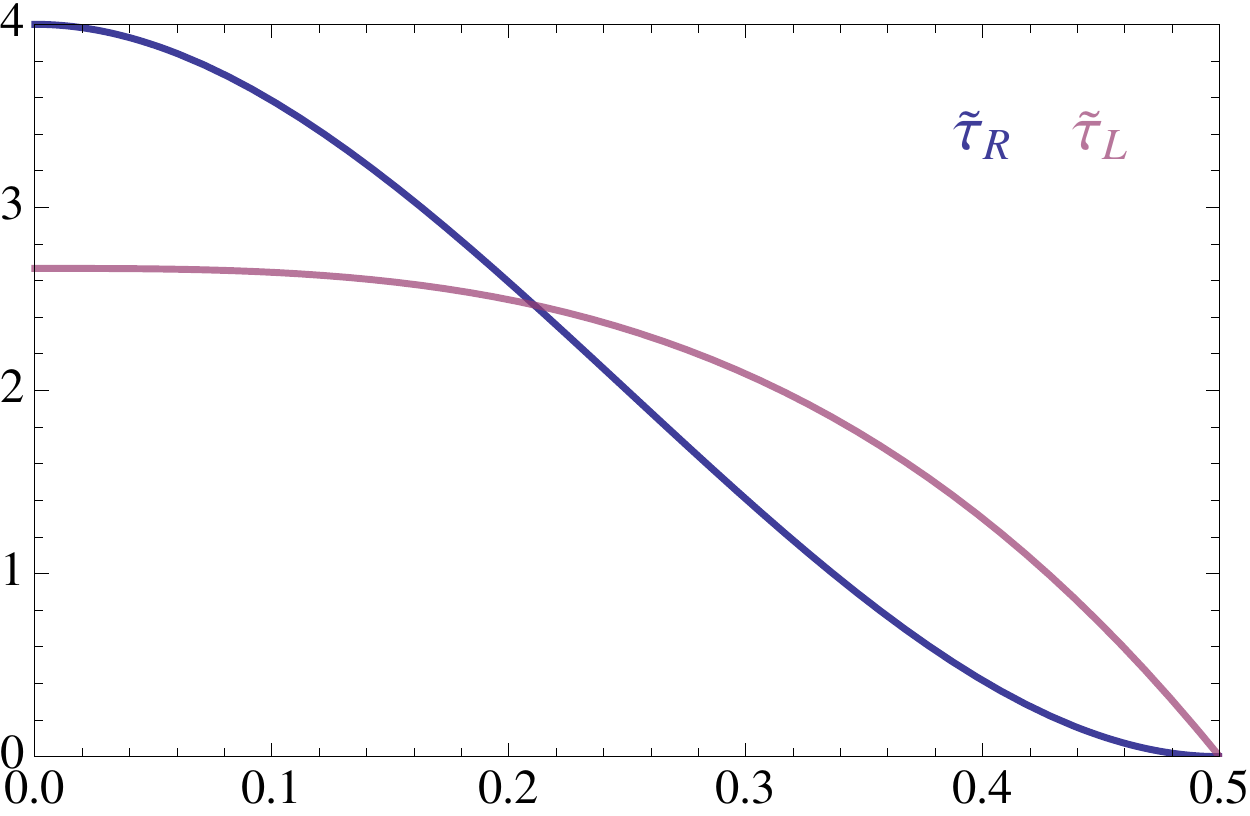}
\put(-363,105){\LARGE  $\frac 1\Gamma \frac{d\Gamma}{dx}$}
\put(-190,-17){\large $x=E_\ell/m_\stau$}
\end{center}
\caption{Distribution of lepton energies from stau decays (neglecting
  terms of order
  $(m_\ell/m_\tau)^2$ and $m_\tau/m_\stau$). Note how heavily
  preferred soft leptons are for right-handed staus.}
\label{fig:staupol}
\end{figure}
   
\item The presence of Majorana particles in both the gluino and
  Higgsino simplified models results in same-sign leptons roughly 50\%
  of the time.\footnote{In the Higgsino case, the fraction of the
    events containing same-sign leptons is less than 50\%,
    since $\sim 20\%$ of the total production rate comes from $\cho^+\cho^-$
    production.}  As the displaced $e\mu$ search requires
  opposite-sign leptons, sensitivity to these scenarios is trivially
  degraded.  In principle, data-driven techniques that utilize the
  same-sign displaced lepton background to predict the backgrounds in
  the opposite-sign regions could wash out a signal (such as the
  method used in \thesearch~for determining the heavy flavor
  backgrounds).  While especially dangerous for gluinos and Higgsinos,
  this control region contamination can even happen in the LQD stop
  model considered in \thesearch, where the long lifetimes make mesino
  oscillation of the stops \cite{Sarid:1999zx} a viable possibility,
  potentially leading to as many as 3 in 8 events possessing leptons
  with the same sign \cite{Evans:2012bf}.
   
\item In both the cases of gluino and Higgsino production, many of the
  events contain additional leptons (from the decays of either tops or
  taus) that are vetoed in the search.  The prevalence of additional
  prompt leptons depends on the particular production and decay modes
  in a given simplified model, but, aside from the case of direct
  production, additional leptons are a generic possibility.  In the
  gluino benchmark model considered here, nearly half of all signal
  events are discarded due to the presence of an additional prompt
  lepton.  Importantly, this veto should be unnecessary, as no major
  backgrounds tend to be produced with additional isolated leptons at
  any appreciable rate.  In most models of interest, an approximate
  $\mathbf Z_2$ symmetry is what provides the displacement.  Thus,
  typically only two genuinely displaced leptons will appear per event,
  and combinatoric ambiguities can be resolved simply by choosing the
  leptons with the highest impact parameters.
   
\item The \search~search uses very tight isolation requirements to
  improve rejection of heavy flavor backgrounds.  These isolation
  requirements are significant enough that the hadronic
  activity is sufficient to reduce the overall efficiency by 10-15\%
  in the case of direct stau production, and 25-35\% in the gluino
  case where there are many additional jets.  In large part, this is
  another side effect of the low lepton $p_T$ arising from
  right-handed polarized $\tau$ decays -- softer leptons require less
  hadronic activity to fail isolation requirements.  At larger
  transverse displacements, the heavy flavor background is greatly
  reduced, so looser isolation criteria, particularly in SR3, would
  serve to enhance sensitivity, especially for longer lifetimes.
   
   \end {itemize}

\section{Models with Displaced Same-Flavor Leptons}
\label{sec:SFmodels}

Although a search for an opposite-sign $e$ and $\mu$ with large impact
parameters is in principle sensitive to displaced stau NLSPs, and more
generally to any new physics that gives rise to displaced decays
exhibiting lepton-flavor universality (e.g., displaced $\cho^+\to W^+
\tilde G$), it is insensitive to models that have displaced
same-flavor leptons originating from different vertices.  Even the RPV
stop benchmark model considered in
\thesearch~\cite{Khachatryan:2014mea} would more generically result in a
same-flavor signature.  In the CMS search, it was assumed that the
$LQD$ RPV operators were lepton-flavor-universal, but, as the known
superpotential couplings (i.e., SM Yukawas) exhibit large hierarchies,
this assumption is not well-motivated.  Hierarchical couplings would
generically produce one dominant stop decay path.  Due to the additional
hadronic activity at these displaced vertices, this RPV stop model is
powerfully constrained by other displaced searches
\cite{Csaki:2015uza,CMS:2014wda,Aad:2015rba}, and will not be
discussed in more detail in this work.

Rather generally, one can frame displaced lepton models as a charged
particle $\phi^\pm$ that decays into an invisible particle $\chi$ and
one of the three flavors of charged leptons, $e$, $\mu$, or $\tau$
with branching fractions, $B_e$, $B_\mu$, $B_\tau$, respectively, that 
sum to unity.  If one has that $B_\tau=0$ and $B_eB_\mu < 0.06$, then
this model has fewer $e^\pm\mu^\mp$ events than the analogous
$B_\tau=1$ case.  Similarly, if one has $B_e=0$ and $B_\mu > B_\tau$,
there are, again fewer $e^\pm\mu^\mp$ events.  Of course, these
oversimplifications  neglect the softer charged leptons arising in
$\tau$ decays and the important effects of $\tau$ helicity.  For
simplicity of discussion, we will use the requirements,
\beq
B_eB_\mu < 0.01 \mbox{ and } B_\tau \lesssim 0.1;
\label{eq:aligncond}
\eeq
to illustrate the parametric requirements for being at most weakly
constrained by \thesearch, i.e., at a level below the
lepton-flavor-universal $\stau$ NLSP models in Section
\ref{sec:staus}.  These conditions are pointed out to highlight the
regions of parameter space where a same-flavor search is
\emph{essential}, as there is little hope that a displaced $e\mu$
search alone will be able to constrain that scenario.  We stress that
even when these conditions are maximally violated, the search we will
propose in Section~\ref{sec:SFsearch} will generically set limits
which, at the very least, would be competitive with
\cite{Khachatryan:2014mea} and will be more sensitive across large
regions of parameter space.  The following subsections will discuss in
detail several models that preferentially give rise to pairs of
displaced single muons; the extension to electrons is trivial.

\subsection{Slepton Co-NLSP}
\label{sec:CoNLSP}

A realistic, minimal possibility within GMSB is that the right-handed
sleptons are all co-NLSPs \cite{Ruderman:2010kj}, each decaying to
their respective SM partner.  In this case, the same-flavor muon and
same-flavor electron signal will appear together, along with the
weakly constrained tau signal.  The proposed same-flavor search
discussed in the next section would be the best handle on long-lived
slepton co-NLSPs, even without invoking one of the 
mechanisms discussed below to produce same-flavor dominated
signatures.

\subsection{Sleptons in Extended Gauge Mediation}
\label{sec:EGMSB}

Models of extended gauge mediation (EGMSB) \cite{Chacko:2001km}, where
one introduces direct couplings into the superpotential between the SM
and messenger superfields, provide a simple mechanism whereby a first-
or second-generation slepton can become the NLSP
\cite{Shadmi:2011hs,Calibbi:2014pza}.  In order to not flood the
casual reader with technical details, here we only present a
streamlined discussion.  Further details are provided in
Appendix~\ref{sec:EGMSBsoft}.

For EGMSB to directly affect right-handed sleptons, one must introduce
couplings in the superpotential of the form
\begin{equation} 
W \supset \kappa_i E^c_i \Phi \tilde\Phi,
\end{equation}
where $\Phi,\, \tilde \Phi$ are messengers with appropriate gauge
quantum numbers.  For simplicity, we will focus on a model where $W
\supset \kappa_i E^c_i \Phi_U \Phi_{\bar D}$, where $\Phi_U$ and $
\Phi_{\bar D}$ have the quantum numbers $(\bar 3,1)_{-\frac23}$ and
$(3,1)_{-\frac13}$, respectively. With the general formulas from
\cite{Evans:2013kxa}, the EGMSB contribution to the slepton mass can
be determined.  This EGMSB-induced splitting can be written
(neglecting the effects of running) as a function of $m_{\tilde \ell}$
and $\kappa_i$,
 \beq
\Delta m_{\tilde \mu} \sim  25 \kappa_2^2 m_{\tilde \ell},
\label{eq:Deltam}
\eeq
for $\kappa_1,\kappa_3 \ll \kappa_2$ and $\Delta m_{\tilde \mu}\ll
m_{\tilde \ell}$; see Appendix~\ref{sec:EGMSBsoft} for details.  With
$\kappa_2 \sim 6\times 10^{-2}$, this will cause an $\order{ 10\mbox{
    GeV}}$ splitting between the smuon and the other right-handed
sleptons.

In order to preferentially generate same-flavor final states, some
alignment is required.  The lightest slepton eigenvalue points in the
$\vec\kappa$ direction within flavor space, so the branching ratios
are simply $B_i =\kappa_i^2/(\kappa_1^2+\kappa_2^2+\kappa_3^2)$.  From
our simple conditions (\ref{eq:aligncond}), we can infer that
\begin{equation} 
\frac{ \kappa_1}{ \kappa_2}\ll 0.1 \mbox{ and } \frac{ \kappa_3}{ \kappa_2}\lesssim 0.3
\end{equation}
forces us into a region of parameter space where a displaced
same-flavor search is essential to constrain this scenario.
 
As this simple model exhibits rank one chiral flavor violation
\cite{Evans:2015swa} in the right-handed sector, it is insulated
against many flavor constraints as compared to an anarchic
scenario. The most constraining flavor observable on the right-handed
slepton mass matrix is $\mu\to e\gamma$ \cite{Adam:2013mnn}, which
typically constrains the product $\kappa_1\kappa_2 \lesssim 10^{-3} $
\cite{Jelinski:2014uba,Calibbi:2014yha}.  For larger splittings and/or
particular choices of EGMSB couplings (e.g.,
\ref{eq:othermodelsplitting}), $\mu\to e\gamma$ could  reduce the viable 
parameter space, but any model with a pure $\tilde \mu$ or $\tilde e$
NLSP will be safe from this constraint.  A general study of flavor
constraints in leptonic $\chi$FV is well beyond the scope of this
work.
   
To summarize, EGMSB models can produce a same-flavor signature by
splitting the $\tilde e$ or $\tilde \mu$ from the other sleptons using
a fairly small, $\order{10^{-1}}$, EGMSB coupling.  The simplified
model requires only a moderate alignment of the flavored coupling
$\vec\kappa$ with the electron ($\kappa_1$) or muon ($\kappa_2$)
direction in order to avoid flavor constraints and give a relatively
pure displaced $e^+e^-$ or $\mu^+\mu^-$ signal.  In principle, this
slepton splitting mechanism is modular and could be combined with
other EGMSB operators, e.g., to alleviate tuning in the Higgs sector.

\subsection{$R$-Parity Violating Decays of Staus via LLE Operators}
\label{sec:LLE}

Another model generically predicting flavor-non-universal slepton
decays arises in the presence of R-parity-violating LLE
operators \cite{Barbier:2004ez}.  With $R$-parity violated, the
following trilinear superpotential terms are now allowed:
  \beq
W\ni   \lambda_{ijk} L_i L_j E^c_k+\lambda'_{ijk} L_i Q_j D^c_k+\lambda''_{ijk} U^c_i D^c_j D^c_k.
\eeq
A stau that is at least partially left-handed can decay via the
$\lambda_{i32} L_i L_3 E^c_2$ operator to give a pure muon final
state, i.e., $\stau_1^+\to \mu^+\nu_i$.  The stau lifetime in these
models can be expressed as
  \beq
  c\tau \approx 1\mbox{ cm} \lp \frac{10^{-7}}{\lambda_{i32}} \rp^2
  \lp \frac{100\mbox{ GeV}}{m_{\stau}}  \rp \sec^2 \theta_\stau ,
\eeq  
where the mixing angle $\theta_\stau=0$ corresponds to a pure
left-handed state, and we have defined $\lambda_{i32}\equiv
\sqrt{\lambda_{132}^2+\lambda_{232}^2}$.
  
From a model-building perspective, it is more generic to have an
(N)LSP stau be right-handed than left-handed, but some models, such as
GGM \cite{Meade:2008wd}, can readily accommodate a spectrum with an
NSLP left-handed stau. However, some degree of left-right mixing is
generic, due to the Yukawa-induced mass term, $(\mu^* m_\tau
\tan\beta) \stau_L \stau_R$.  If the $\lambda_{i32}$ coupling is the
only non-zero RPV coupling, then the stau decay will proceed to the
100\% muon final state, even for $\theta_\stau \sim \pi/2$.  More
generally, from the criteria of (\ref{eq:aligncond}), the
$\lambda_{i32}$ coupling must dominate by
 \beqa
\lambda_{i32}&\gg& 10 \sqrt{ (\lambda_{123}^2+\lambda_{133}^2) \tan^2\theta_\stau +  \lambda_{131}^2+\lambda_{231}^2} \\
\lambda_{i32}&\gtrsim& 3\sqrt{\lambda_{233}^2 (\tan^2 \theta_\stau+2)+\lambda_{133}^2}
\eeqa  
in order for an opposite-flavor search to have little to no sensitivity.
Additionally, the LQD couplings $\lambda'_{3ij}$ must also be small to
evade constraints from the displaced jet searches
\cite{CMS:2014wda,Aad:2015rba}.  Since the small RPV couplings yield
long lifetimes, we note that $\snu \to \stau+\{\mbox{soft}\}$
transitions will occur much more rapidly than $\snu \to \ell^+\ell'^-$
decays, unless $m_\snu-m_\stau< 1$ GeV \cite{Kraml:2007sx}.
   
There are typically no flavor constraints in this model due to the
small sizes of the RPV couplings; the only exception is proton decay
when UDD operators are simultaneously introduced.  For particular
flavor structures, some of these bounds require that
$\abs{\lambda_{ijk}\lambda''_{i'j'k'}}\lesssim 10^{-26}$
\cite{Barbier:2004ez}.  Imposing baryon number conservation removes
all such issues.  Alternatively, as long as first-generation particles
are not heavily involved, UDD coefficients could still be as large as
$10^{-3}$ without any conflict with proton decay constraints
\cite{Barbier:2004ez}.  Amusingly, it would be possible for displaced
lepton signatures to live alongside a prompt paired-dijet signature of
RPV stops.

\subsection{Lepton-Flavored Dark Matter from Freezein}
\label{sec:LFDM}

Models of flavored dark matter, where the dark matter is charged under
the flavor symmetries of either the quarks or the leptons, can give
rise to novel signatures at the LHC \cite{Agrawal:2011ze}.  In these
models, the lifetime of the decaying particle is generically directly
related to the cosmological abundance of dark matter.  Long lifetimes
at colliders require couplings which are typically much smaller than
those required for DM to originate from thermal freezeout to the SM.
On the other hand, very small couplings are naturally predicted by
models of {\em freezein}, where dark matter is produced by the
out-of-equilibrium decays of a particle in the thermal bath
\cite{Hall:2009bx}.

Minimal models of lepton-flavored freezein DM can be written in terms
of a fermionic DM flavor multiplet $\chi_i$ and a charged scalar $\zeta$,
\begin{equation} 
 \mathcal{L}\supset y^{LDM}_{ij}\ell^c_i  \zeta^-  \chi_i + m_{\chi,ij} \chi_i \bar \chi_j +\hc  +m^2_\zeta \zeta^+\zeta^-,
\end{equation}
or a scalar DM flavor multiplet $S_i$ and a charged fermion $\psi$,
 \beq
 \mathcal{L}\supset y^{LDM}_{ij}\ell^c_i  \psi  S_j +m_\psi \psi \bar\psi +\hc + m^2_{S, ij} S_i^\dagger S_j.
 \label{eq:fermportal}
\eeq
The charged particle is present in the thermal bath, and decays
via the small flavored Yukawa coupling $y ^ {LDM}_{ij}$.  When $y\lll
1$, the resulting out-of-equilibrium decays produce a relic abundance
of DM that is directly proportional to the width of the charged parent.
Note that unlike all other models discussed in this work so far, the
new charged particle of the model in (\ref{eq:fermportal}),
$\psi$, is a fermion, and thus has a higher production cross-section
for a given mass. For this reason we will specialize to that model
throughout the rest of this subsection.  For freezein in a standard
thermal cosmology, the values of $y$ that yield acceptable relic
abundances imply that $\psi$ will be detector-stable, unless the dark
matter mass is low enough to present serious issues with structure
formation, i.e., free-streaming and Tremaine-Gunn constraints
\cite{Boyarsky:2008xj,Boyarsky:2008ju}.  However, an alternative
generic possibility is that, at the time of dark matter freezein,
characterized by the temperature $T_{FI}\sim m_\psi/4$, the energy
density of the universe is dominated by the coherent oscillations of a
massive field, e.g., an inflaton or a heavy modulus
\cite{Chung:1998rq,Giudice:2000ex}.  This non-thermal phase of
evolution terminates when the massive field decays to radiation (e.g.,
the SM).  Some non-thermal epoch is required in any inflationary
cosmology in order to populate the thermal bath of the SM after
inflation, and indeed, the prime example of such an epoch is
post-inflationary reheating. The coupling strengths necessary for
freezein during a matter-dominated era to produce the correct DM
relic abundance today are much larger than the couplings required by
standard radiation-dominated freezein, as the higher initial DM
density is diluted by the entropy released when the heavy particle
decays.  These coupling strengths can provide lifetimes relevant for
displaced decays at colliders \cite{Co:2015pka}.  We discuss the
details of this mechanism in Appendix~\ref{sec:MDFI}.

For the purposes of this work, we will use a simple model with
  \beq
 \mathcal{L}\supset y_i \ell^c_i  \psi  S  + m_\psi \psi \bar\psi +\hc + m^2_{S} S^\dagger S,
 \label{eq:fermportal2}
\eeq
where the Yukawa coupling $y_i$ is flavor-aligned with one species of
lepton.  For a lifetime $c\tau$ yielding a displaced collider
signatures and mass $m_\psi$, one can typically choose $m_S$ and
$T_{RH}$ such that the dark matter relic abundance matches the
observed value today (see Appendix~\ref{sec:MDFI}).  For simplicity of
illustration, we will always choose the dark matter to be effectively
massless for the purposes of LHC kinematics, i.e., $m_S\ll m_\psi$,
although in some instances (lower $\psi$ mass and/or shorter
lifetimes) this may imply that the $S$ relic abundance represents only
a portion of the dark matter density today.
 
As $y_i\lesssim 10^{-7}$, there are no constraints from precision flavor
observables.  From our conditions (\ref{eq:aligncond}), we can infer
that
 \beq
\frac{ y_1}{y_2}\ll 0.1 \mbox{ and } \frac{ y_3}{ y_2}\lesssim 0.3
\eeq
forces us into a region of parameter space where a displaced
same-flavor search is essential to constrain this scenario.

\section{A Search for Displaced Same-Flavor Leptons}
\label{sec:SFsearch}

In this section, we will construct a simple search for same-flavor
leptons with large impact parameters.  The heavy stable charged
particle searches have been projected to 13 TeV elsewhere
\cite{Feng:2015wqa}, and while there are new results using 2.4
fb$^{-1}$ of data at 13 TeV \cite{CMS:2015kdx}, we will not recast
these in this work.  While we have made some suggestions on how to
improve the sensitivity of the existing disappearing track searches to
this kinked track scenario, estimating backgrounds for these searches
is beyond the scope of this work, so we will make no attempt to design
13 TeV versions of these searches.
 
Estimating the backgrounds to a search for displaced same-flavor
leptons is challenging, and, especially in the case of leptons coming
from heavy flavor, requires data-driven techniques.  To approximate
the backgrounds at 13 TeV for a displaced same-flavor lepton search,
we will utilize \thesearch's 8 TeV background projections, shown in
Figure 1 of Ref.~\cite{Khachatryan:2014mea}, which are in very good
agreement with the data.  In order to use these backgrounds directly,
we will mirror the cuts of \thesearch~(table \ref{tab:cuts}), adding a
$0.3<\Delta \phi_{\mu\mu} < 2.8$ cut to remove backgrounds from cosmic
muons and cosmic muon bundles (we assume this has a negligible effect
on all other backgrounds\footnote{In fact, this is a conservative
  assumption, as the $\Delta \phi$ cut could potentially help suppress
  heavy flavor and $Z\to\tau\tau$ backgrounds even further, at minimal
  cost to signal acceptance.}).  While existing Run II studies (e.g.,
\cite{Khachatryan:2015uqb}) rely on lepton triggers with $p_T$
thresholds well below the lepton acceptance cuts of
\cite{Khachatryan:2014mea}, these thresholds will almost certainly
increase with higher luminosity.  For the purpose of this sensitivity
study, we choose to continue with the Run I cuts, instead of
confronting backgrounds we cannot reliably estimate.

There are several backgrounds relevant for the $e\mu$ channel
\cite{Khachatryan:2014mea}: heavy flavor, $Z\to\tau\tau$, top, and
other electroweak processes.  As all of these backgrounds can contain
$b$s, $c$s, and/or $\tau$s which can give a genuine displacement due
to their long lifetimes, it is a priori unclear what fraction of the
background has a genuinely large lepton impact parameter and what
fraction is due to track mis-reconstruction or detector effects
creating an artificial displacement from prompt leptons.  As both 
$t\bar t$ and  electroweak backgrounds are very small in the
$e\mu$ search, we assume that prompt sources of same-flavor leptons,
notably $Z\to \ell^+\ell^-$, can be neglected or controlled, e.g., by
cutting out a $Z$ window in the lepton invariant mass.
 
Estimating the 8 TeV same-flavor backgrounds from the data presented
in the CMS opposite-flavor search~(Figure 1 of
Ref.~\cite{Khachatryan:2014mea}) requires several assumptions and
approximations.  First, we assume that, for each background $x$, the
two lepton displacements are uncorrelated and the population of
background events can be factorized, i.e.,
\beqa
\label{eq:probcomb} P^x_{e\mu}(d_e,d_\mu)&=&2 P^x_e(d_e) P^x_\mu(d_\mu) \\
P^x_{ee}(d_{e_1},d_{e_2})&=& P^x_e(d_{e_1}) P^x_e(d_{e_2}) \\
P^x_{\mu\mu}(d_{\mu_1},d_{\mu_2})&=&  P^x_\mu(d_{\mu_1}) P^x_\mu(d_{\mu_2}).
\eeqa
We also assume that prior to the application of displacement cuts
and selection efficiencies \cite{CMSemuEfficiency}, all backgrounds
are flavor universal.  Guided by the assumption that genuine
displacement of $b$s, $c$s, or $\tau$ parents dominate the backgrounds
at smaller displacement, we assume that the background shape in the
first several bins can be fit as
\begin{equation} 
P^x_\ell(d)= \epsilon_\ell (d) A^x_\ell e^{-\alpha^x_\ell d},
\label{eq:probfunc}
\end{equation}
where $\epsilon_\ell$ is the displacement-dependent lepton selection
efficiency \cite{CMSemuEfficiency}, and the fit parameters $A^x_\ell$
and $\alpha^x_\ell$ for each background, $x$, depend only on the
lepton species.  This exponential assumption is supported by the data
in Figure 1 of Ref.~\cite{Khachatryan:2014mea}.  However, due to the
preselection requirement of $d_\ell >0.1$ mm, the $d_\ell \leq0.1$ mm
data is not presented at all in the search.  In order to approximate
these regions, we use the first four $d_0$ bins to derive
$\alpha^x_\ell$ in (\ref{eq:probfunc}) for each of the three main
backgrounds $x$ ($Z$, HF and top).\footnote{For simplicity, we neglect
  the negligibly small ``other EW'' backgrounds.  When we extrapolate
  to 13 TeV, we assume these backgrounds remain negligibly small.}
With this exponential we can extrapolate an expression for the missing
$0\leq d_0\leq 0.1$ mm bin in each background for both electron and
muon samples (when the other lepton has $d_0 > 0.1$ mm).  Then, using
the ABCD method across the selections $d_0 \leq 0.1$ mm and $d_0 >
0.1$ mm for both electrons and muons, we can estimate the $e$ and
$\mu$ $d_0 \leq 0.1$ mm bin separately for each of the three
background channels.\footnote{As we assume genuine displacements
  dominate the backgrounds, we expect this estimate would not produce
  the true contents of the $d_0 \leq 0.1$ mm bins, but capture only
  those effects that scale approximately like exponentials which have
  not become negligibly small for $d_0>0.1$ mm.  In particular, we
  would expect $t\bar t$ to have a very large population from prompt
  leptons.}  Using this information and factoring out the
identification efficiencies, we can derive normalizations $A_e^x$ and
$A_\mu^x$ in (\ref{eq:probfunc}) for the full distributions under the
assumption that the total truth-level background events are
flavor-universal (i.e., the same number of electrons and muons are
found in each sample).  With this factor, we normalize the CMS
background distributions in Figure 1 of Ref.~\cite{Khachatryan:2014mea}
and, using these normalized distributions as the $P^x_\ell(d)$ in
(\ref{eq:probcomb}), have enough information to make an estimate of
the 8 TeV {\em same}-flavor backgrounds in the signal regions of the
CMS search.  We apply a systematic uncertainty of 30\% (10\%) to our
estimates of the HF ($Z$ and top) backgrounds.  As a cross-check, we
compare our resulting estimate for the 8 TeV $e\mu$ backgrounds to the
published background estimates in Table~\ref{tab:SRs}.  Our estimates
agree with the expected experimental backgrounds to within 10\% of the
published results.  The residual disagreement, which is too small to
substantially affect our conclusions, can be understood as a breakdown
of our assumption that the two lepton displacements are uncorrelated.

 \begin{table}[!t]
\begin{center}
\begin{tabular}{|l|ccc|}
\hline
 \bf Sample & \bf  SR1 &  \bf SR2 &  \bf SR3 \\
 \hline  \hline
$e^\pm\mu^\mp$ 8 TeV (CMS actual) & $18.0\pm3.8$ & $1.01\pm0.31$ & $0.051\pm0.018$  \\
$e^\pm\mu^\mp$ 8 TeV (our estimate) & $19.8\pm4.1$ & $0.92\pm0.28$ & $0.055\pm0.024$  \\
 \hline  \hline
$e^\pm\mu^\mp$ 13 TeV & $34.1\pm6.5$ & $1.49\pm0.44$ & $0.086\pm0.038$  \\
$e^+e^-$ 13 TeV & $25.2\pm3.6$ & $1.43\pm0.33$ & $0.31\pm0.06$  \\
$\mu^+\mu^-$ 13 TeV & $13.0\pm3.1$ & $0.50\pm0.15$ & $0.012\pm0.006$  \\
  \hline 
\end{tabular}
\end{center}
\caption{Projected backgrounds estimated using the methods described
  in the text.  Our 13 TeV extrapolations assume 20 fb$^{-1}$, a 30\%
  systematic uncertainty on the heavy flavor backgrounds, and 10\% 
  systematic uncertainties on all other backgrounds.}
\label{tab:SRs}
\end{table}

To project these background estimates to 13 TeV, we again must make
several assumptions and approximations.  For top and $Z$ backgrounds,
we assume these are dominated by near-threshold production, so we
simply rescale these by the ratio of the inclusive cross-sections,
$\sigma_X(13\mbox{ \small TeV})/\sigma_X(8\mbox{ \small TeV})$, and
neglect effects of altered lepton kinematics.  This cross-section
ratio is 3.28 for top and 1.74 for $Z$ production
\cite{Campbell:2006wx}.  At 8 TeV the HF backgrounds are dominant in
all signal regions, and at 13 TeV we expect this to remain true.
However, there are multiple competing effects that can influence the
scaling of the HF background.  First, the $b\bar b$ cross-section
rises by 1.53 (we do not separately model the charm contribution for
simplicity) \cite{Campbell:2006wx}.  Additionally, the $b\bar b$
kinematics change so that more $b$s are boosted.  Boosted $b$s produce
harder leptons and survive to longer displacements before decaying,
but also result in leptons with smaller opening angles and produce
harder hadrons that can foil isolation.  Whether more boost of the
parent $B$ meson translates to more isolated displaced leptons is
unclear a priori.  In Monte Carlo $b\bar b$ samples, we examined the
probability to find an isolated, displaced lepton of sufficient $p_T$
from a heavy flavor decay.  This probability was found to be
approximately independent of the boost of the parent $B$ meson.
Although this information was determined from Monte Carlo and thus
should be viewed with some caution, we took this as sufficient
evidence that, for the purposes of this study, we could neglect the
effects of altered $b\bar b$ kinematics and rescale the heavy flavor
background by the cross-section alone.  In doing this rescaling for
each background, we are implicitly assuming that the tails of the
distributions also scale simply with the cross-section; however, as
the background estimates are rather small in SR2 and SR3 where these
tails are most relevant, only an egregious underestimate would result
in a qualitative change to our projected limits.
 
\begin{figure}[!t]
\begin{center}
\includegraphics[scale=0.6]{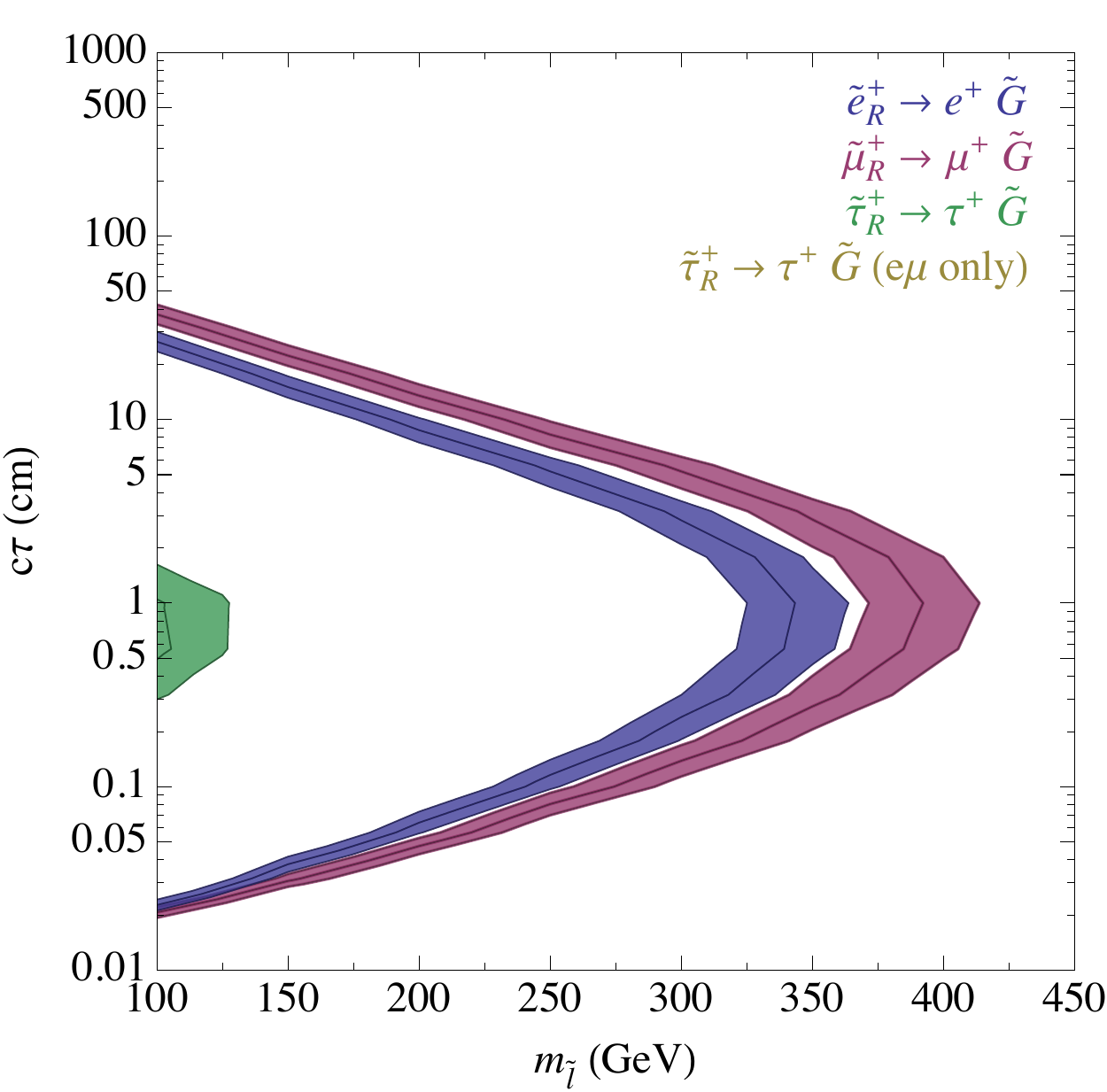}
\includegraphics[scale=0.6]{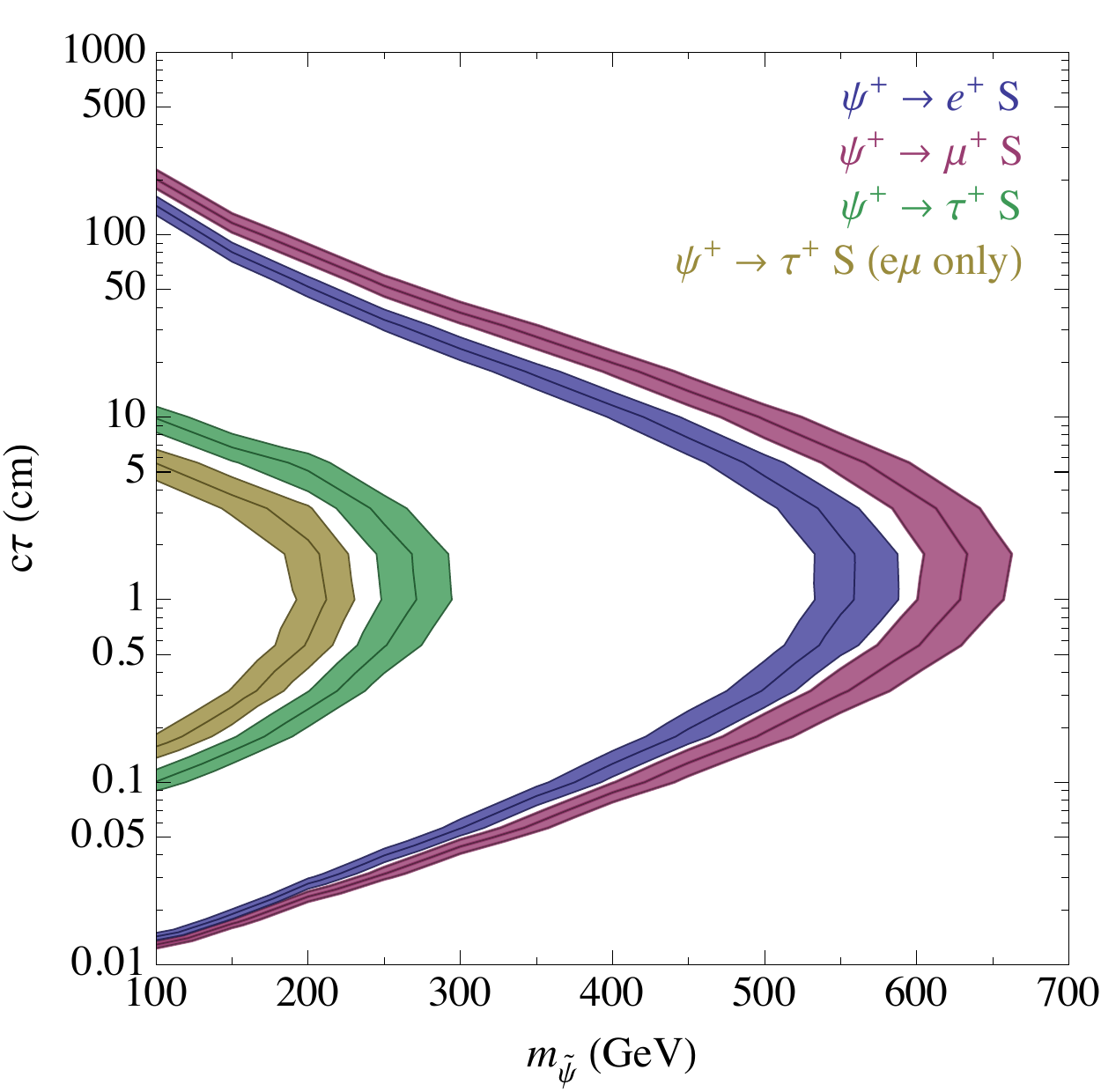}
\end{center}
\caption{{\bf Left:} 13 TeV reach for direct production of a single
  species of slepton that decays as $\tilde \ell\to \ell \tilde G$.
  Using a 13 TeV version of \thesearch\ without the same-flavor
  channels has no sensitivity.  The width of the band reflects a 25\%
  modeling uncertainty. {\bf Right:} 13 TeV reach for direct
  production of a unit charge singlet fermion that decays to a single
  species of lepton and a very light dark matter particle.  In gold,
  we present limits on $\tau$-flavored dark matter without including
  the same-flavor channels (which are weaker by almost 100 GeV).  We
  show a 25\% modeling uncertainty.}
\label{fig:direct13}
\end{figure}

Rescaling the individual distributions from 8 TeV to 13 TeV and
projecting the same HF ($Z$ and top) systematic uncertainty of 30\%
(10\%) present in the 8 TeV data, we derive estimates for the
different signal regions (Table \ref{tab:SRs}).  Using these
background projections, we can estimate the 13 TeV sensitivity to
models of direct slepton production with EGMSB-like decay chains
$\tilde e^\pm \to e \tilde G$ and $\tilde \mu^\pm \to \mu \tilde G$.
In addition to combining all nine 13 TeV search regions (table
\ref{tab:SRs}) to project limits on a $\stau_R$ NLSP, we also show the
limits from the $e\mu$ channel alone to illustrate the improvement a
combination gives to the reach.  Lastly, we show the reach for a
lepton-flavored dark matter motivated model with an
$SU(2)_L$ singlet  charged fermion that decays as
$\psi^\pm \to e/\mu/\tau S$ to a light scalar dark matter $S$.
 This model, which has been discussed elsewhere \cite{Bai:2014osa} 
 with larger $\psi \ell S$ couplings, was constructed in FeynRules 
 \cite{Christensen:2008py} with all limits presented using leading-order 
 cross-sections. All results are shown in Figure~\ref{fig:direct13}.

While we chose to mirror the 8 TeV search in order to get a more
reliable modeling of the background, we note that all of the
same-flavor models typically predict very hard leptons (unlike in the
$\stau$ cases).  Considering not only lower $p_T$ thresholds essential
for sensitivity to staus, but also a higher lepton $p_T$ threshold
signal region, e.g., SR1' with $p_{T,\ell}>50$ GeV, could vastly
reduce backgrounds in SR1 while having minimal impact on the benchmark
signal models.  This additional search region could greatly increase
sensitivity at lower $c\tau$ values.

\section{Discussion and Conclusions}
\label{sec:conclusions}

Probing all feasible lifetimes for NLSP particles in GMSB is paramount
in the search for new physics at the LHC.  The very generic GMSB
scenario containing an NLSP $\stau_R$ is currently under-constrained
for many macroscopic stau lifetimes.  The only existing search able to
target the displaced leptons from $\stau_R$ decays is the CMS search
for $e^\pm\mu^\mp$ with large impact parameters
\cite{Khachatryan:2014mea}, while HSCP
\cite{Chatrchyan:2013oca,ATLAS:2014fka} and disappearing track
searches \cite{Aad:2013yna,CMS:2014gxa} can target the long-lived
sleptons themselves.  With the exception of the HSCP searches, the
experimental analyses did not consider NLSP staus as one of their
benchmark signal models, so the cuts were not tailored to probe the
specific signatures of long-lived $\stau$s.  Only the HSCP search
currently places limits beyond those of LEP on the direct production
of a $\stau_R$ NLSP.

In Section \ref{sec:staus}, we recast an HSCP search, two disappearing
track searches, and \thesearch\ to place constraints on direct stau
production, as well as on simplified models with displaced staus
originating from cascade decays initiated by Higgsinos, stops, and
gluinos.  While we find meaningful constraints on these models,
several modifications to the searches were discussed in detail that
could improve sensitivity to long-lived staus.  Our most important
suggestion for the disappearing track searches and \thesearch\ is
simply to include NLSP staus as a benchmark model.  We found the
recasting recommendations provided in the CMS searches to be
invaluable for our recasting efforts.  If the ATLAS disappearing track
search were to provide efficiency maps or similar resources, it would
greatly improve the reliability of any recast of their results.  It may be 
possible to improve sensitivity to staus if the ATLAS disappearing
track search were to check for energy deposits that originate from the 
terminus of the disappearing track and/or arrive later than typical by 
employing their calorimeter's exceptional pointing and timing 
capabilities.  Similarly, CMS could use their calorimeter's timing 
information to permit delayed energy deposits to live within their 
strict isolation cone.  
More generally for the disappearing 
track searches at both experiments, an extension or related analysis 
that attempts to reconstruct a kinked track signature could greatly 
improve sensitivity to sleptons.

For \thesearch, which uses the most recently designed experimental
strategy, we discussed in Secs.~\ref{sec:CMSemu}
and~\ref{sec:SFsearch} several avenues to improve sensitivity to
displaced $\stau_R$ NLSPs and similar signatures.  We briefly
summarize these proposed improvements and suggest a few other
possibilities that could enhance the sensitivity:
\begin{itemize}

\item Leptons from boosted right-handed $\tau$ decays are typically
  soft, and thus lowering the $p_T$ thresholds as much as possible for one
  or both species of leptons can greatly improve signal acceptance.

\item The stringent isolation requirements could be relaxed in the
  higher displacement (and thus lower background) signal regions in
  order to increase the signal acceptance.  Again, as leptons from
  right-handed $\tau$ decays are typically soft, relaxing the
  isolation criteria can have a notable impact on signal acceptance.

\item The flavor-universal decay of the $\stau$ results in not only
  $e^\pm \mu^\mp$ final states, but also $e^+e^-$ and $\mu^+\mu^-$.  A
  combination of all channels can improve reach, especially because of
  the lower expected backgrounds in the case of $\mu^+\mu^-$.

\item The veto on additional leptons seems unnecessary and
  can reduce acceptance in noisier production channels, e.g.,
  gluino-initiated decay chains.

\item The presence of Majorana particles such as gluinos or neutral
  Higgsinos in the decay chain can give rise to same-sign lepton
  signatures (as can mesino oscillation).  Not only can the inclusion
  of same-sign lepton bins extend the reach, but the effects of the
  same-sign lepton signals should be considered in the context of
  control region contamination.

\item There are 1.5 orders of magnitude in $c\tau$ between the peaks
  in sensitivity for the disappearing track searches and \thesearch,
  with a noticeable deficiency in the range $c\tau=3$ -- $5$ cm. For this
  reason, it is very important to be able to extend the search regions
  beyond the $d_0<2$ cm range.  If electron reconstruction cannot be
  extended to higher impact parameters, extending the range of muon
  reconstruction alone could still notably increase sensitivity to
  longer lifetimes ($c\tau\sim 10$ cm), especially as these high
  displacement regions are likely to remain low in background
  (although cosmic muon backgrounds may become more important).

\item Including highly displaced hadronic taus in $e\tau_h$,
  $\mu\tau_h$, and even $\tau_h\tau_h$ channels, would improve the
  reach.  Determining the feasibility of such a search is beyond the
  scope of this work, but we note this possibility as one of the most
  robust, if challenging, ways to extend sensitivity to long-lived $\stau_R$s.

\end{itemize}

While long-lived $\stau_R$s are a particularly well-motivated signal
model, it is worth noting on more general grounds that the current LHC
search program has a gap in coverage for same-flavor solitary
 leptons with large impact parameters.  While
HSCP searches and to a lesser extent displaced track searches provide
good coverage at longer lifetimes, at shorter lifetimes these
signatures can be efficiently hidden from standard prompt BSM
searches, thanks to the tight lepton quality criteria and cosmic muon
vetoes employed by these analyses.  Of the large and increasing number
of LHC searches for displaced objects, only \thesearch\, is in
principle sensitive to solitary displaced
leptons, and would miss any model that preferentially yields
same-flavor leptons.  We have discussed several classes of theories
which can give rise to displaced same-flavor lepton signatures, such
as extended GMSB, RPV SUSY, and lepton-flavored dark matter, and have
proposed specific extensions to existing search strategies to enhance
discovery prospects for these signatures.  Additionally, the
same-flavor signature would be the best handle on models of GMSB with
long-lived co-NLSP sleptons, and would provide valuable additional
sensitivity to stau NLSPs alone.  As displaced leptons are both a
well-motivated exotic detector object and one of the least constrained
by current searches, closing this gap is a key step in maximizing the
physics potential of the LHC as Run II goes forward.

\section*{Acknowledgments}
\noindent
We thank F.~Bishara, J.~P.~Chou, M.~Diamond, C.~Frugiuele, A.~Haas, 
C.~Hill, Y.~Kats, S.~Knapen, A.~Monteux, and D.~Shih for useful conversations.  
We thank L.~Quertenmont and W.~Wulsin for answering questions about 
specific searches.  We thank D.~Lamprea for assistance with Resummino.  
We are especially grateful to J.~Antonelli for useful discussions and
comments on the draft.

\appendix

\section{Validation of Recasting Procedures}
\label{sec:val}

     \begin{figure}[!t]
\begin{center}
\includegraphics[scale=0.6]{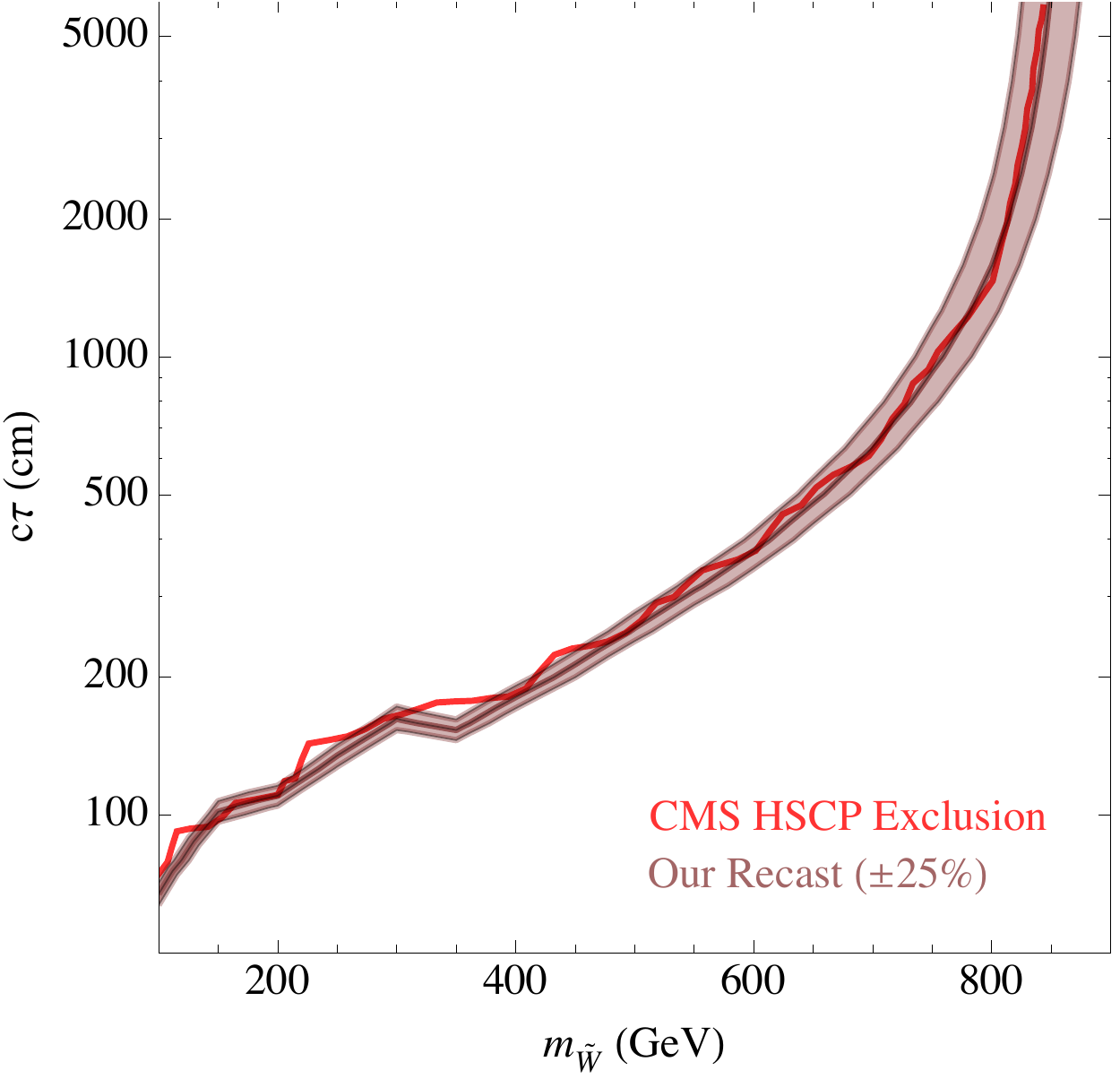}
\includegraphics[scale=0.6]{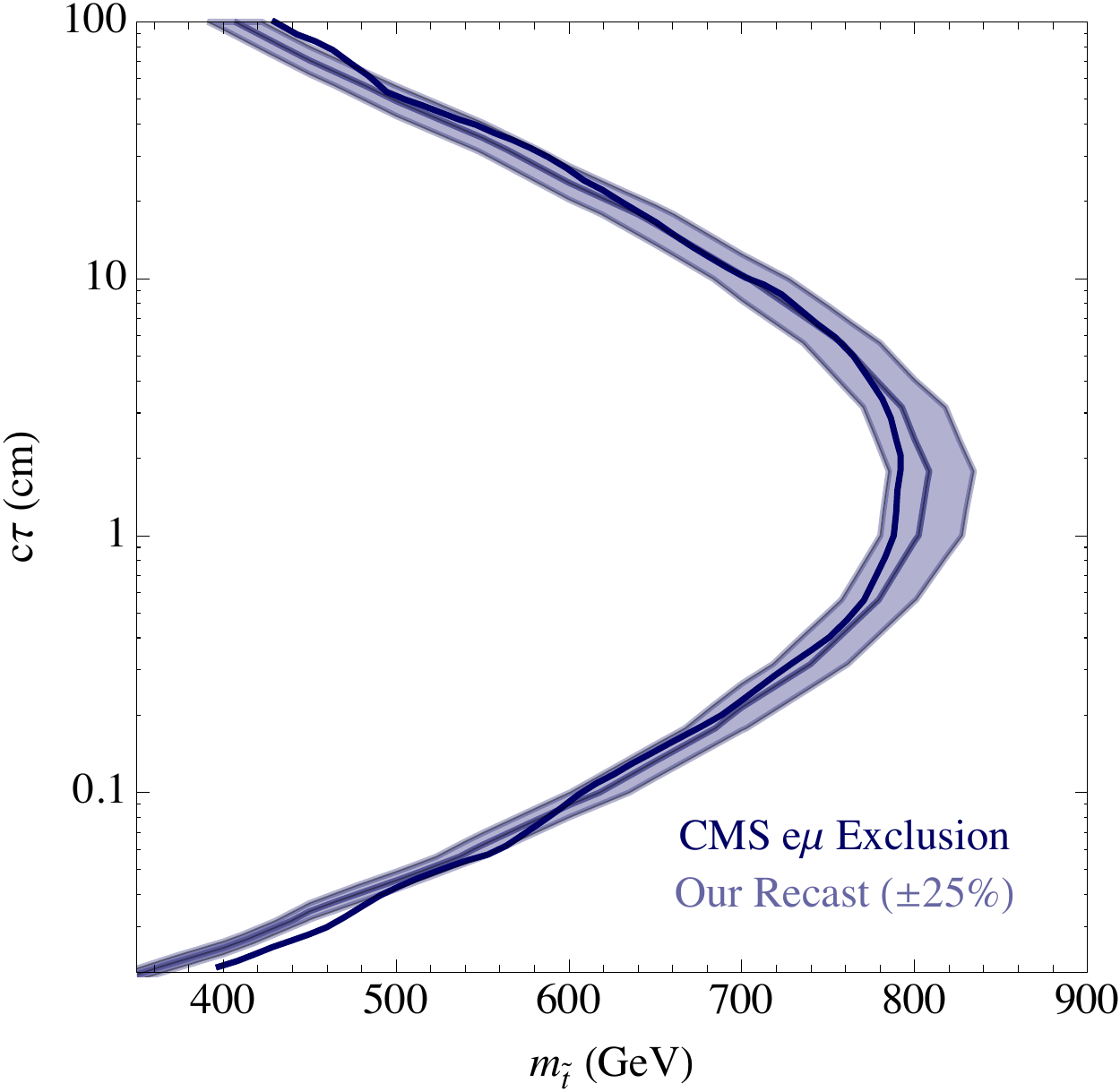} \\
\includegraphics[scale=0.6]{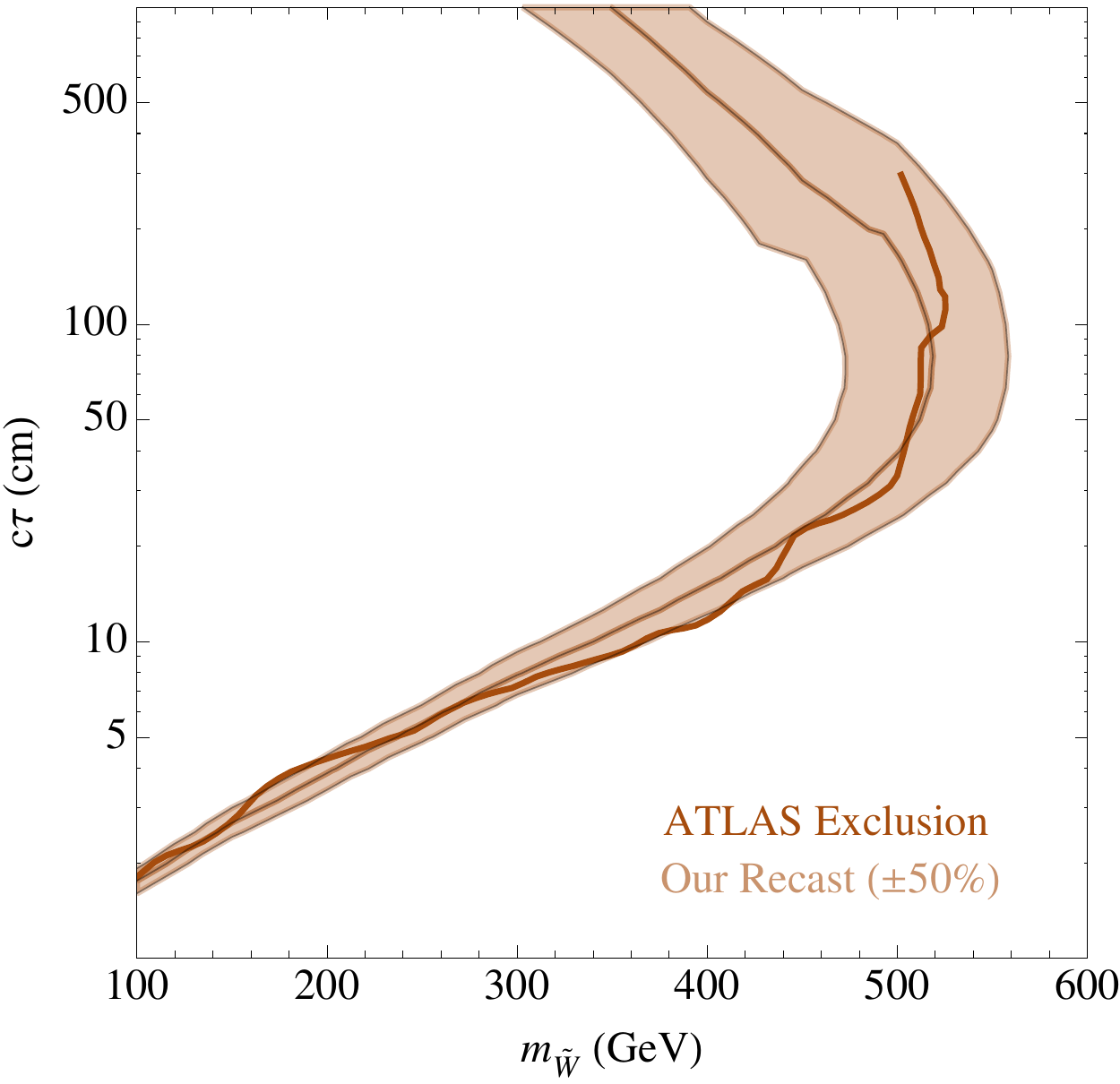}
\includegraphics[scale=0.6]{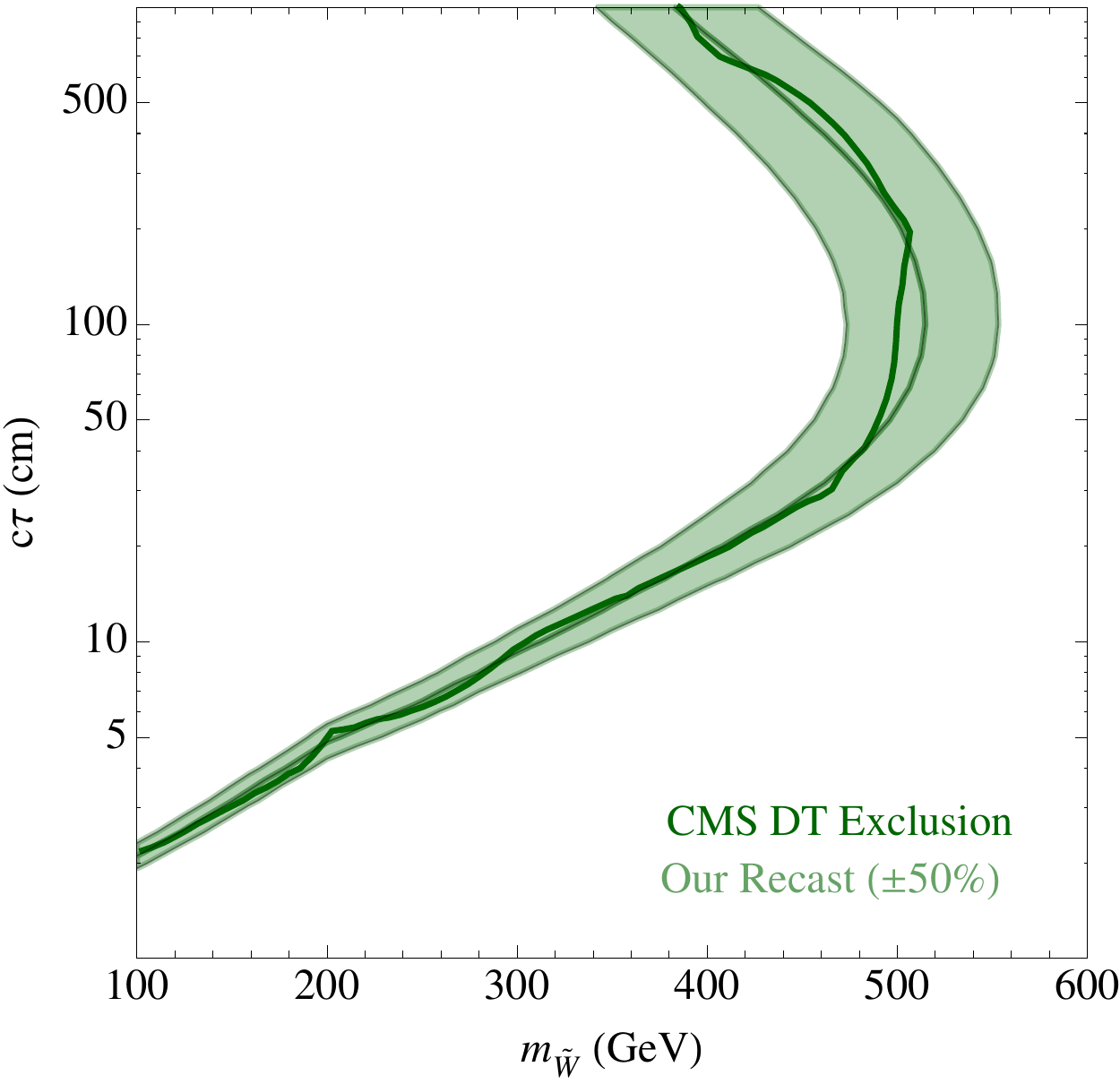}
\end{center}
\caption{{\bf Upper Left:} Validation of the CMS search for heavy
  stable charged particles \cite{Khachatryan:2015lla}.  The width 
  of our recast exclusion band reflects a 25\% modeling 
  uncertainty.  {\bf Upper
    Right:} Validation of \thesearch\ \cite{Khachatryan:2014mea} for
  the displaced supersymmetry benchmark model
  \cite{Graham:2012th}. {\bf Lower Left:} Validation of the ATLAS
  disappearing tracks search \cite{Aad:2013yna}.  {\bf Lower Right:}
  Validation of the CMS disappearing tracks search
  \cite{CMS:2014gxa}.}
\label{fig:DTval}
\end{figure}

In Figure~\ref{fig:DTval}, we present our validation results for each
of the four searches we consider in detail
\cite{Khachatryan:2015lla,Khachatryan:2014mea,Aad:2013yna,CMS:2014gxa}.
In the case of \thesearch, the benchmark signal model is stop pair
production with displaced $R$-parity-violating decays $\st \to \ell_i
b$, with equal branching fractions to each of the three species of
leptons.  The other three searches consider an AMSB wino model.  Both
the CMS HSCP and CMS disappearing track (DT) searches agree
excellently across the entire parameter space.  In the case of the
ATLAS disappearing track search, agreement is very good for most of
the parameter space, but we observe $\order{50\%}$ discrepancies
between our recast result and the experimental result at higher values
of $c\tau$.  The \search\ search agrees very well in the
region where it is most sensitive, 300 $\mu$m $\lesssim c\tau \lesssim
50$ cm, but exhibits significant deviations on the tails of
sensitivity.  It is likely the case that we are slightly
underestimating sensitivity for lifetimes near 1 m or 100 $\mu$m, but
this discrepancy has no qualitative impact on the results.

For the CMS HSCP search and \thesearch, we apply the recommended 25\%
modeling uncertainty.  For both disappearing track searches we apply a
50\% modeling uncertainty, primarily because of the additional
uncertainty introduced by the decay products originating from the
displaced secondary stau vertex.

\begin{figure}[!t]
\begin{center}
\includegraphics[scale=0.6]{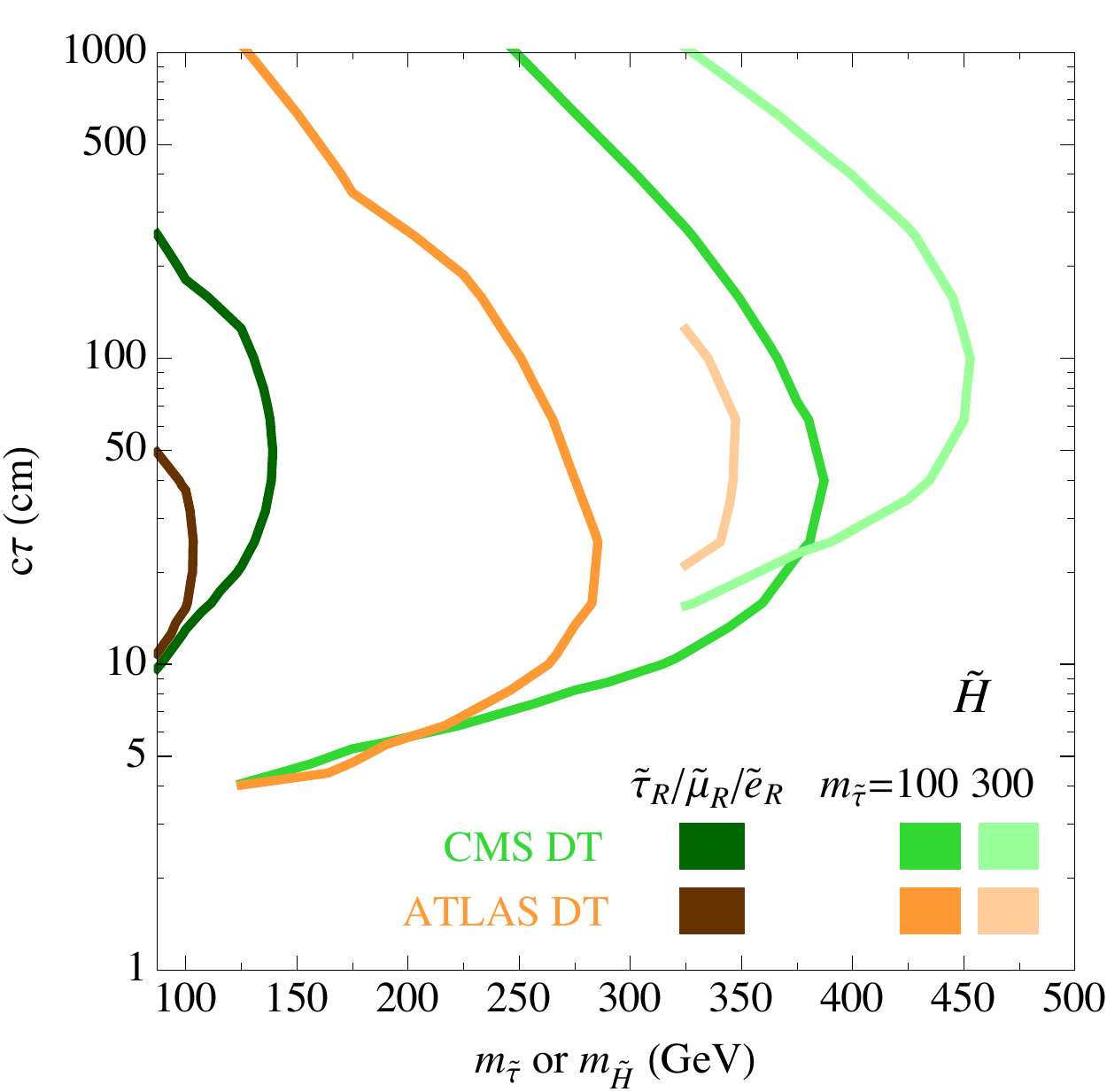}
\includegraphics[scale=0.6]{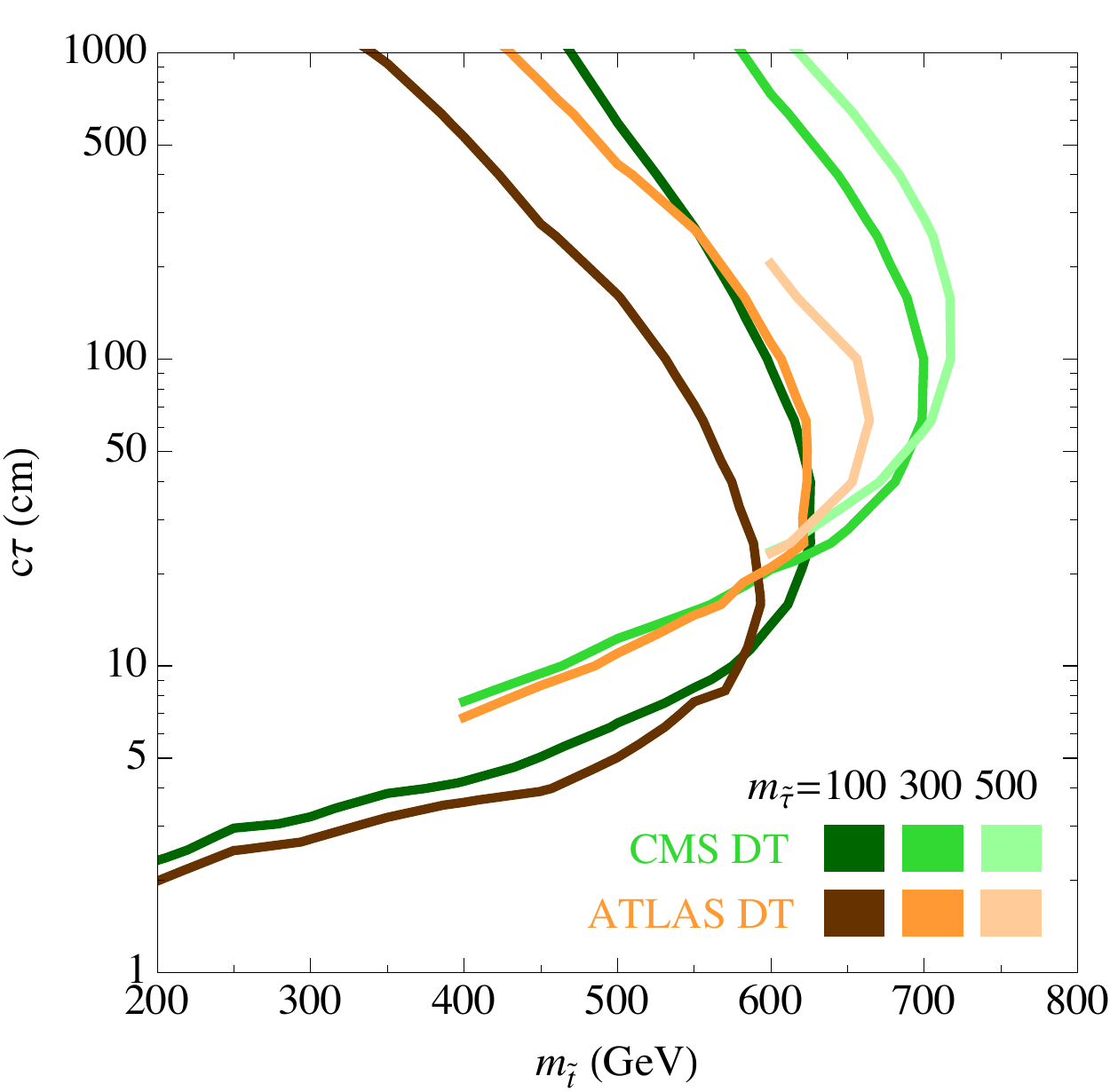}
\end{center}
\caption{{\bf Left:} Comparison of the disappearing track searches at
  ATLAS \cite{Aad:2013yna} (orange) and CMS \cite{CMS:2014gxa} (green)
  for the case of direct production of three flavors of sleptons
  (dark) and Higgsinos (see Figure \ref{fig:direct}).  {\bf Right:}
  Comparison of the disappearing track searches for the case of stops
  (see Figure \ref{fig:stopgo} right).}
\label{fig:DTcomp}
\end{figure}

In Section~\ref{sec:GMSBstaus}, for lucidity in presentation we
display at a given $(m, c\tau)$ only the stronger of the two limits
from the ATLAS and CMS disappearing track searches.  In
Figure~\ref{fig:DTcomp}, we show the sensitivity of the two
disappearing track searches separately to several of the simplified
signal models considered in Section~\ref{sec:GMSBstaus}.  In all
scenarios, ATLAS has a markedly reduced sensitivity at longer
lifetimes.  This can be understood easily as ATLAS vetoes tracks that
reach the muon chamber whereas CMS does not.  Due to the presence of
additional prompt leptons in the Higgsino-initiated simplified model,
ATLAS vetoes more events and finds weaker limits compared to CMS.  At
shorter lifetimes, the ATLAS search typically performs slightly better
than the CMS search, but the two set nearly identical limits after our
modeling uncertainties are taken into account.

\section{Details of the Extended Gauge Mediation Model}
\label{sec:EGMSBsoft}

In this appendix, we present a more detailed discussion of the EGMSB
model presented in Section~\ref{sec:EGMSB}.  In gauge mediation, a SM
singlet superfield $X$ acquires a VEV and an $F$-term, i.e.,
$\veva{X}=M + \theta^2 F$, thereby breaking SUSY (for a nice review,
see \cite{Giudice:1998bp}).  In minimal GMSB, $N$ vector-like
messenger superfields $\Phi_i, \bar\Phi_i$ have a superpotential
coupling to $X$, $W= X \Phi_i \bar \Phi_i$, which gives mass $M$ to
the messengers.  The messengers are charged under the SM gauge group
and communicate SUSY breaking to the MSSM fields via gauge loops.  All
MSSM gauginos get soft masses at one loop,
\beq
M_{\lambda_r}(M) = g_r^2 N_{\mathrm{eff}} \tilde \Lambda,
\eeq
and all scalars get contributions to their soft masses-squared at two loops,
\beq
\tilde m_A^2(M) = 2 N_{\mathrm{eff}} \sum_r c_r g_r^4 \tilde \Lambda^2 .
\label{eq:GMSBsoftsq}
\eeq
Here we have defined $\Lambda \equiv \frac FM$, $\tilde \Lambda \equiv
\frac1{16\pi^2} \Lambda$, and the effective number of messengers
$N_{\mathrm{eff}}$ (a $5\oplus \bar5$ of SU(5) contributes 1 and a
$10\oplus \bar{10}$ contributes 3).  The quadratic Casimir of a MSSM
field $c_r$ under the SM gauge group $r$ is, for hypercharge,
normalized according to its embedding in SU(5) or larger grand unified
theory, $c_1=\frac35 Y^2$. 

In EGMSB, one introduces additional direct couplings in the superpotential
between the MSSM and messenger superfields. To directly
affect right-handed sleptons, these couplings should take the form
\beq
W \supset \kappa_i E^c_i \Phi \tilde\Phi  \hspace{1 cm} \mbox{ with }  \; (  \Phi,\tilde  \Phi ) = \lp\Phi_{\bar E}, \Phi_{S}\rp, \, \lp\Phi_{L_1}, \Phi_{L_2}\rp \mbox{ or } \lp\Phi_{U}, \Phi_{\bar D}\rp 
\eeq
where $\Phi_{A}$ is a messenger with the same gauge quantum numbers as
the superfield $A$.   That is, in the notation
$(SU(3)_C,SU(2)_L)_{U(1)_Y}$, we have: $\bar E=(1,1)_{-1}$,
$L_i=(1,2)_{- \frac 12}$, $U=(\bar 3,1)_{-\frac23}$, $\bar
D=(3,1)_{-\frac13}$, and $S=(1,1)_{0}$.  Two distinct $L$ fields are
required lest the coupling be identically zero.  With the general
formulas from \cite{Evans:2013kxa}, the EGMSB contribution to the
slepton mass-squared  can be written
\begin{equation} 
 \delta \tilde m^2_{E,ij} = \left[ (d_E+2)\kappa_k^*\kappa_k - 2 C^{E\Phi\tilde \Phi}_r g^2_r - \frac{16\pi^2}3 h\!\lp\!\frac \Lambda M\! \rp\lp\frac \Lambda M \rp^2 \right] d_E \kappa_i^* \kappa_j  \tilde \Lambda^2,
 \label{eq:EGMSBsoftsq}
\end{equation}
where $d_E$ sums the number of messenger fields with direct
superpotential couplings to $E$, the coefficient $C^{E\Phi\tilde
  \Phi}_r$ is the sum of the Casimirs of the operators in the
superpotential coupling, $C^{E\Phi\tilde \Phi}_r = c^{E}_r+c^{
  \Phi}_r+c^{\tilde \Phi}_r$, and the function $h(x)$ is given by
$h(x)= 1+\frac45 x^2 + \order{x^4}$ \cite{Craig:2012xp}.
Contributions to all other soft masses and the trilinear $A$-terms are
suppressed by $y_\tau$.  For small values of $\kappa_i\ll g_r$, the
first term in (\ref{eq:EGMSBsoftsq}) can be neglected.
 
   \begin{table}[t]
\begin{center}
\begin{tabular}{|l||cc|c|c|} \hline
{\bf Operator} & $N_{\mathrm{eff}}=\abs{\Delta b}$ & $N_{max}$& $d_E$  & $C^{E\Phi\tilde \Phi} $  \\ \hline \hline
$E^c_i \Phi_{\bar E} \Phi_{S}$ &  $3 N$ & $2$ & $N$ & $(\frac65,\,0,0)$ \\
$E^c_i \Phi_{L_1} \Phi_{L_2}$ & $2 N$ & $3$ & $2N$ & $(\frac9{10},\frac32,0)$ \\
$E^c_i \Phi_{U} \Phi_{\bar D}$ & $4N $ & $1$ & $3N$ & $(\frac{14}{15},0,\frac 83)$ \\
 \hline 
\end{tabular}
\caption{Model parameters influenced by the choice of EGMSB operator. $\abs{\Delta b}$ is the contribution of the messengers to the SU(5) beta function.  $N_{max}$ is the maximum number of messengers consistent with perturbative unification, i.e., $\abs{\Delta b}\leq6$.  $C^{E\Phi\tilde \Phi} $ is the sum of the quadratic Casimirs of the three fields within the operator $E\Phi\tilde \Phi$ for the groups (U(1), SU(2), SU(3)) respectively.
\label{tab:EGMSB}
}
\end{center}  
\end{table}

From the above expressions we can get an idea for the parameter scales
involved over the masses and lifetimes of interest.  For simplicity of
discussion, we will neglect the effects of running.  The slepton NLSP
lifetime (\ref{eq:GMSB4life}) can be inverted to give an expression
for the SUSY-breaking scale in terms of the slepton mass and lifetime,
\beq
F = \lp100 \mbox{ TeV}\rp^2 \lp\frac{c\tau}{100\, \mu m}\rp^\frac12 \lp\frac{m_{\tilde \ell}}{100 \mbox{ GeV}}\rp^\frac52,
\label{eq:lifeinv}
\eeq
and the GMSB contribution to the slepton mass (\ref{eq:GMSBsoftsq})
can be inverted to give 
$ \Lambda = 16 \pi^2 m_{\tilde \ell} ( 1/g_1^2)
\sqrt{5/6N_{\mathrm{eff}}}$, or
\beq
\frac{\Lambda}M= \frac{\Lambda^2}F \sim \frac{1}{3N_{\mathrm{eff}}} \lp\frac{100\, \mu m}{c\tau}\rp^\frac12 \lp\frac{100 \mbox{ GeV}}{m_{\tilde \ell}}\rp^\frac12,
 \label{eq:softinv}
\eeq
where in the last equation we have approximated $g_1(M)\sim0.5$.  Now,
focusing on the $\kappa_i E^c_i \Phi_{U} \Phi_{\bar D}$ model with
$N=1$ for simplicity, we can use (\ref{eq:EGMSBsoftsq}) and
Table~\ref{tab:EGMSB} to see that the EGMSB contribution to the
slepton mass-squared is
\beq
 \delta m^2_{\tilde \ell,ij} = - \left[ \frac{28}{5} g^2_1 + 16 g^2_3 + 16\pi^2 h\!\lp\!\frac \Lambda M\! \rp\lp\frac \Lambda M \rp^2 \right] \kappa_i^* \kappa_j  \tilde \Lambda^2.
\eeq
Then, substituting (\ref{eq:lifeinv}), (\ref{eq:softinv}), and using
$g_3(M)\sim 1$, we can approximate this expression as
\beq
 \delta m^2_{\tilde \ell,ij} \sim -50 \left[ 
 1 + \frac 1{16}\lp\frac{100\, \mu m}{c\tau}\rp \lp\frac{100 \mbox{ GeV}}{m_{\tilde \ell}}\rp \right]  \kappa_i^* \kappa_j  m_{\tilde \ell}^2.
\eeq
Taking for definiteness the $\vec\kappa$ direction in flavor space to
be aligned with the muon, we can
approximate the small splitting in slepton mass (as opposed to mass
squared) between $\tilde \mu$ and the other sleptons as
\beq
\Delta m_{\tilde \mu} \sim  25  \kappa_2^2 m_{\tilde \ell},
\eeq
which, at the lowest masses of interest, gives $\order{10\mbox{ GeV}}$
splittings for $\kappa_2 \sim 6 \times 10^{-2}$.

An analogous expression can be derived for the other messenger
models. At the other extreme using the same basic approximations, the
$\kappa_i E^c_i \Phi_{\bar E} \Phi_{S}$ model gives
\beq
\Delta m_{\tilde \mu} \sim  \frac43 \left[  1 + \frac{1}{N^2} \lp\frac{100\, \mu m}{c\tau}\rp \lp\frac{100 \mbox{ GeV}}{m_{\tilde \ell}}\rp \right]  \kappa_2^2 m_{\tilde \ell},
\label{eq:othermodelsplitting}
\eeq
which is more sensitive to the lifetime, but typically requires
$\kappa_2 \sim 0.2-0.3$ for an $\order{10\mbox{ GeV}}$ splitting.

\section{Freezein During an Early Matter-Dominated Era}
\label{sec:MDFI}

In this appendix, we provide a brief discussion of dark matter freezein
during an early period of matter domination.  Such epochs
of matter domination are created by the coherent oscillations of a
heavy modulus or inflaton, $\phi$, with a late decay into relativistic
species.  Expansion during this time is non-adiabatic due to the
entropy injection from the decay of the heavy species.  This period of matter
domination lasts until a time, $t\sim\Gamma_\phi^{-1}$, at which point
enough of the heavy species have decayed so that the universe enters a
radiation-dominated era.  The lifetime of $\phi$ determines the reheat
 temperature (see e.g.~\cite{Kofman:1997yn} for a review),
\beq
T_{RH} = \lp\frac{90}{8\pi^3 g_*}\rp^{1/4}\sqrt{\Gamma_\phi M_{pl}},
\eeq
which is the temperature where the universe transitions into a
standard radiation-dominated adiabatic expansion.  Importantly, at
times earlier than $T_{RH}$, the temperature of the thermal bath is
higher than $T_{RH}$ \cite{Chung:1998rq}.\footnote{We  
assume that the decay products of the heavy scalar field have
 reached thermal equilibrium at some temperature above the 
 scale where freezein becomes relevant.}
 
For simplicity, we will consider DM freezein during the period of
reheating following inflation, and take $\phi$ to be the inflaton; periods
of modulus-domination yield quantitatively similar results in our
region of interest.  The Boltzmann equations describing the evolution
of the energy density stored in the inflaton field, $\rho_\phi$, and
the energy density stored in the relativistic species, $\rho_R$, are
 \beqa
 \label{eq:BZM1}
 \dot \rho_\phi + 3 H \rho_\phi \!\!&=&\!\! -\Gamma_\phi \rho_\phi \\
 \dot \rho_R + 4 H \rho_R \!\!&=&\!\! \Gamma_\phi \rho_\phi.
  \label{eq:BZM2}
 \eeqa
Here, $a$ is the scale factor, and $H= \dot a/a$ is the Hubble 
parameter.  We define the co-moving quantities
  \beq
\Phi\equiv \rho_\phi a^3 \qq \mbox{and} \qq R\equiv\rho_R a^4.
\eeq
Using the Friedmann equation
   \beq
   \label{eq:hubble}
  H^2 = \frac{8\pi}{3}\frac{1}{M_{pl}^2a^4} \lp a\Phi+R \rp,
\eeq
the Boltzmann equations (\ref{eq:BZM1}\,-\,\ref{eq:BZM2}) can be
rewritten as
  \beqa
 \label{eq:BZM3}
\Phi' \!\!&=&\!\! -\mathcal C \frac{a}{\sqrt{a\Phi+R}}\Phi \\
R' \!\!&=&\!\!\mathcal C \frac{a^2}{\sqrt{a\Phi+R}}\Phi,
  \label{eq:BZM4}
 \eeqa
 where we have defined
 \beq
\mathcal C \equiv \sqrt{\frac{3}{8\pi}} M_{pl} \Gamma_\phi.
\eeq
Initial conditions are determined from the end of inflation, which
occurs at some scale $a_i$, with inflaton energy density
$\rho_\phi(a_i) \equiv \rho_{\phi,i}$, while $\rho_R (a_i) = 0$.  In
practice, the late-time behavior of the system is insensitive to the
exact values of $a_i$ and $ \rho_{\phi,i}$.

In general, these equations must be solved numerically.  However, it
is useful to construct an approximate analytic solution to this system
as follows.  First, we approximate the transition from
matter-domination to radiation-domination as an instantaneous energy
transfer at $a_{RH}$, so that $\rho_\phi(a_{RH}^-)=\rho_R(a_{RH}^+)$.
A second simplifying assumption is to neglect the  
effect of
$\Gamma_\phi$ on $\Phi$ until this instantaneous transfer. 
Thirdly, we take the Hubble parameter to
depend only on $\Phi$ prior to $T_{RH}$, neglecting the small
contribution of the radiation.
As long as we are not concerned with the detailed behavior 
near the transition from matter-domination to radiation-domination, 
i.e., near $T_{RH}$, these are excellent approximations. 
With these approximations,
Eqs.~(\ref{eq:BZM3}\,-\,\ref{eq:BZM4}) can be simply integrated to
yield, at leading order,
\beq
 R(a)\approx  \frac{2\mathcal C}5   \Phi_i^{1/2} \lp a^{5/2}- a_i^{5/2} \rp 
     \approx  \frac{2\mathcal C}5   \Phi_{RH}^{1/2} \lp a^{5/2}- a_i^{5/2} \rp,
\eeq
where our approximations allow us to further express $\Phi_{RH}$ simply in
terms of the reheating temperature and $a_{RH}$.  From this expression,
it is evident that for $a \gg a_i$, dependence on the detailed
choice of $a_i$ drops out.

The radiation energy density defines an expression for temperature.  
In the matter-dominated regime, we have
\beq
 T(a)= \frac1a  \lp\frac{30 \mathcal C}{\pi^2 g_*(T)}\rp^{1/4} \left[ \frac25   \Phi_i^{1/2} \lp a^{5/2}- a_i^{5/2} \rp \right]^{1/4}.
 \label{eq:temp}
\eeq
From this relation, it is easy to see that the temperature scales approximately as
\beq
 T\propto a^{-3/8}.
 \label{eq:MDaTscale}
\eeq

We are interested in the freezein production of DM during the
matter-dominated epoch prior to $T_{RH}$.  
The Boltzmann equation governing the freezein of scalar
lepton-flavored dark matter through the decay of a fermionic charged
parent, $\psi$, in thermal equilibrium can be written as
  \beq
  \dot n_S + 3Hn_S = \int \!\! d\Pi_\psi  d\Pi_S  d\Pi_\ell  (2\pi)^4 \delta^4(\sum p_i)\abs{\bar{\mathcal M}(\psi\to S\ell)}^2 f_\psi(1-f_\ell)(1+f_S),
  \label{eq:BZMS}
\eeq
where we have taken all other dark matter interaction rates, including
the inverse process $\ell S\to\psi$, to be negligible.  To simplify
(\ref{eq:BZMS}), we note that $f_S\ll 1 $ and approximate $(1-
f_\ell)\approx 1$, giving \cite{Hall:2009bx}
    \beq
  \dot n_S + 3Hn_S \approx \int \!\! d\Pi_\psi 2m_\psi \Gamma_\psi f_\psi = \frac{g_\psi \Gamma_\psi m_\psi}{2\pi^2}  \int_{m_\psi}^\infty \sqrt{E^2 - m_\psi^2} f_\psi dE ,
  \label{eq:BZMS2}
\eeq
where $g_\psi\!=\!4$ is the number of internal degrees of freedom for
$\psi$.  Since temperatures where $T\sim
m_\psi/\{\mathrm{few}\}$ dominate freezein, using a Maxwell-Boltzmann
approximation for $f_\psi$  is not an unreasonable approximation, but will
yields a slightly higher dark matter density than that obtained by
using Fermi-Dirac statistics.  Defining the quantity $\mathcal S\equiv
n_S a^3/\Gamma_\psi$, we can simplify (\ref{eq:BZMS2}) to
  \beq
 \Gamma_\psi \mathcal S' =\frac{a^2}{H} \frac{g_\psi\Gamma_\psi m_\psi^2 T}{2\pi^2} K_1\!\lp\frac{m_\psi}{T}\rp  I_{FD}\! \lp\frac{m_\psi}{T} \rp,
  \label{eq:BZMSfinal}
\eeq
where $K_1(x)$ is a modified Bessel function of the second kind, and
the monotonic function
\beq
   I_{FD}(x) = \frac{1}{x K_1(x)}\lp \int_{x}^\infty \frac{\sqrt{u^2 - x^2} }{1+e^u} du\rp \approx \left\{ 
   \begin{tabular}{ll}
   $\frac{\pi^2}{12}\approx0.822$ & $x\ll1$\\
  1 & $x\gg1$
  \end{tabular}
     \right. 
\eeq
encapsulates the departure from the Maxwell-Boltzmann approximation.

Using (\ref{eq:hubble}) and (\ref{eq:temp}), equation
(\ref{eq:BZMSfinal}) can be numerically integrated as
\beq
  \mathcal S(a') = \frac{g_\psi m_\psi^2 }{2\pi^2} \int_{a_i}^{a'} \frac{a^2T(a)}{H(a)}  K_1\!\lp\frac{m_\psi}{T(a)}\rp  I_{FD}\! \lp\frac{m_\psi}{T(a)} \rp da.
\eeq
As long as $T_{RH}$ is sufficiently smaller than $m_\psi$, our
simplifying assumptions about the transition from matter-domination to
radiation-domination are reliable approximations.  After $a_{RH}$, the
universe expands adiabatically, so $\mathcal S$ is constant during this era. 
We can then relate
this quantity to the present day dark matter abundance,
\beq
\Omega_{DM} = \frac{m_S n_{S,0}}{\rho_{crit,0}} =  m_S \Gamma_\psi \frac{\mathcal S(a_0)}{a_0^3  \rho_{crit,0}   } =  m_S \Gamma_\psi \frac{\mathcal S(a_{RH})}{a_0^3  \rho_{crit,0}}  =  m_S \Gamma_\psi \frac{\mathcal S(T_{RH},m_\psi,\frac{a_{RH}}{a_i})}{ \rho_{crit,0}   },
\label{eq:DMfinal}
\eeq
where $a_0\equiv1$ is the scale factor today and
$\rho_{crit,0}=3.80\times 10^{-47}$ GeV$^{4}$ is the critical energy
density of the universe.  In the last equality, $\mathcal S(a_{RH})$
has been written explicitly in terms of all parameters on which it
depends.  As long as the matter-dominated era is sufficiently
long, i.e., $a_{RH}/a_i \gg (m_\psi/T_{RH})^{8/3}$, $\mathcal S$ is
insensitive to $a_i$.  Assuming this condition on $a_i$ holds, we can write
(\ref{eq:DMfinal}) as
\beq
 \frac{ \Omega_{DM}}{ \Omega_{DM,obs}} = \lp \frac{\mathcal S(T_{RH},m_\psi)}{5.2\times 10^{-31} \mbox{ GeV}^2} \rp \lp \frac{m_S}{1 \mbox{ MeV}}\rp \lp \frac{1 \mbox{ cm}}{c\tau_\psi} \rp. 
\label{eq:DMsimp}
\eeq
The scaling relations derived in Ref.~\cite{Co:2015pka} 
illustrate that  $\mathcal S(T_{RH},m_\psi) \propto T_{RH}^7/m_\psi^9$.  
 Noting this relation, we can extract the correct numerical 
 factors to express (\ref{eq:DMsimp}) approximately as
   \beq
 \frac{ \Omega_{DM}}{ \Omega_{DM,obs}} \approx  \lp \frac{20}{m_\psi/T_{RH}} \rp^7  \lp \frac{500\mbox{ GeV}}{m_\psi} \rp^2 \lp \frac{m_S}{1 \mbox{ MeV}}\rp \lp \frac{1 \mbox{ cm}}{c\tau_\psi} \rp.
\label{eq:DMsimp2}
\eeq 
Equation (\ref{eq:DMsimp2}) can fix one of the remaining
four parameters: $c\tau_\psi$, $m_\psi$, $m_S$, or $T_{RH}$.
 Thus, displaced decays at the LHC imply a relatively low reheat
temperature, $T_{RH} \lesssim$ TeV.

From (\ref{eq:DMsimp2}), it would appear that for specific collider
parameters $c\tau_\psi$ and $m_\psi$, one can always choose $T_{RH}$
and $m_S$ to produce the correct dark matter relic abundance.
However, if at some point in the early universe the number density of
dark matter becomes too large, the neglected rate for the inverse
process $\ell S\to\psi$ will become important.  To estimate the range
of validity of the the above calculation, we will require that the DM
number density satisfies
  \beq
n_{S}(T) < k_{crit} n_{S,eq}(T) = k_{crit} \frac{\zeta(3)}{\pi^2} T^3,
  \label{eq:thermeqcond}
\eeq
where $\zeta(3)\approx 1.202$ is the Riemann zeta function,
$n_{S,eq}(T)$ is the equilibrium number density of a relativistic
scalar particle in thermal equilibrium, and $k_{crit}<1$ is a 
measure of when (\ref{eq:BZMS}) ceases to be reliable.  
 As the freezein mechanism does not produce a thermal
distribution for the dark matter and $f_\psi(p)\leq f^{eq}(p)$ 
\cite{Monteux:2015qqa}, a numerical study beyond the 
scope of this work would be required to determine precisely 
where these rates become comparable. 

As the bulk of freezein happens near $T \sim m_\psi/\{\mathrm{few}\}$, we are
interested in (\ref{eq:thermeqcond}) applied near $T_{FI} \approx
m_\psi/4$. 
After freezein ($T < T_{FI} \approx m_\psi/4$), production
of dark matter is negligible, so the number density simply redshifts
with the expanding universe,
\beq
 n_S(a_{FI}) =n_S(a_{RH})  \frac{a_{RH}^3}{a_{FI}^3} =n_S(a_{RH}) \lp \frac{T_{FI}}{T_{RH}}\rp^{8},
  \label{eq:comovethermpast}
\eeq
where we have used (\ref{eq:MDaTscale}).  Of course,
$n_S(a_{RH})$ can be directly related to the dark matter density today,
   \beq
n_S(a_{RH}) = \frac{ \Omega_{DM}}{m_S} \frac{\rho_{crit,0}}{s_0} s_{RH}  = \frac{ \Omega_{DM}}{m_S} \frac{\rho_{crit,0}}{s_0} \frac{2\pi^2}{45} g_{*S}(T_{RH}) T_{RH}^3,
  \label{eq:comovethermtoday}
\eeq
where $s_0 = 2.22\times 10^{-38}$ GeV$^3$ is the entropy density
today.  Combining (\ref{eq:thermeqcond}-\ref{eq:comovethermtoday}), we
derive
 \beq
       k_{crit} > \frac{2\pi^4}{45 \zeta(3)} \frac{ \Omega_{DM}}{m_S} \frac{\rho_{crit,0}}{s_0}  g_{*S}(T_{RH}) \lp \frac{T_{FI}}{T_{RH}}\rp^{5}
\label{eq:kcritcondn},
\eeq
as the region where (\ref{eq:BZMS}) is reliable.  While this condition
is by necessity simplified, for much of the parametric range of
interest for collider phenomenology ($100$ GeV$\lesssim m_\phi\lesssim
1$ TeV; $100\,\mu$m $\lesssim c\tau \lesssim 1$ m), it is possible to
choose $T_{RH}$ so that the dark matter relic abundance is satisfied
for $m_S\ll m_\psi$, while satisfying (\ref{eq:kcritcondn}).
However, at low $\psi$ masses and short $\psi$ lifetimes, requiring
$m_S\ll m_\psi$ can lead to difficulty with (\ref{eq:kcritcondn}).
This will result in a net reduction of the DM relic abundance relative 
to (\ref{eq:DMsimp2}), due to the additional depletion of the DM.  
When this is the case, $S$ will make up only a fraction of the current
abundance, and some other particle(s) must constitute the rest.
Of course, the dark matter could also have $m_S\sim m_\psi$, so that 
its mass substantially influences collider kinematics.   Although 
interesting possibilities, in order to simplify our presentation, we will 
take $m_S\ll m_\psi$ throughout this work.

\small{
\bibliography{Meso}}

\providecommand{\href}[2]{#2}\begingroup\raggedright\begin{thebibliography}{10}

\bibitem{Halkiadakis:2014qda}
E.~Halkiadakis, G.~Redlinger, and D.~Shih, ``{Status and Implications of
  Beyond-the-Standard-Model Searches at the LHC},''
  \href{http://dx.doi.org/10.1146/annurev-nucl-102313-025632}{{\em Ann. Rev.
  Nucl. Part. Sci.} {\bfseries 64} (2014) 319--342},
\href{http://arxiv.org/abs/1411.1427}{{\ttfamily arXiv:1411.1427 [hep-ex]}}.

\bibitem{Evans:2015caa}
J.~A. Evans, ``{Flavors of Supersymmetry Beyond Vanilla},''
\href{http://arxiv.org/abs/1509.08504}{{\ttfamily arXiv:1509.08504 [hep-ph]}}.

\bibitem{Barbier:2004ez}
R.~Barbier {\em et~al.}, ``{$R$-parity violating supersymmetry},''
  \href{http://dx.doi.org/10.1016/j.physrep.2005.08.006}{{\em Phys.Rept.}
  {\bfseries 420} (2005) 1},
\href{http://arxiv.org/abs/hep-ph/0406039}{{\ttfamily arXiv:hep-ph/0406039
  [hep-ph]}}.

\bibitem{Fan:2011yu}
J.~Fan, M.~Reece, and J.~T. Ruderman, ``{Stealth Supersymmetry},''
  \href{http://dx.doi.org/10.1007/JHEP11(2011)012}{{\em JHEP} {\bfseries 1111}
  (2011) 012},
\href{http://arxiv.org/abs/1105.5135}{{\ttfamily arXiv:1105.5135 [hep-ph]}}.

\bibitem{Evans:2013jna}
J.~A. Evans, Y.~Kats, D.~Shih, and M.~J. Strassler, ``{Toward Full LHC Coverage
  of Natural Supersymmetry},''
\href{http://arxiv.org/abs/1310.5758}{{\ttfamily arXiv:1310.5758 [hep-ph]}}.

\bibitem{Fan:2015mxp}
J.~Fan, R.~Krall, D.~Pinner, M.~Reece, and J.~T. Ruderman, ``{Stealth
  Supersymmetry Simplified},''
\href{http://arxiv.org/abs/1512.05781}{{\ttfamily arXiv:1512.05781 [hep-ph]}}.

\bibitem{Giudice:1998bp}
G.~F. Giudice and R.~Rattazzi, ``{Theories with gauge mediated supersymmetry
  breaking},'' \href{http://dx.doi.org/10.1016/S0370-1573(99)00042-3}{{\em
  Phys. Rept.} {\bfseries 322} (1999) 419--499},
\href{http://arxiv.org/abs/hep-ph/9801271}{{\ttfamily arXiv:hep-ph/9801271
  [hep-ph]}}.

\bibitem{Arvanitaki:2012ps}
A.~Arvanitaki, N.~Craig, S.~Dimopoulos, and G.~Villadoro, ``{Mini-Split},''
  \href{http://dx.doi.org/10.1007/JHEP02(2013)126}{{\em JHEP} {\bfseries 02}
  (2013) 126},
\href{http://arxiv.org/abs/1210.0555}{{\ttfamily arXiv:1210.0555 [hep-ph]}}.

\bibitem{Gupta:2007ui}
S.~K. Gupta, B.~Mukhopadhyaya, and S.~K. Rai, ``{Right-chiral sneutrinos and
  long-lived staus: Event characteristics at the large hadron collider},''
  \href{http://dx.doi.org/10.1103/PhysRevD.75.075007}{{\em Phys. Rev.}
  {\bfseries D75} (2007) 075007},
\href{http://arxiv.org/abs/hep-ph/0701063}{{\ttfamily arXiv:hep-ph/0701063
  [hep-ph]}}.

\bibitem{Liu:2015bma}
Z.~Liu and B.~Tweedie, ``{The Fate of Long-Lived Superparticles with Hadronic
  Decays after LHC Run 1},''
  \href{http://dx.doi.org/10.1007/JHEP06(2015)042}{{\em JHEP} {\bfseries 06}
  (2015) 042},
\href{http://arxiv.org/abs/1503.05923}{{\ttfamily arXiv:1503.05923 [hep-ph]}}.

\bibitem{Csaki:2015uza}
C.~Csaki, E.~Kuflik, S.~Lombardo, O.~Slone, and T.~Volansky, ``{Phenomenology
  of a Long-Lived LSP with R-Parity Violation},''
\href{http://arxiv.org/abs/1505.00784}{{\ttfamily arXiv:1505.00784 [hep-ph]}}.

\bibitem{Zwane:2015bra}
N.~Zwane, ``{Long-Lived Particle Searches in R-Parity Violating MSSM},''
\href{http://arxiv.org/abs/1505.03479}{{\ttfamily arXiv:1505.03479 [hep-ph]}}.

\bibitem{Aad:2014yea}
{ ATLAS} collaboration, G.~Aad {\em et~al.}, ``{Search for long-lived neutral
  particles decaying into lepton jets in proton-proton collisions at $
  \sqrt{s}=8 $ TeV with the ATLAS detector},''
  \href{http://dx.doi.org/10.1007/JHEP11(2014)088}{{\em JHEP} {\bfseries 11}
  (2014) 088},
\href{http://arxiv.org/abs/1409.0746}{{\ttfamily arXiv:1409.0746 [hep-ex]}}.

\bibitem{CMS:2014hka}
{ CMS} collaboration, V.~Khachatryan {\em et~al.}, ``{Search for long-lived
  particles that decay into final states containing two electrons or two muons
  in proton-proton collisions at $\sqrt{s} =$ 8 TeV},''
  \href{http://dx.doi.org/10.1103/PhysRevD.91.052012}{{\em Phys. Rev.}
  {\bfseries D91} no.~5, (2015) 052012},
\href{http://arxiv.org/abs/1411.6977}{{\ttfamily arXiv:1411.6977 [hep-ex]}}.

\bibitem{Aad:2015rba}
{ ATLAS} collaboration, G.~Aad {\em et~al.}, ``{Search for massive, long-lived
  particles using multitrack displaced vertices or displaced lepton pairs in pp
  collisions at $\sqrt{s}$ = 8 TeV with the ATLAS detector},''
\href{http://arxiv.org/abs/1504.05162}{{\ttfamily arXiv:1504.05162 [hep-ex]}}.

\bibitem{Aad:2012zx}
{ ATLAS} collaboration, G.~Aad {\em et~al.}, ``{Search for long-lived, heavy
  particles in final states with a muon and multi-track displaced vertex in
  proton-proton collisions at $\sqrt{s}=7$ TeV with the ATLAS detector},''
  \href{http://dx.doi.org/10.1016/j.physletb.2013.01.042}{{\em Phys. Lett.}
  {\bfseries B719} (2013) 280--298},
\href{http://arxiv.org/abs/1210.7451}{{\ttfamily arXiv:1210.7451 [hep-ex]}}.

\bibitem{Khachatryan:2014mea}
{ CMS} collaboration, V.~Khachatryan {\em et~al.}, ``{Search for Displaced
  Supersymmetry in events with an electron and a muon with large impact
  parameters},'' \href{http://dx.doi.org/10.1103/PhysRevLett.114.061801}{{\em
  Phys. Rev. Lett.} {\bfseries 114} no.~6, (2015) 061801},
\href{http://arxiv.org/abs/1409.4789}{{\ttfamily arXiv:1409.4789 [hep-ex]}}.

\bibitem{Aad:2012tfa}
{ ATLAS} collaboration, G.~Aad {\em et~al.}, ``{Observation of a new particle
  in the search for the Standard Model Higgs boson with the ATLAS detector at
  the LHC},'' \href{http://dx.doi.org/10.1016/j.physletb.2012.08.020}{{\em
  Phys.Lett.} {\bfseries B716} (2012) 1},
\href{http://arxiv.org/abs/1207.7214}{{\ttfamily arXiv:1207.7214 [hep-ex]}}.

\bibitem{Chatrchyan:2012ufa}
{ CMS} collaboration, S.~Chatrchyan {\em et~al.}, ``{Observation of a new boson
  at a mass of 125~GeV with the CMS experiment at the LHC},''
  \href{http://dx.doi.org/10.1016/j.physletb.2012.08.021}{{\em Phys.Lett.}
  {\bfseries B716} (2012) 30},
\href{http://arxiv.org/abs/1207.7235}{{\ttfamily arXiv:1207.7235 [hep-ex]}}.

\bibitem{Draper:2011aa}
P.~Draper, P.~Meade, M.~Reece, and D.~Shih, ``{Implications of a 125 GeV Higgs
  for the MSSM and Low-Scale SUSY Breaking},''
  \href{http://dx.doi.org/10.1103/PhysRevD.85.095007}{{\em Phys. Rev.}
  {\bfseries D85} (2012) 095007},
\href{http://arxiv.org/abs/1112.3068}{{\ttfamily arXiv:1112.3068 [hep-ph]}}.

\bibitem{Craig:2012xp}
N.~Craig, S.~Knapen, D.~Shih, and Y.~Zhao, ``{A Complete Model of Low-Scale
  Gauge Mediation},'' \href{http://dx.doi.org/10.1007/JHEP03(2013)154}{{\em
  JHEP} {\bfseries 1303} (2013) 154},
\href{http://arxiv.org/abs/1206.4086}{{\ttfamily arXiv:1206.4086 [hep-ph]}}.

\bibitem{Abdullah:2012tq}
M.~Abdullah, I.~Galon, Y.~Shadmi, and Y.~Shirman, ``{Flavored Gauge Mediation,
  A Heavy Higgs, and Supersymmetric Alignment},''
  \href{http://dx.doi.org/10.1007/JHEP06(2013)057}{{\em JHEP} {\bfseries 06}
  (2013) 057},
\href{http://arxiv.org/abs/1209.4904}{{\ttfamily arXiv:1209.4904 [hep-ph]}}.

\bibitem{Byakti:2013ti}
P.~Byakti and T.~S. Ray, ``{Burgeoning the Higgs mass to 125 GeV through
  messenger-matter interactions in GMSB models},''
  \href{http://dx.doi.org/10.1007/JHEP05(2013)055}{{\em JHEP} {\bfseries 1305}
  (2013) 055},
\href{http://arxiv.org/abs/1301.7605}{{\ttfamily arXiv:1301.7605 [hep-ph]}}.

\bibitem{Evans:2013kxa}
J.~A. Evans and D.~Shih, ``{Surveying Extended GMSB Models with $m$$_{h}$=125
  GeV},'' \href{http://dx.doi.org/10.1007/JHEP08(2013)093}{{\em JHEP}
  {\bfseries 1308} (2013) 093},
\href{http://arxiv.org/abs/1303.0228}{{\ttfamily arXiv:1303.0228 [hep-ph]}}.

\bibitem{Calibbi:2013mka}
L.~Calibbi, P.~Paradisi, and R.~Ziegler, ``{Gauge Mediation beyond Minimal
  Flavor Violation},'' \href{http://dx.doi.org/10.1007/JHEP06(2013)052}{{\em
  JHEP} {\bfseries 1306} (2013) 052},
\href{http://arxiv.org/abs/1304.1453}{{\ttfamily arXiv:1304.1453 [hep-ph]}}.

\bibitem{Knapen:2013zla}
S.~Knapen and D.~Shih, ``{Higgs Mediation with Strong Hidden Sector
  Dynamics},'' \href{http://dx.doi.org/10.1007/JHEP08(2014)136}{{\em JHEP}
  {\bfseries 1408} (2014) 136},
\href{http://arxiv.org/abs/1311.7107}{{\ttfamily arXiv:1311.7107 [hep-ph]}}.

\bibitem{Ding:2013pya}
R.~Ding, T.~Li, F.~Staub, and B.~Zhu, ``{Focus Point Supersymmetry in Extended
  Gauge Mediation},'' \href{http://dx.doi.org/10.1007/JHEP03(2014)130}{{\em
  JHEP} {\bfseries 1403} (2014) 130},
\href{http://arxiv.org/abs/1312.5407}{{\ttfamily arXiv:1312.5407 [hep-ph]}}.

\bibitem{Basirnia:2015vga}
A.~Basirnia, D.~Egana-Ugrinovic, S.~Knapen, and D.~Shih, ``{125 GeV Higgs from
  Tree-Level $A$-terms},''
  \href{http://dx.doi.org/10.1007/JHEP06(2015)144}{{\em JHEP} {\bfseries 06}
  (2015) 144},
\href{http://arxiv.org/abs/1501.00997}{{\ttfamily arXiv:1501.00997 [hep-ph]}}.

\bibitem{Fischler:2013tva}
W.~Fischler and W.~Tangarife, ``{Vector-like Fields, Messenger Mixing and the
  Higgs mass in Gauge Mediation},''
  \href{http://dx.doi.org/10.1007/JHEP05(2014)151}{{\em JHEP} {\bfseries 1405}
  (2014) 151},
\href{http://arxiv.org/abs/1310.6369}{{\ttfamily arXiv:1310.6369 [hep-ph]}}.

\bibitem{Liu:2013vaa}
C.~Liu and Z.-h. Zhao, ``{A Realization of Effective SUSY with Strong
  Unification},'' \href{http://dx.doi.org/10.1103/PhysRevD.89.057701}{{\em
  Phys. Rev.} {\bfseries D89} no.~5, (2014) 057701},
\href{http://arxiv.org/abs/1312.7389}{{\ttfamily arXiv:1312.7389 [hep-ph]}}.

\bibitem{Allanach:2015cia}
B.~Allanach, M.~Badziak, C.~Hugonie, and R.~Ziegler, ``{Light Sparticles from a
  Light Singlet in Gauge Mediation},''
  \href{http://dx.doi.org/10.1103/PhysRevD.92.015006}{{\em Phys. Rev.}
  {\bfseries D92} no.~1, (2015) 015006},
\href{http://arxiv.org/abs/1502.05836}{{\ttfamily arXiv:1502.05836 [hep-ph]}}.

\bibitem{Delgado:2015bwa}
A.~Delgado, M.~Garcia-Pepin, and M.~Quiros, ``{GMSB with Light Stops},''
  \href{http://dx.doi.org/10.1007/JHEP08(2015)159}{{\em JHEP} {\bfseries 08}
  (2015) 159},
\href{http://arxiv.org/abs/1505.07469}{{\ttfamily arXiv:1505.07469 [hep-ph]}}.

\bibitem{Batra:2003nj}
P.~Batra, A.~Delgado, D.~E. Kaplan, and T.~M.~P. Tait, ``{The Higgs mass bound
  in gauge extensions of the minimal supersymmetric standard model},''
  \href{http://dx.doi.org/10.1088/1126-6708/2004/02/043}{{\em JHEP} {\bfseries
  02} (2004) 043},
\href{http://arxiv.org/abs/hep-ph/0309149}{{\ttfamily arXiv:hep-ph/0309149
  [hep-ph]}}.

\bibitem{Cheung:2012zq}
C.~Cheung and H.~L. Roberts, ``{Higgs Mass from D-Terms: a Litmus Test},''
  \href{http://dx.doi.org/10.1007/JHEP12(2013)018}{{\em JHEP} {\bfseries 12}
  (2013) 018},
\href{http://arxiv.org/abs/1207.0234}{{\ttfamily arXiv:1207.0234 [hep-ph]}}.

\bibitem{Craig:2012bs}
N.~Craig and A.~Katz, ``{A Supersymmetric Higgs Sector with Chiral D-terms},''
  \href{http://dx.doi.org/10.1007/JHEP05(2013)015}{{\em JHEP} {\bfseries 05}
  (2013) 015},
\href{http://arxiv.org/abs/1212.2635}{{\ttfamily arXiv:1212.2635 [hep-ph]}}.

\bibitem{Lu:2013cta}
X.~Lu, H.~Murayama, J.~T. Ruderman, and K.~Tobioka, ``{A Natural Higgs Mass in
  Supersymmetry from Non-Decoupling Effects},''
  \href{http://dx.doi.org/10.1103/PhysRevLett.112.191803}{{\em Phys. Rev.
  Lett.} {\bfseries 112} (2014) 191803},
\href{http://arxiv.org/abs/1308.0792}{{\ttfamily arXiv:1308.0792 [hep-ph]}}.

\bibitem{McGarrie:2014xxa}
M.~McGarrie, G.~Moortgat-Pick, and S.~Porto, ``{Confronting Higgs couplings
  from D-term extensions and Natural SUSY at the LHC and ILC},''
  \href{http://dx.doi.org/10.1140/epjc/s10052-015-3361-5}{{\em Eur. Phys. J.}
  {\bfseries C75} no.~4, (2015) 150},
\href{http://arxiv.org/abs/1411.2040}{{\ttfamily arXiv:1411.2040 [hep-ph]}}.

\bibitem{Bertuzzo:2014sma}
E.~Bertuzzo and C.~Frugiuele, ``{A natural SM-like 126 GeV Higgs via
  non-decoupling D-terms},''
\href{http://arxiv.org/abs/1412.2765}{{\ttfamily arXiv:1412.2765 [hep-ph]}}.

\bibitem{Abbiendi:2005gc}
{ OPAL} collaboration, G.~Abbiendi {\em et~al.}, ``{Searches for gauge-mediated
  supersymmetry breaking topologies in $e^+e^-$ collisions at LEP2},''
  \href{http://dx.doi.org/10.1140/epjc/s2006-02524-8}{{\em Eur. Phys. J.}
  {\bfseries C46} (2006) 307--341},
\href{http://arxiv.org/abs/hep-ex/0507048}{{\ttfamily arXiv:hep-ex/0507048
  [hep-ex]}}.

\bibitem{Chatrchyan:2013oca}
{ CMS} collaboration, S.~Chatrchyan {\em et~al.}, ``{Searches for long-lived
  charged particles in pp collisions at $\sqrt{s}$ = 7 and 8 TeV},''
  \href{http://dx.doi.org/10.1007/JHEP07(2013)122}{{\em JHEP} {\bfseries 07}
  (2013) 122},
\href{http://arxiv.org/abs/1305.0491}{{\ttfamily arXiv:1305.0491 [hep-ex]}}.

\bibitem{ATLAS:2014fka}
{ ATLAS} collaboration, G.~Aad {\em et~al.}, ``{Searches for heavy long-lived
  charged particles with the ATLAS detector in proton-proton collisions at $
  \sqrt{s}=8 $ TeV},'' \href{http://dx.doi.org/10.1007/JHEP01(2015)068}{{\em
  JHEP} {\bfseries 01} (2015) 068},
\href{http://arxiv.org/abs/1411.6795}{{\ttfamily arXiv:1411.6795 [hep-ex]}}.

\bibitem{Heisig:2012zq}
J.~Heisig and J.~Kersten, ``{Long-lived staus from strong production in a
  simplified model approach},''
  \href{http://dx.doi.org/10.1103/PhysRevD.86.055020}{{\em Phys. Rev.}
  {\bfseries D86} (2012) 055020},
\href{http://arxiv.org/abs/1203.1581}{{\ttfamily arXiv:1203.1581 [hep-ph]}}.

\bibitem{Aad:2013yna}
{ ATLAS} collaboration, G.~Aad {\em et~al.}, ``{Search for charginos nearly
  mass degenerate with the lightest neutralino based on a disappearing-track
  signature in pp collisions at $ \sqrt{s}=8 $ TeV with the ATLAS detector},''
  \href{http://dx.doi.org/10.1103/PhysRevD.88.112006}{{\em Phys. Rev.}
  {\bfseries D88} no.~11, (2013) 112006},
\href{http://arxiv.org/abs/1310.3675}{{\ttfamily arXiv:1310.3675 [hep-ex]}}.

\bibitem{CMS:2014gxa}
{ CMS} collaboration, V.~Khachatryan {\em et~al.}, ``{Search for disappearing
  tracks in proton-proton collisions at $ \sqrt{s}=8 $ TeV},''
  \href{http://dx.doi.org/10.1007/JHEP01(2015)096}{{\em JHEP} {\bfseries 01}
  (2015) 096},
\href{http://arxiv.org/abs/1411.6006}{{\ttfamily arXiv:1411.6006 [hep-ex]}}.

\bibitem{Dimopoulos:1996vz}
S.~Dimopoulos, M.~Dine, S.~Raby, and S.~D. Thomas, ``{Experimental signatures
  of low-energy gauge mediated supersymmetry breaking},''
  \href{http://dx.doi.org/10.1103/PhysRevLett.76.3494}{{\em Phys. Rev. Lett.}
  {\bfseries 76} (1996) 3494--3497},
\href{http://arxiv.org/abs/hep-ph/9601367}{{\ttfamily arXiv:hep-ph/9601367
  [hep-ph]}}.

\bibitem{Ambrosanio:1997rv}
S.~Ambrosanio, G.~D. Kribs, and S.~P. Martin, ``{Signals for gauge mediated
  supersymmetry breaking models at the CERN LEP-2 collider},''
  \href{http://dx.doi.org/10.1103/PhysRevD.56.1761}{{\em Phys. Rev.} {\bfseries
  D56} (1997) 1761--1777},
\href{http://arxiv.org/abs/hep-ph/9703211}{{\ttfamily arXiv:hep-ph/9703211
  [hep-ph]}}.

\bibitem{Ruderman:2010kj}
J.~T. Ruderman and D.~Shih, ``{Slepton co-NLSPs at the Tevatron},''
  \href{http://dx.doi.org/10.1007/JHEP11(2010)046}{{\em JHEP} {\bfseries 11}
  (2010) 046},
\href{http://arxiv.org/abs/1009.1665}{{\ttfamily arXiv:1009.1665 [hep-ph]}}.

\bibitem{Khachatryan:2015lla}
{ CMS} collaboration, V.~Khachatryan {\em et~al.}, ``{Constraints on the pMSSM,
  AMSB Model and on Other Models from the Search for Long-Lived Charged
  Particles in Proton-Proton Collisions at $\sqrt{s}$ = 8 TeV},''
  \href{http://dx.doi.org/10.1140/epjc/s10052-015-3533-3}{{\em Eur. Phys. J.}
  {\bfseries C75} no.~7, (2015) 325},
\href{http://arxiv.org/abs/1502.02522}{{\ttfamily arXiv:1502.02522 [hep-ex]}}.

\bibitem{Aad:2015asa}
{ ATLAS} collaboration, G.~Aad {\em et~al.}, ``{Search for pair-produced
  long-lived neutral particles decaying in the ATLAS hadronic calorimeter in
  $pp$ collisions at $\sqrt{s}$ = 8 TeV},''
  \href{http://dx.doi.org/10.1016/j.physletb.2015.02.015}{{\em Phys. Lett.}
  {\bfseries B743} (2015) 15--34},
\href{http://arxiv.org/abs/1501.04020}{{\ttfamily arXiv:1501.04020 [hep-ex]}}.

\bibitem{Aad:2015uaa}
{ ATLAS} collaboration, G.~Aad {\em et~al.}, ``{Search for long-lived, weakly
  interacting particles that decay to displaced hadronic jets in proton-proton
  collisions at $\sqrt{s}=8$ TeV with the ATLAS detector},''
  \href{http://dx.doi.org/10.1103/PhysRevD.92.012010}{{\em Phys. Rev.}
  {\bfseries D92} no.~1, (2015) 012010},
\href{http://arxiv.org/abs/1504.03634}{{\ttfamily arXiv:1504.03634 [hep-ex]}}.

\bibitem{Alwall:2011uj}
J.~Alwall, M.~Herquet, F.~Maltoni, O.~Mattelaer, and T.~Stelzer, ``{MadGraph 5
  : Going Beyond},'' \href{http://dx.doi.org/10.1007/JHEP06(2011)128}{{\em
  JHEP} {\bfseries 1106} (2011) 128},
\href{http://arxiv.org/abs/1106.0522}{{\ttfamily arXiv:1106.0522 [hep-ph]}}.

\bibitem{Hagiwara:2012vz}
K.~Hagiwara, T.~Li, K.~Mawatari, and J.~Nakamura, ``{TauDecay: a library to
  simulate polarized tau decays via FeynRules and MadGraph5},''
  \href{http://dx.doi.org/10.1140/epjc/s10052-013-2489-4}{{\em Eur. Phys. J.}
  {\bfseries C73} (2013) 2489},
\href{http://arxiv.org/abs/1212.6247}{{\ttfamily arXiv:1212.6247 [hep-ph]}}.

\bibitem{Sjostrand:2007gs}
T.~Sjostrand, S.~Mrenna, and P.~Z. Skands, ``{A Brief Introduction to PYTHIA
  8.1},'' \href{http://dx.doi.org/10.1016/j.cpc.2008.01.036}{{\em
  Comput.Phys.Commun.} {\bfseries 178} (2008) 852},
\href{http://arxiv.org/abs/0710.3820}{{\ttfamily arXiv:0710.3820 [hep-ph]}}.

\bibitem{Khachatryan:2014gga}
{ CMS} collaboration, V.~Khachatryan {\em et~al.}, ``{Performance of the CMS
  missing transverse momentum reconstruction in pp data at $\sqrt{s}$ = 8
  TeV},'' \href{http://dx.doi.org/10.1088/1748-0221/10/02/P02006}{{\em JINST}
  {\bfseries 10} no.~02, (2015) P02006},
\href{http://arxiv.org/abs/1411.0511}{{\ttfamily arXiv:1411.0511
  [physics.ins-det]}}.

\bibitem{Kramer:2012bx}
M.~Kramer, A.~Kulesza, R.~van~der Leeuw, M.~Mangano, S.~Padhi, T.~Plehn, and
  X.~Portell, ``{Supersymmetry production cross sections in $pp$ collisions at
  $\sqrt{s}=7$ TeV},''
\href{http://arxiv.org/abs/1206.2892}{{\ttfamily arXiv:1206.2892 [hep-ph]}}.

\bibitem{Beenakker:1996ed}
W.~Beenakker, R.~Hopker, and M.~Spira, ``{PROSPINO: A Program for the
  production of supersymmetric particles in next-to-leading order QCD},''
\href{http://arxiv.org/abs/hep-ph/9611232}{{\ttfamily arXiv:hep-ph/9611232
  [hep-ph]}}.

\bibitem{Fuks:2013vua}
B.~Fuks, M.~Klasen, D.~R. Lamprea, and M.~Rothering, ``{Precision predictions
  for electroweak superpartner production at hadron colliders with
  Resummino},'' \href{http://dx.doi.org/10.1140/epjc/s10052-013-2480-0}{{\em
  Eur. Phys. J.} {\bfseries C73} (2013) 2480},
\href{http://arxiv.org/abs/1304.0790}{{\ttfamily arXiv:1304.0790 [hep-ph]}}.

\bibitem{CMSHSCPeffmap}
{ CMS} collaboration, ``{Supplementary material from constraints on the pMSSM,
  AMSB Model and on Other Models from the Search for Long-Lived Charged
  Particles in Proton-Proton Collisions at $\sqrt{s}$ = 8 TeV},''
  \url{http://hepdata.cedar.ac.uk/view/ins1343509} (2015).

\bibitem{Randall:1998uk}
L.~Randall and R.~Sundrum, ``{Out of this world supersymmetry breaking},''
  \href{http://dx.doi.org/10.1016/S0550-3213(99)00359-4}{{\em Nucl. Phys.}
  {\bfseries B557} (1999) 79--118},
\href{http://arxiv.org/abs/hep-th/9810155}{{\ttfamily arXiv:hep-th/9810155
  [hep-th]}}.

\bibitem{Giudice:1998xp}
G.~F. Giudice, M.~A. Luty, H.~Murayama, and R.~Rattazzi, ``{Gaugino mass
  without singlets},''
  \href{http://dx.doi.org/10.1088/1126-6708/1998/12/027}{{\em JHEP} {\bfseries
  12} (1998) 027},
\href{http://arxiv.org/abs/hep-ph/9810442}{{\ttfamily arXiv:hep-ph/9810442
  [hep-ph]}}.

\bibitem{CMS:2014bra}
{ CMS} collaboration, ``{Search for Displaced SUSY in Dilepton Final States},''
  \href{http://cds.cern.ch/record/1706154}{CMS-PAS-B2G-12-024} (2014).

\bibitem{CMSemuEfficiency}
{ CMS} collaboration, ``{Displaced SUSY Parametrisation Study For User},''
  \url{https://twiki.cern.ch/twiki/bin/view/CMSPublic/DisplacedSusyParametrisa%
tionStudyForUser} (2014).

\bibitem{Fox:2002bu}
P.~J. Fox, A.~E. Nelson, and N.~Weiner, ``{Dirac gaugino masses and supersoft
  supersymmetry breaking},'' {\em JHEP} {\bfseries 0208} (2002) 035,
\href{http://arxiv.org/abs/hep-ph/0206096}{{\ttfamily arXiv:hep-ph/0206096
  [hep-ph]}}.

\bibitem{Aad:2014gfa}
{ ATLAS} collaboration, G.~Aad {\em et~al.}, ``{Search for nonpointing and
  delayed photons in the diphoton and missing transverse momentum final state
  in 8 TeV $pp$ collisions at the LHC using the ATLAS detector},''
  \href{http://dx.doi.org/10.1103/PhysRevD.90.112005}{{\em Phys. Rev.}
  {\bfseries D90} no.~11, (2014) 112005},
\href{http://arxiv.org/abs/1409.5542}{{\ttfamily arXiv:1409.5542 [hep-ex]}}.

\bibitem{CMS:2015sjc}
``{Search for long-lived neutral particles in the final state of delayed
  photons and missing energy in proton-proton collisions at $\sqrt s$ = 8
  TeV},'' \url{http://cds.cern.ch/record/2063495} (2015).

\bibitem{Tsai:1971vv}
Y.-S. Tsai, ``{Decay Correlations of Heavy Leptons in $e^+ e^-\to
  \ell^+\ell^-$},'' \href{http://dx.doi.org/10.1103/PhysRevD.13.771,
  10.1103/PhysRevD.4.2821}{{\em Phys. Rev.} {\bfseries D4} (1971) 2821}.
[Erratum: Phys. Rev.D13,771(1976)].

\bibitem{Hagiwara:1989fn}
K.~Hagiwara, A.~D. Martin, and D.~Zeppenfeld, ``{Tau Polarization Measurements
  at LEP and SLC},''
\href{http://dx.doi.org/10.1016/0370-2693(90)90120-U}{{\em Phys. Lett.}
  {\bfseries B235} (1990) 198--202}.

\bibitem{Agashe:2014kda}
{ Particle Data Group} collaboration, K.~A. Olive {\em et~al.}, ``{Review of
  Particle Physics},''
\href{http://dx.doi.org/10.1088/1674-1137/38/9/090001}{{\em Chin. Phys.}
  {\bfseries C38} (2014) 090001}.

\bibitem{Sarid:1999zx}
U.~Sarid and S.~D. Thomas, ``{Mesino -- anti-mesino oscillations},''
  \href{http://dx.doi.org/10.1103/PhysRevLett.85.1178}{{\em Phys.Rev.Lett.}
  {\bfseries 85} (2000) 1178},
\href{http://arxiv.org/abs/hep-ph/9909349}{{\ttfamily arXiv:hep-ph/9909349
  [hep-ph]}}.

\bibitem{Evans:2012bf}
J.~A. Evans and Y.~Kats, ``{LHC Coverage of RPV MSSM with Light Stops},''
  \href{http://dx.doi.org/10.1007/JHEP04(2013)028}{{\em JHEP} {\bfseries 1304}
  (2013) 028},
\href{http://arxiv.org/abs/1209.0764}{{\ttfamily arXiv:1209.0764 [hep-ph]}}.

\bibitem{CMS:2014wda}
{ CMS} collaboration, V.~Khachatryan {\em et~al.}, ``{Search for long-lived
  neutral particles decaying to quark-antiquark pairs in proton-proton
  collisions at $\sqrt{s} =$ 8 TeV},''
  \href{http://dx.doi.org/10.1103/PhysRevD.91.012007}{{\em Phys. Rev.}
  {\bfseries D91} no.~1, (2015) 012007},
\href{http://arxiv.org/abs/1411.6530}{{\ttfamily arXiv:1411.6530 [hep-ex]}}.

\bibitem{Chacko:2001km}
Z.~Chacko and E.~Ponton, ``{Yukawa deflected gauge mediation},''
  \href{http://dx.doi.org/10.1103/PhysRevD.66.095004}{{\em Phys. Rev.}
  {\bfseries D66} (2002) 095004},
\href{http://arxiv.org/abs/hep-ph/0112190}{{\ttfamily arXiv:hep-ph/0112190
  [hep-ph]}}.

\bibitem{Shadmi:2011hs}
Y.~Shadmi and P.~Z. Szabo, ``{Flavored Gauge-Mediation},''
  \href{http://dx.doi.org/10.1007/JHEP06(2012)124}{{\em JHEP} {\bfseries 06}
  (2012) 124},
\href{http://arxiv.org/abs/1103.0292}{{\ttfamily arXiv:1103.0292 [hep-ph]}}.

\bibitem{Calibbi:2014pza}
L.~Calibbi, A.~Mariotti, C.~Petersson, and D.~Redigolo, ``{Selectron NLSP in
  Gauge Mediation},'' \href{http://dx.doi.org/10.1007/JHEP09(2014)133}{{\em
  JHEP} {\bfseries 09} (2014) 133},
\href{http://arxiv.org/abs/1405.4859}{{\ttfamily arXiv:1405.4859 [hep-ph]}}.

\bibitem{Evans:2015swa}
J.~A. Evans, D.~Shih, and A.~Thalapillil, ``{Chiral Flavor Violation from
  Extended Gauge Mediation},''
  \href{http://dx.doi.org/10.1007/JHEP07(2015)040}{{\em JHEP} {\bfseries 07}
  (2015) 040},
\href{http://arxiv.org/abs/1504.00930}{{\ttfamily arXiv:1504.00930 [hep-ph]}}.

\bibitem{Adam:2013mnn}
{ MEG} collaboration, J.~Adam {\em et~al.}, ``{New constraint on the existence
  of the $\mu^+ \to e^+\gamma$ decay},''
  \href{http://dx.doi.org/10.1103/PhysRevLett.110.201801}{{\em Phys. Rev.
  Lett.} {\bfseries 110} (2013) 201801},
\href{http://arxiv.org/abs/1303.0754}{{\ttfamily arXiv:1303.0754 [hep-ex]}}.

\bibitem{Jelinski:2014uba}
T.~Jelinski and J.~Pawelczyk, ``{Masses and FCNC in Flavoured GMSB scheme},''
\href{http://arxiv.org/abs/1406.4001}{{\ttfamily arXiv:1406.4001 [hep-ph]}}.

\bibitem{Calibbi:2014yha}
L.~Calibbi, P.~Paradisi, and R.~Ziegler, ``{Lepton Flavor Violation in Flavored
  Gauge Mediation},''
  \href{http://dx.doi.org/10.1140/epjc/s10052-014-3211-x}{{\em Eur. Phys. J.}
  {\bfseries C74} no.~12, (2014) 3211},
\href{http://arxiv.org/abs/1408.0754}{{\ttfamily arXiv:1408.0754 [hep-ph]}}.

\bibitem{Meade:2008wd}
P.~Meade, N.~Seiberg, and D.~Shih, ``{General Gauge Mediation},''
  \href{http://dx.doi.org/10.1143/PTPS.177.143}{{\em Prog. Theor. Phys. Suppl.}
  {\bfseries 177} (2009) 143--158},
\href{http://arxiv.org/abs/0801.3278}{{\ttfamily arXiv:0801.3278 [hep-ph]}}.

\bibitem{Kraml:2007sx}
S.~Kraml and D.~T. Nhung, ``{Three-body decays of sleptons in models with
  non-universal Higgs masses},''
  \href{http://dx.doi.org/10.1088/1126-6708/2008/02/061}{{\em JHEP} {\bfseries
  02} (2008) 061},
\href{http://arxiv.org/abs/0712.1986}{{\ttfamily arXiv:0712.1986 [hep-ph]}}.

\bibitem{Agrawal:2011ze}
P.~Agrawal, S.~Blanchet, Z.~Chacko, and C.~Kilic, ``{Flavored Dark Matter, and
  Its Implications for Direct Detection and Colliders},''
  \href{http://dx.doi.org/10.1103/PhysRevD.86.055002}{{\em Phys. Rev.}
  {\bfseries D86} (2012) 055002},
\href{http://arxiv.org/abs/1109.3516}{{\ttfamily arXiv:1109.3516 [hep-ph]}}.

\bibitem{Hall:2009bx}
L.~J. Hall, K.~Jedamzik, J.~March-Russell, and S.~M. West, ``{Freeze-In
  Production of FIMP Dark Matter},''
  \href{http://dx.doi.org/10.1007/JHEP03(2010)080}{{\em JHEP} {\bfseries 03}
  (2010) 080},
\href{http://arxiv.org/abs/0911.1120}{{\ttfamily arXiv:0911.1120 [hep-ph]}}.

\bibitem{Boyarsky:2008xj}
A.~Boyarsky, J.~Lesgourgues, O.~Ruchayskiy, and M.~Viel, ``{Lyman-alpha
  constraints on warm and on warm-plus-cold dark matter models},''
  \href{http://dx.doi.org/10.1088/1475-7516/2009/05/012}{{\em JCAP} {\bfseries
  0905} (2009) 012},
\href{http://arxiv.org/abs/0812.0010}{{\ttfamily arXiv:0812.0010 [astro-ph]}}.

\bibitem{Boyarsky:2008ju}
A.~Boyarsky, O.~Ruchayskiy, and D.~Iakubovskyi, ``{A Lower bound on the mass of
  Dark Matter particles},''
  \href{http://dx.doi.org/10.1088/1475-7516/2009/03/005}{{\em JCAP} {\bfseries
  0903} (2009) 005},
\href{http://arxiv.org/abs/0808.3902}{{\ttfamily arXiv:0808.3902 [hep-ph]}}.

\bibitem{Chung:1998rq}
D.~J.~H. Chung, E.~W. Kolb, and A.~Riotto, ``{Production of massive particles
  during reheating},'' \href{http://dx.doi.org/10.1103/PhysRevD.60.063504}{{\em
  Phys. Rev.} {\bfseries D60} (1999) 063504},
\href{http://arxiv.org/abs/hep-ph/9809453}{{\ttfamily arXiv:hep-ph/9809453
  [hep-ph]}}.

\bibitem{Giudice:2000ex}
G.~F. Giudice, E.~W. Kolb, and A.~Riotto, ``{Largest temperature of the
  radiation era and its cosmological implications},''
  \href{http://dx.doi.org/10.1103/PhysRevD.64.023508}{{\em Phys. Rev.}
  {\bfseries D64} (2001) 023508},
\href{http://arxiv.org/abs/hep-ph/0005123}{{\ttfamily arXiv:hep-ph/0005123
  [hep-ph]}}.

\bibitem{Co:2015pka}
R.~T. Co, F.~D'Eramo, L.~J. Hall, and D.~Pappadopulo, ``{Freeze-In Dark Matter
  with Displaced Signatures at Colliders},''
  \href{http://dx.doi.org/10.1088/1475-7516/2015/12/024}{{\em JCAP} {\bfseries
  1512} no.~12, (2015) 024},
\href{http://arxiv.org/abs/1506.07532}{{\ttfamily arXiv:1506.07532 [hep-ph]}}.

\bibitem{Feng:2015wqa}
J.~L. Feng, S.~Iwamoto, Y.~Shadmi, and S.~Tarem, ``{Long-Lived Sleptons at the
  LHC and a 100 TeV Proton Collider},''
\href{http://arxiv.org/abs/1505.02996}{{\ttfamily arXiv:1505.02996 [hep-ph]}}.

\bibitem{CMS:2015kdx}
{ CMS} collaboration, ``{Search for Long-lived Charged Particles in
  Proton-Proton Collisions at $\sqrt{s}=13$ TeV},''
  \url{http://cds.cern.ch/record/2114818} (2015).

\bibitem{Khachatryan:2015uqb}
{ CMS} collaboration, V.~Khachatryan {\em et~al.}, ``{Measurement of the top
  quark pair production cross section in proton-proton collisions at
  $\sqrt{s}=13$ TeV},''
\href{http://arxiv.org/abs/1510.05302}{{\ttfamily arXiv:1510.05302 [hep-ex]}}.

\bibitem{Campbell:2006wx}
J.~M. Campbell, J.~W. Huston, and W.~J. Stirling, ``{Hard Interactions of
  Quarks and Gluons: A Primer for LHC Physics},''
  \href{http://dx.doi.org/10.1088/0034-4885/70/1/R02}{{\em Rept. Prog. Phys.}
  {\bfseries 70} (2007) 89},
\href{http://arxiv.org/abs/hep-ph/0611148}{{\ttfamily arXiv:hep-ph/0611148
  [hep-ph]}}.

\bibitem{Bai:2014osa}
Y.~Bai and J.~Berger, ``{Lepton Portal Dark Matter},''
  \href{http://dx.doi.org/10.1007/JHEP08(2014)153}{{\em JHEP} {\bfseries 08}
  (2014) 153},
\href{http://arxiv.org/abs/1402.6696}{{\ttfamily arXiv:1402.6696 [hep-ph]}}.

\bibitem{Christensen:2008py}
N.~D. Christensen and C.~Duhr, ``{FeynRules - Feynman rules made easy},''
  \href{http://dx.doi.org/10.1016/j.cpc.2009.02.018}{{\em Comput. Phys.
  Commun.} {\bfseries 180} (2009) 1614--1641},
\href{http://arxiv.org/abs/0806.4194}{{\ttfamily arXiv:0806.4194 [hep-ph]}}.

\bibitem{Graham:2012th}
P.~W. Graham, D.~E. Kaplan, S.~Rajendran, and P.~Saraswat, ``{Displaced
  Supersymmetry},'' \href{http://dx.doi.org/10.1007/JHEP07(2012)149}{{\em JHEP}
  {\bfseries 07} (2012) 149},
\href{http://arxiv.org/abs/1204.6038}{{\ttfamily arXiv:1204.6038 [hep-ph]}}.

\bibitem{Kofman:1997yn}
L.~Kofman, A.~D. Linde, and A.~A. Starobinsky, ``{Towards the theory of
  reheating after inflation},''
  \href{http://dx.doi.org/10.1103/PhysRevD.56.3258}{{\em Phys. Rev.} {\bfseries
  D56} (1997) 3258--3295},
\href{http://arxiv.org/abs/hep-ph/9704452}{{\ttfamily arXiv:hep-ph/9704452
  [hep-ph]}}.

\bibitem{Monteux:2015qqa}
A.~Monteux and C.~S. Shin, ``{Thermal Goldstino Production with Low Reheating
  Temperatures},'' \href{http://dx.doi.org/10.1103/PhysRevD.92.035002}{{\em
  Phys. Rev.} {\bfseries D92} (2015) 035002},
\href{http://arxiv.org/abs/1505.03149}{{\ttfamily arXiv:1505.03149 [hep-ph]}}.

\end{thebibliography}\endgroup

\end{document}